\documentclass[pdflatex,sn-mathphys-num]{sn-jnl}

\usepackage{graphicx}%
\usepackage{multirow}%
\usepackage{amsmath,amssymb,amsfonts}%
\usepackage{amsthm}%
\usepackage{mathrsfs}%
\usepackage[title]{appendix}%
\usepackage{xcolor}%
\usepackage{textcomp}%
\usepackage{manyfoot}%
\usepackage{booktabs}%
\usepackage{algorithm}%
\usepackage{algorithmicx}%
\usepackage{algpseudocode}%
\usepackage{listings}%

\theoremstyle{thmstyleone}%
\theoremstyle{thmstyletwo}%
\theoremstyle{thmstylethree}%

\newcommand{\noi}{\noindent}
\newcommand{\mybox}[2]{{\color{$\#$ 1}\fbox{\normalcolor$\#$ 2}}}

\begin{document}

\title[Article Title]{Computational Physics Applied

to Photonic Devices}


\author[1]{\fnm{Gian-Luca} \sur{Oppo}}\email{g.l.oppo@strath.ac.uk}
\equalcont{This author contributed equally to this work.}

\affil[1]{\orgdiv{Department of Physics}, \orgname{University of Strathclyde}, \orgaddress{\street{107 Rottenrow}, \city{Glasgow}, \postcode{G4 0NG}, \state{Scotland}, \country{U.K}}}


\abstract{We all know that the first laser device was realised by Theodore Maiman at Hughes Labs in 1960. Less known is that the very first computer simulations of the relaxation oscillations displayed by Maiman's laser were also performed in 1960 on a digital IBM 704 computer. The reason is that lasers and almost all photonic devices are described by nonlinear equations that are more often than not impossible to be solved analytically, i.e. on a piece of paper. Since then the development and applications of lasers and photonic devices has progressed hand in hand with computer simulations and numerical programming. In this review we introduce and numerically solve the model equations for a variety of devices, lasers, lasers with modulated parameters, lasers with injection, Kerr resonators, saturable absorbers and optical parametric oscillators. By using computer simulations we demonstrate stability and instability of nonlinear solutions in these photonic devices via pitchfork, saddle-node, Hopf and Turing bifurcations; bistability, nonlinear oscillations, deterministic chaos, Turing patterns, conservative solitons; bright, dark and grey cavity solitons; frequency combs, spatial disorder, spatio-temporal chaos, defect mediated turbulence and even rogue waves. There has been a one-to-one correspondence between computer simulations of all these nonlinear features and laboratory experiments with applications in ultrafast optical communications, optical memories, neural networks, frequency standards, optical clocks, future GPS, astronomy and quantum technologies. All of this has been made possible by 'novel insights into spatio-temporal dynamics of lasers, nonlinear and quantum optical systems, achieved through the development and application of powerful techniques for small-scale computing' (2011 Occhialini Medal and Prize of the Institute of Physics and Societa' Italiana di Fisica).}




\maketitle



\section*{List of Acronyms}
\begin{itemize}
\item[•] AH = Andronov-Hopf (introduced in Subsection \ref{subsec:Bifurc})
\item[•] BEC = Bose-Einstein Condensate (introduced in Section \ref{sec:NLSE})
\item[•] CGLE = Complex Ginzburg-Landau Equation (introduced in Subsection \ref{subsect:Chaos})
\item[•] CS = Cavity Soliton (introduced in Section \ref{sec:STDkerr})
\item[•] CW = Continuous Wave (introduced in Subsection \ref{subsec:DarkNLSE})
\item[•] DMT = Defect Mediated Turbulence (introduced in Subsection \ref{subsect:Chaos})
\item[•] DOPO = Degenerate Optical Parametric Oscillator (introduced in Section \ref{sec:TDopo})
\item[•] DW = Domain Wall (introduced in Section \ref{sec:STDopo})
\item[•] GPE = Gross-Pitaevskii Equation (introduced in Section \ref{sec:NLSE})
\item[•] HSS = Homogeneous Stationary States (introduced in Subsection \ref{subsec:Turing})
\item[•] HOT = Higher Order Terms (introduced in Appendix I, Section \ref{sec:AppI})
\item[•] LCS = Laser Cavity Solitons (introduced in Subsection \ref{subsec:LNested})
\item[•] LEF = Linewidth Enhancement Factor (introduced in Subsection \ref{subsec:RO})
\item[•] LLE = Lugiato-Lefever Equation (introduced in Section \ref{sec:STDkerr})
\item[•] NLSE = Non-Linear Schr\"odinger Equation (introduced in Section \ref{sec:NLSE})
\item[•] ODE = Ordinary Differential Equations (introduced in Section \ref{sec:intro})
\item[•] OPO = Optical Parametric Oscillator (introduced in Section \ref{sec:TDopo})
\item[•] PDC = Parametric Down Conversion (introduced in Section \ref{sec:TDopo})
\item[•] PDE = Partial Differential Equations (introduced in Section \ref{sec:intro})
\item[•] RO = Relaxation Oscillations (introduced in Section \ref{sec:intro})
\item[•] SHE = Swift-Hohenberg Equation (introduced in Subsection \ref{subsec:Turing})
\item[•] SIF = Societa' Italiana di Fisica (introduced in Section \ref{sec:intro})

\end{itemize} 

\newpage
\section{Introduction}\label{sec:intro}
The year 1960, when I was one year old, has been pivotal for the development of laser physics and nonlinear optics. On one side Theodore Maiman at Hughes Laboratories realised the first laser (Light Amplification by Stimulated Emission of Radiation) by using a ruby rod \cite{Maiman60}. On the other hand H. Statz and G. deMars reported on the theory of maser (later laser) rate equations \cite{Statz60}. The rate equations model of Statz and deMars was introduced to describe Relaxation Oscillations (RO) to equilibrium observed experimentally in Ruby Masers \cite{Kikuchi59} but also in Quartz and MgO Masers \cite{Chester58}. It was later shown in \cite{Tang63} that the rate equations model of Statz and deMars would equally apply to RO observed in solid-state lasers \cite{Collins60} as lasers were considered to be just masers emitting in the visible part of the electromagnetic spectrum. The work of Statz and deMars is fundamental in two important ways: 1) it shows that models for lasers (and for all photonic devices considered here) contain nonlinear terms and 2) numerical integrations/simulations are required to obtain plots that reproduce the experimentally observed behaviour. The second issue is somewhat less well known that the first one. All the plots displaying the RO of the energy and population density in the medium were obtained by using an analogue computer in \cite{Statz60} and a digital IBM 704 computer in \cite{Tang63}. As a matter of fact A. Yariv had used an IBM 704 computer at Bell Labs to describe modulations of the emitted power in masers in 1960 as reported in the same book of the Statz and deMars original paper \cite{Singer60}. The understanding of the mechanisms behind RO in lasers led to the discovery and realization of giant optical pulses via the Q-switching technique by R. Hellwarth in 1961 \cite{Hellwarth61}.

These historical considerations demonstrate that optical nonlinearity and computational physics, the topics of this review article, are key ingredients for an accurate description of the behaviour of lasers and many other photonic devices. I would like to start with some apologies. The research fields covered here span across photonics, laser physics, nonlinear optics, computational physics, dynamical systems, complex systems and quantum optics. The list of references could be enormous and distract the reader from the central topics of the review. Moreover, on the occasion of the 2011 Occhialini Medal and Prize of the UK Institute of Physics and the Societa' Italiana di Fisica (SIF) in L'Aquila \cite{Occhialini11}, the SIF President, Prof. Luisa Cifarelli aksed me to write a review paper about my area of research. The motivation of the 2011 Occhialini Medal and Prize reads: 'For novel insights into spatio-temporal dynamics of lasers, nonlinear and quantum optical systems, achieved through the development and application of powerful techniques for small-scale computing'. For these reasons, many of the references listed in this review paper are taken from my own research work over the last 40 years of activity. I apologise for the hundreds of papers and book chapters that are not cited here. Many of these missed references can be found however in the cited work listed in the bibliography below. Moreover, as I concentrate on the computational aspects of the description of photonic devices, detailed derivation of the model equations are left to the existent rich literature (see e.g. \cite{LugiatoBook,NarducciAbrahamBook,Oppo24}). 

The review paper is organised as follows. Few elements of dynamical systems theory, including bifurcations, Turing instabilities, deterministic chaos and spatio-temporal turbulence are reviewed in Section \ref{sec:background} to provide the background to many stationary and dynamical states that are observed in the theory and experiments of photonic devices. The remainder of the work is separated into two parts: Part A and Part B deal with temporal and spatio-temporal descriptions of photonic devices, respectively. The reason behind this choice is that the systems in Part A are described by Ordinary Differential Equations (ODEs) while those in Part B by Partial Differential Equations (PDEs). In Section \ref{sec:TDlaser} we study nonlinear laser models leading to relaxation oscillations, onset of deterministic chaos, and with external modulation or injection. We then extend this analysis to passive and Kerr cavities (see Section \ref{sec:TDkerr}) and to optical parametric oscillators (see Section \ref{sec:TDopo}). This completes Part A. In Part B we consider systems with variables that depend on more than one coordinate (also known as independent variables). This is why the mathematical description moves from a purely temporal evolution (see Part A) to a spatio-temporal evolution described by PDEs. We start with pure propagation of light in a Kerr medium described by the Nonlinear Schr\"odinger equation in Section \ref{sec:NLSE} leading to the formation of bright and dark solitons. 
Nonlinear Kerr cavities, the renown Lugiato-Lefever equation and cavity solitons are reviewed in Section \ref{sec:STDkerr}. Spatio-temporal phenomena in spatially extended lasers and optical paramateric oscillators are described in Sections \ref{sec:STDlaser} and \ref{sec:STDopo}. Final remarks are presented in Section \ref{sec:Conclusions} while examples of numerical codes in Python for ODEs and PDEs are provided in Appendices I and II, respectively. Note that all numerical codes used in this review \cite{Note1} have been run on 'small-scale computing' architectures, i.e. today's standard laptops, deliberately avoiding expensive, planet warming, time wasting and often useless large scale computations that are typical of 'ab initio' 
approaches to photonics.

\section{Basic Elements of the Theory of Dynamical Systems}\label{sec:background}
Differential equations describing the evolution of photonic devices are intrinsically nonlinear. This is trivially demonstrated in any everyday laser device such as for example a laser pointer. By changing a parameter, the input energy from a battery, the laser pointer shows two possible 'equilibrium' states, the laser off and laser on states. This is due to nonlinearity and this is why the evolution of the physical variables describing photonic devices is rooted in the theory of nonlinear dynamical systems. Several of the states of operation of photonic devices starting from those of lasers, are described by bifurcations, i.e. drastic changes in the behaviour of a system of nonlinear differential equations upon the variation of a control parameter. In both Part A and Part B of this review paper, we make extensive use of basic bifurcations (transcritical, pitchfork, saddle-node, Andronov-Hopf, period doubling) and of chaotic states of ODEs  from dynamical systems theory \cite{StrogatzBook}, as well as Turing instabilities and spatio-temporal turbulence that are characteristic of PDEs.   

\subsection{Bifurcations}\label{subsec:Bifurc}
\begin{figure}[h]
\centering
\includegraphics[width=0.49\linewidth]{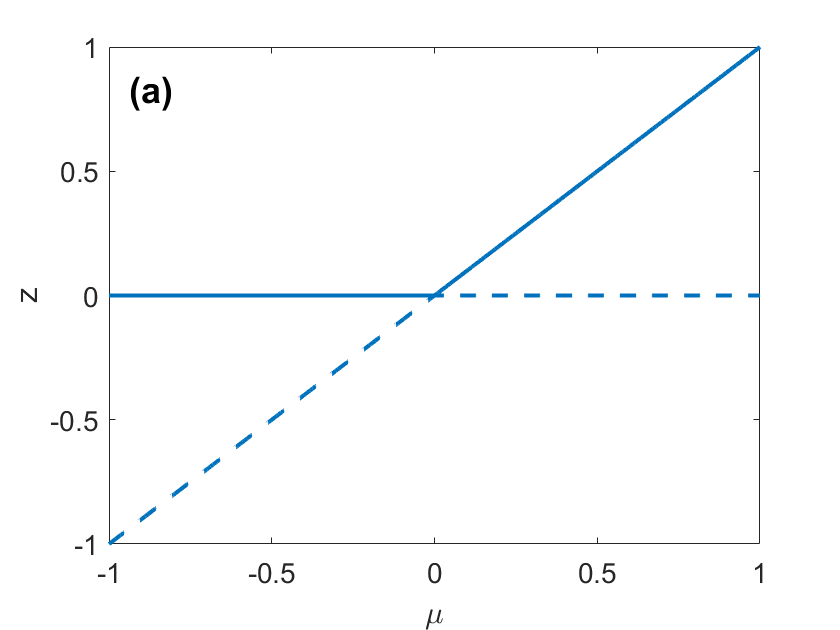}
\includegraphics[width=0.49\linewidth]{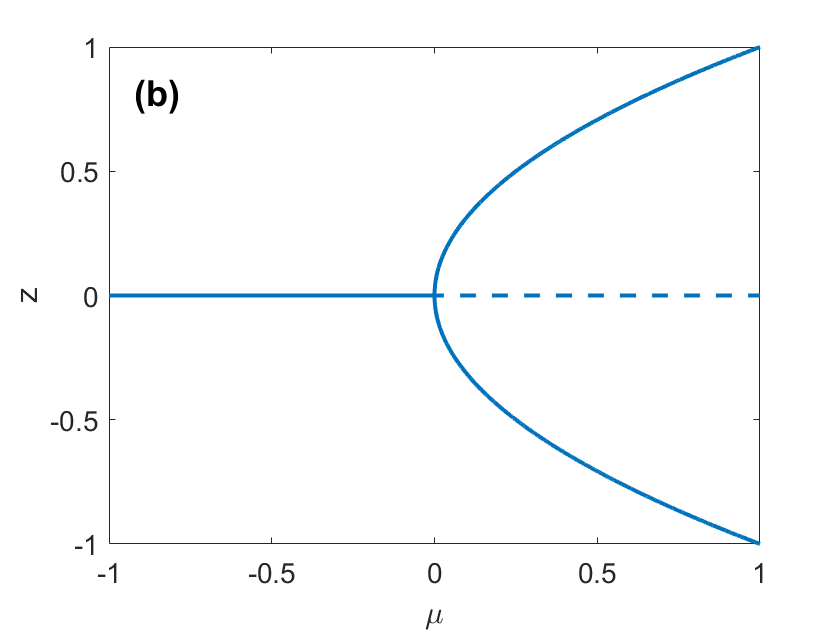}
\caption{(Color online) (a) Bifurcation diagram of a transcritical bifurcation. (b) Bifurcation diagram of a supercritical pitchfork bifurcation. Solid (dashed) lines correspond to stable (unstable) stationary states.}
\label{fig:TranPitch}
\end{figure}

{\it {\bf Transcritical Bifurcation.}}
In a transcritical bifurcation two stationary states exchange their stability when a control parameter is varied. Both before and after the bifurcation, there is one stable and one unstable stationary state but their stability is exchanged at the bifurcation point. The unstable fixed point becomes stable and vice versa. The normal form (i.e. one of the simplest mathematical model displaying the essential features) of a transcritical bifurcation is
\begin{equation}
\frac{dz}{dt} = \mu z - z^2 
\label{TransBif}
\end{equation}
where $z(t)$ is a real variable that is a function of the time-coordinate $t$ while $\mu$ is a real control parameter. The two stationary states ($d{\bar z}/dt=0$) are ${\bar z}=0$ and ${\bar z}=\mu$. For $\mu<0$, the stationary state ${\bar z}=0$ is stable and the stationary state ${\bar z}=\mu$ is unstable. For $\mu>0$, the stationary state ${\bar z}=0$ is unstable and the stationary state ${\bar z}=\mu$ is stable. This can be seen by the linear stability analysis of the stationary states where a perturbation $\delta z = z - {\bar z}$ of a generic stationary state ${\bar z}$ evolves as
\begin{equation}
\frac{d(\delta z)}{dt} = \mu (\delta z) - 2 {\bar z} (\delta z) 
\end{equation}
where we have neglected all terms of the kind $(\delta z)^n$ with $n>1$ with $n$ being a positive integer (linearization). Clearly a perturbation $\delta z$ of the stationary state ${\bar z} = 0$ decays (grows) in time for $\mu<0$ ($\mu>0$). A perturbation $\delta z$ of the stationary state ${\bar z} = \mu$ grows (decays) in time for $\mu<0$ ($\mu>0$). See Fig. \ref{fig:TranPitch}(a) for a bifurcation diagram of the transcritical bifurcation showing the stability of the stationary states.\\

\noi {\it {\bf Pitchfork Bifurcation.}}
In a pitchfork bifurcation there is a single stationary state before the bifurcation and three stationary states (one being the original state present before the bifurcation) after the bifurcation. In the supercritical case, the single stationary state is stable (unstable) before (after) the bifurcation while two new stable stationary states appear at the bifurcation point. In the subcritical case, the single stationary state is unstable (stable) before (after) the bifurcation while two new unstable stationary states appear at the bifurcation point.

The normal form of a supercritical pitchfork bifurcation is
\begin{equation}
\frac{dz}{dt} = \mu z - z^3 
\label{SupPitchBif}
\end{equation}
where $z(t)$ is a real variable that is a function of the time-coordinate $t$ while $\mu$ is a real control parameter. Before the bifurcation at $\mu=0$ there is only one stationary state ${\bar z}=0$. After the bifurcation there are two additional stationary states ${\bar z} = \pm \sqrt{\mu}$ that only exist for $\mu>0$. The linear stability analysis of the stationary states for a perturbation $\delta z = z - {\bar z}$ of a generic stationary state ${\bar z}$ provides
\begin{equation}
\frac{d(\delta z)}{dt} = \mu (\delta z) - 3 {\bar z}^2 (\delta z) 
\end{equation}
where we have neglected all terms of the kind $(\delta z)^n$ with $n>1$. Clearly a perturbation $\delta z$ of the stationary state ${\bar z} = 0$ decays (grows) in time for $\mu<0$ ($\mu>0$). A perturbation $\delta z$ of the stationary states ${\bar z} = \pm \sqrt{\mu}$ decays in time for $\mu>0$. See Fig. \ref{fig:TranPitch}(b) for a bifurcation diagram of the supercritical pitchfork bifurcation.

The normal form of a subcritical pitchfork bifurcation is
\begin{equation}
\frac{dz}{dt} = \mu z + z^3 
\label{SubPitchBif}
\end{equation}
with simple stability considerations leading to the stationary state ${\bar z} = 0$ being stable (unstable) in time for $\mu<0$ ($\mu>0$) while the stationary states ${\bar z} = \pm \sqrt{-\mu}$ exist and are unstable for $\mu<0$. 
In Section \ref{sec:TDlaser}, we are going to see that the laser threshold is a pitchfork bifurcation that looks like a transcritical bifurcation when considering the light intensity as a dynamical variable.\\

\begin{figure}[t]
\centering
\includegraphics[width=0.51\linewidth]{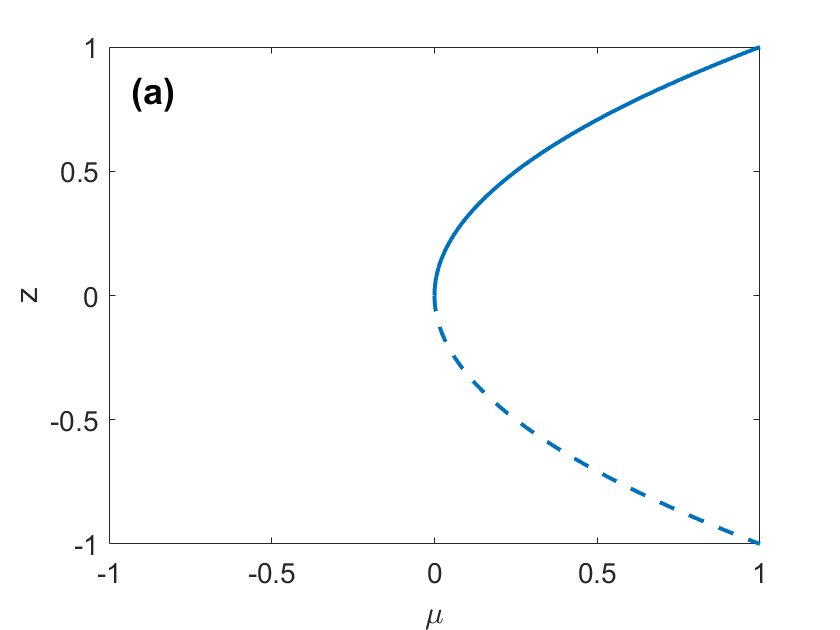}
\includegraphics[width=0.46\linewidth]{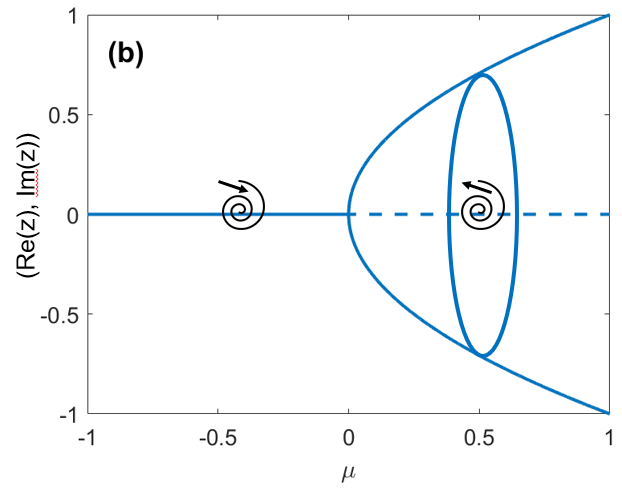}
\caption{(Color online) (a) Bifurcation diagram of a saddle-node bifurcation. (b) Bifurcation diagram of an Andronov-Hopf bifurcation. The plane perpendicular to the figure is the $(Re(z), Im(z))$ plane. For $\mu<0$ the trajectory spirals towards the stationary state $z=0$, for $\mu>0$ the trajectory spirals out of the stationary state $z=0$ and relaxes onto the limit cycle (blue oval curve) corresponding to a periodic oscillation. Solid (dashed) lines correspond to stable (unstable) stationary states.}
\label{fig:SN_HB}
\end{figure}

\noi {\it {\bf Saddle-Node Bifurcation.}}
In a saddle-node (or blue-sky) bifurcation, two stationary states suddenly appear when changing a control parameter $\mu$. One of the stationary state is stable (node), while the other is unstable (saddle). The normal form of a saddle-node bifurcation is:
\begin{equation}
\frac{dz}{dt} = \mu - z^2 
\label{SNBif}
\end{equation}
with obvious meaning of all symbols. For $\mu<0$ there are no stationary states, while for $\mu>0$ there are two stationary states ${\bar z} = \pm \sqrt{\mu}$. The linear stability analysis of these stationary states provides:
\begin{equation}
\frac{d(\delta z)}{dt} = - 2 {\bar z} (\delta z) 
\end{equation}
showing that a perturbation $\delta z$ of the stationary state ${\bar z} = + \sqrt{\mu}$ decays in time indicating stability while a perturbation $\delta z$ of the stationary state ${\bar z} = - \sqrt{\mu}$ grows in time indicating instability. See Fig. \ref{fig:SN_HB}(a) for a bifurcation diagram of the saddle-node bifurcation.
In Section \ref{sec:TDkerr}, we are going to see that the bending of the cavity resonance curve in Kerr resonators introduces saddle-node bifurcations leading to optical bistability.\\

\noi {\it {\bf Andronov-Hopf Bifurcation.}}
An Andronov-Hopf (AH) bifurcation occurs when the variation of a parameter of the system causes the set of solutions (trajectories) to change from being attracted to (or repelled by) a stationary solution, to become attracted to (or repelled by) an oscillatory, periodic solution. Since the minimum size of the phase space (minimum number of independent variables) in a continuous time dynamical system for oscillatory behaviour is two, the AH bifurcation is a two-dimensional analogue of the pitchfork bifurcation seen above, including possible supercritical and subcritical features. 

The normal form of a supercritical AH bifurcation is:
\begin{equation}
\frac{dz}{dt} = (\mu + i \omega) z - |z|^2 \, z 
\label{AHBif}
\end{equation}
where $z(t)$ is now a time dependent complex variable, $\mu$ and $\omega$ are real parameters, $|z|^2$ is the amplitude square of $z$ and $i$ is the imaginary unit. There is a stationary solution for $z=Re(z)+iIm(z)$ where both the real and imaginary part of $z$ are zero. By using the amplitude $|z|=A>0$ and phase $\phi(t)$, $z= A \exp(i \phi)$ one can rewrite Eq. (\ref{AHBif}) as:
\begin{align}
\frac{dA}{dt} &= \left( \mu - A^2 \right) \, A \nonumber \\
\frac{d\phi}{dt} &=& \omega \label{AHphaseBif}
\end{align}
which shows that for $A \ne 0$ the solution oscillates in time as the phase is given by $\phi = \phi_0 + \omega t$. For $\mu>0$ there is a stationary solution with amplitude $A=\sqrt{\mu}$. The linear stability analysis of the amplitude equation tells us that this solution is stable and attractive. Hence the solution $z=\sqrt{\mu} \exp(i (\phi_0 + \omega t))$ corresponds to an asymptotic oscillatory state of the system, generally referred to as a 'limit cycle'. The bifurcation diagram is presented in Fig. \ref{fig:SN_HB}(b). By changing the sign in front of the nonlinear term in Eq. (\ref{AHBif}) one can study the subcritical AH bifurcation.

To our knowledge the first mathematical description of an AH instability is due to James Clerk Maxwell in his 1868 paper the stability of governors ('part of a machine by means of which the velocity of the machine is kept nearly uniform, notwithstanding variations in the driving power or the resistance') \cite{Maxwell_1868}. Historically, this paper is considered the be the first systematic contribution to control theory and self-oscillations. \\

\noi {\it {\bf Period Doubling Bifurcation.}}
In dynamical systems theory, a period-doubling bifurcation occurs when a change in a system parameter causes a new periodic trajectory to emerge from an existing periodic trajectory, for example from an oscillation due to an AH bifurcation. The new periodic trajectory has a period twice that of the original one. With the doubled period, it takes twice as long for the numerical values visited by the system to repeat themselves.

A subnormal form of the period doubling bifurcation is the system of dynamical equations for a damped, forced pendulum with torque:
\begin{align}
\frac{d\theta}{dt} &= \Omega \nonumber \\
\frac{d\Omega}{dt} &= -\sin(\theta) -\alpha \Omega + \beta + \gamma \cos(\omega t) \label{FPDT}
\end{align}
where $\theta$ and $\Omega$ are two angular variables representing the position of the pendulum and its angular velocity, respectively, $\omega$ is the frequency of the forcing and is considered to be equal to one here for convenience. The parameters $\alpha$, $\beta$ and $\gamma$ are the damping, the torque, and the amplitude of the external forcing, respectively. A familiar system described by these equations is a vandalized swing in a children park. The equilibrium position in the absence of forcing ($\gamma=0$) is not $\theta=0$ but $\theta=\arcsin(\beta)$. Without external forcing and without damping ($\alpha=\gamma=0$) the model is just the angular description of a pendulum, historically the first oscillator of real technological importance. Galileo Galilei was the first to record the period of a swinging lamp high in a cathedral in Pisa in 1582. In 1657, Christian Huygens constructed the first pendulum clock, a vast improvement in timekeeping over all previous techniques. 

The numerical integration of Eqs. (\ref{FPDT}) can easily and accurately be done by using a Runge-Kutta method of the 4th order. Appendix I in Section \ref{sec:AppI} describes how to implement this standard method of computational physics for the determination of solutions of systems of nonlinear ODEs similar to Eqs. (\ref{FPDT}). Here we focus on a small set of dynamical oscillations of Eqs. (\ref{FPDT}) by considering $\omega=1$, $\alpha=\beta=0.1$ and changing the amplitude of the external forcing $\gamma$.  
\begin{figure}[h]
\centering
\includegraphics[width=0.49\linewidth]{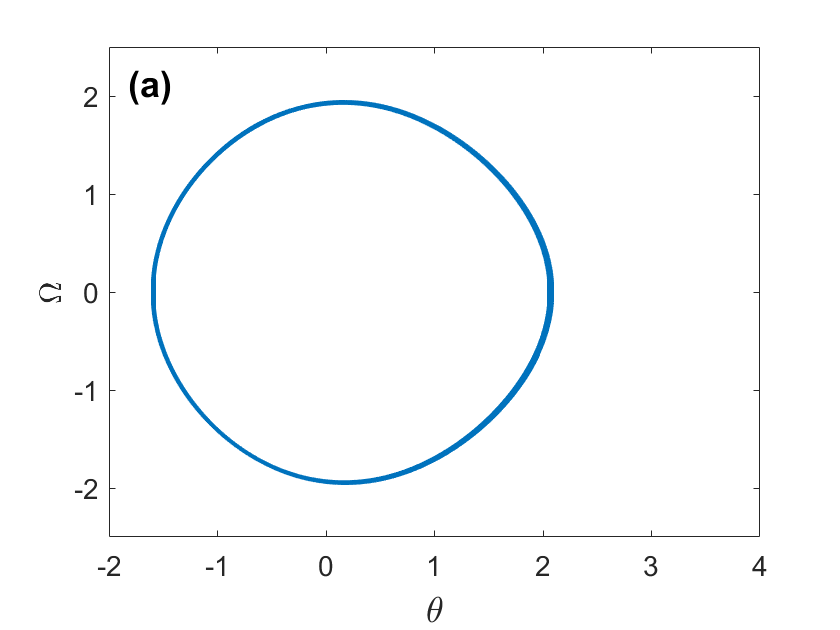}
\includegraphics[width=0.49\linewidth]{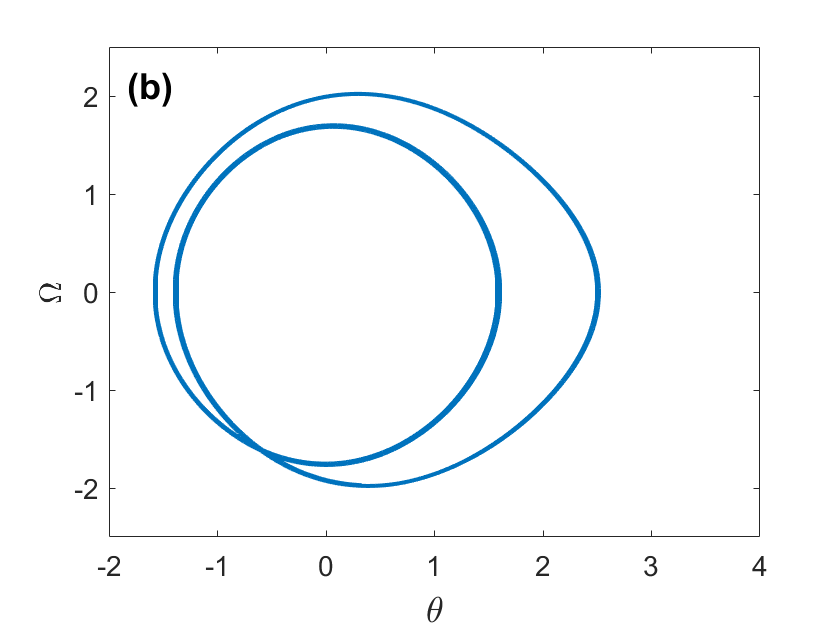}
\caption{(Color online) (a) Limit cycle oscillation of Eqs. (\ref{FPDT}) for $\gamma=0.752$, and $\alpha=\beta=0.1$. (b) Same as (a) but for $\gamma=0.768$. Period doubled orbit. }
\label{fig:PD}
\end{figure}
In Fig. \ref{fig:PD}(a) we can see that the external forcing typically induce steady long-term (asymptotic) oscillations in the swing pendulum after transients have been eliminated. By evaluating the Fourier transforms of the time traces of the variables, one can obtain power spectra from which frequency components, their nonlinear shifts from $\omega$ (the frequency of the external forcing) and their broadening can be evaluated. For the value $\gamma=0.752$ of Fig. \ref{fig:PD}(a), for example the frequency of the oscillations is very close to $\omega$ with a spectral broadening due to the damping $\alpha$, as expected. Things however change when increasing the value of $\gamma$ through a period doubling bifurcation. For $\gamma=0.768$, for example, a full oscillation of the swing pendulum takes twice the period of the oscillation observed for $\gamma=0.752$ as seen in Fig. \ref{fig:PD}(b). Note that this figure is a projection of a three dimensional orbit with no real crossings on the two dimensional projection plane. Eqs. (\ref{FPDT}) are non-autonomous ODEs where the time appears explicitly. By introducing a variable $\psi=\omega t$ and an auxiliary third differential equation $(d \psi / dt) = \omega$ the non-autonomous system is transformed into an autonomous system of three ODEs.

In Section \ref{sec:TDlaser} we will see the generation of chaotic states of laser emissions through a sequence of period doubling bifurcations\\

\subsection{Turing Instabilities}
\label{subsec:Turing}
Up to now we have focused on bifurcations in a single or a system of ODEs. These bifurcations have important roles also in PDEs that are crucial to the photonic systems described in Part B of this paper. There is however a bifurcation due to the instability of homogeneous stationary states that has no counterpart in ODEs. This is the Turing instability named after the renowned mathematician Alan Turing and his famous 1952 paper 'The Chemical Basis of Morphogenesis' \cite{Turing52}. In this paper Turing demonstrates that two chemical species diffusing into each other at different speeds can reach an equilibrium state where the distribution of the two reagents is not uniform but modulated in space. This phenomenon leads to the formation of so called 'Turing patterns' and is due to an instability of the homogeneous state when changing some control parameter. The final structures can be stationary patterns (regular spatial modulations), oscillating patterns or even spatio-temporally disordered structures. To make the description of Turing instabilities as simple as possible we investigate the one variable Swift-Hohenberg Equation (SHE).

The SHE is a single real variable, $u$, PDE, probably the simplest system to display pattern formation, introduced by J. Swift and P. Hohenberg to describe fluctuations close to a convective instability in fluids \cite{Swift77}. In one spatial dimension the SHE reads:
\begin{equation}
\partial_t u = \epsilon u  -  u^3 - \left( \partial_x^2+K_{crit}^2 \right)^2 u = 
\left( \epsilon - K_{crit}^4 \right) u - u^3 - \partial_x^4 u - 2 K_{crit}^2 \partial_x^2  u
\label{SHE}
\end{equation} 
where $u(x,t)$ is a function of the spatial coordinate $x$ and time $t$, $\partial_t$ is the partial derivative with respect to time, $\partial_x^2$ and $\partial_x^4$ the second and forth partial derivatives with respect to $x$, while $\epsilon$ and $K_{crit}$ are control parameters. For simplicity we select $K_{crit}=1$.
Homogeneous Stationary States (HSS) are obtained by setting all partial derivatives to zero and are given by $u_0=0$ and $u_{\pm}= \pm \sqrt{(\epsilon-1)}$. The $u_{\pm}$ states exist only if $\epsilon$ is bigger than one and one can check that there is a pitchfork bifurcation at $\epsilon=1$ where the zero HSS becomes unstable to the newly formed $\pm$ HSSs. In order to establish a Turing instability of the zero HSS to a spatially periodic structure we need to generalise the linear stability analysis of ODEs seen in the previous sections to bifurcations of PDEs.

\begin{figure}[h]
\centering
\includegraphics[width=0.49\linewidth]{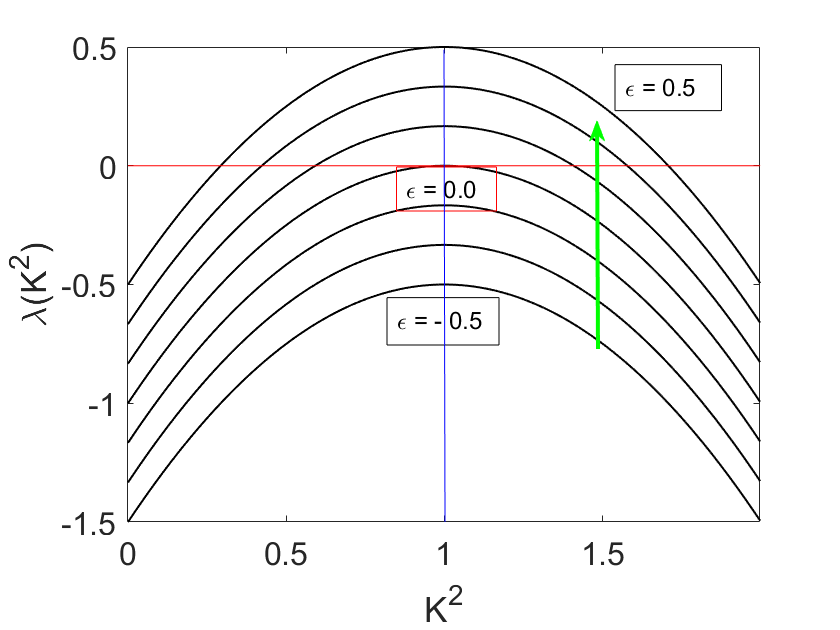}
\includegraphics[width=0.49\linewidth]{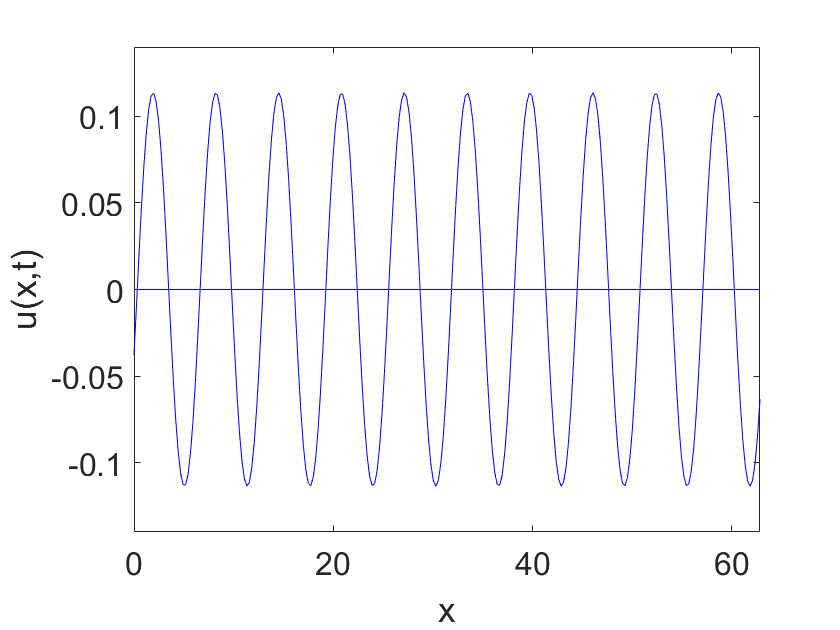}
\caption{(Color online) (a) Dispersion relation of the SHE, Eq. (\ref{lambdaSHE}), for different values of the control parameter $\epsilon$. (b) Stable Turing pattern of the SHE (\ref{SHE}) for $\epsilon=0.01$. The horizontal line is the unstable HSS $u_0=0$. }
\label{fig:DispRe_TP}
\end{figure}

We consider perturbations of the zero HSS of the form $\delta u = a \, \exp(\lambda t) \cos(K x)$ with $|a| \ll 1$ and with $\lambda$ (the temporal eigenvalue) and $K$ (the spatial wavevector) to be determined. By entering this expression in Eq. (\ref{SHE}) we obtain
\begin{equation}
\lambda a = \left( \epsilon - 1  - K^4 + 2 K^2 \right) a \;\;\;\;\;\;\; \Longleftrightarrow \;\;\;\;\;\; 
\lambda(K^2) = \epsilon - 1  - K^4 + 2 K^2
\label{lambdaSHE}
\end{equation}
where we have kept only the linear terms in $a$. Stability is obtained when the temporal eigenvalue $\lambda$ is negative. Note that these eigenvalues are now function of the square of the wavevector $K^2$ (dispersion relation). We can immediately see that for $K^2=0$, the zero solution $u_0=0$ is stable for $\epsilon<1$ since there is a pitchfork bifurcation at $\epsilon=1$ as seen above. In Fig. \ref{fig:DispRe_TP}(a) $\lambda(K^2)$ is plotted versus $K^2$ for different values of the parameter $\epsilon$.  
When increasing $\epsilon$ from negative to positive values there appears a band of wavevectors $K$ of the upside down parabola of the dispersion relation, corresponding to positive values of $\lambda$, i.e. unstable. This means that perturbations of the HSS $u_0=0$ with wavevectors $K$ in this region will grow in time. The instability at finite values of the wavevector $K$ is a Turing instability. The first wavevector to experience the Turing instability is known as the critical wavevector $K_{crit}$ and is obtained by finding the maximum of the dispersion relation parabola, $d \lambda/ d (K^2)=-2K^2+2=0$ leading to {$K_{crit}^2=1$. The growth of the $K$ modes corresponding to $\lambda>0$ saturates when the nonlinear term $-u^3$ in Eq. (\ref{SHE}) becomes the leading one. The long term (asymptotic) solution that is temporally stationary but spatially periodic is a Turing pattern whose spatial scale is given by $\Lambda=2\pi/K_{crit}= 2\pi$ (see Fig. \ref{fig:DispRe_TP}(b)). 

Fig. \ref{fig:DispRe_TP}(b) has been obtained via numerical simulations of Eq. (\ref{SHE}) using techniques for PDE developed in computational physics. In Appendix II in Section \ref{sec:AppII} we provide details and examples (such as the Lugiato-Lefever equation of Section \ref{sec:STDkerr}) of the split-step method used for all PDEs describing photonic devices in this paper.  
\subsection{Chaos and Turbulence} \label{subsect:Chaos}
As well as multiple stationary states, bifurcations, oscillations, period doubled oscillations and Turing patterns, ODEs and PDEs can display irregular temporal and spatio-temporal behaviours. The most well known disordered state is deterministic chaos as described in the next subsection. Spatio-temporal systems display a wider variety of disordered states, from spatio-temporal chaos to turbulence as explained in the second subsection here.\\

\noi {\it {\bf Deterministic Chaos.}}
In system described by ODEs, the onset and persistence of deterministic chaos has represented a major breakthrough since the 1960s. Roughly speaking deterministic chaos in (non-stochastic) ODEs corresponds to aperiodic solutions displaying a high sensitivity to initial conditions and a continuum background in the power spectrum. The relevance of high sensitivity to initial conditions in physical systems was already described by James Clerk Maxwell when presenting his kinetic theory of gases \cite{MaxwellChaos} and by Henri Poincare' and his three celestial body problem \cite{Poincare_1890}. However it is only with the precise numerical simulations of a system of nonlinear ODEs that Edward Lorenz demonstrated the full characteristics of deterministic chaos in his seminal work in 1963 \cite{Lorenz63}. We start from the Lorenz system of nonlinear ODEs:
\begin{align}
\frac{dX}{dt} &= -\sigma(X-Y) \nonumber \\
\frac{dY}{dt} &= \rho X - Y - XZ  \label{Lorenz}\\ 
\frac{dZ}{dt} &= -\beta Z + XY \nonumber
\end{align}
where $X$, $Y$, $Z$ are variables that depend on the time coordinate $t$, with $\sigma$, $\rho$ and $\beta$ being parameters. The Lorenz equations were introduced to describe the properties of a two-dimensional layer of fluid uniformly warmed from below and cooled from above, for atmospheric convection \cite{Lorenz63}. 
\begin{figure}[h]
\centering
\includegraphics[width=0.49\linewidth]{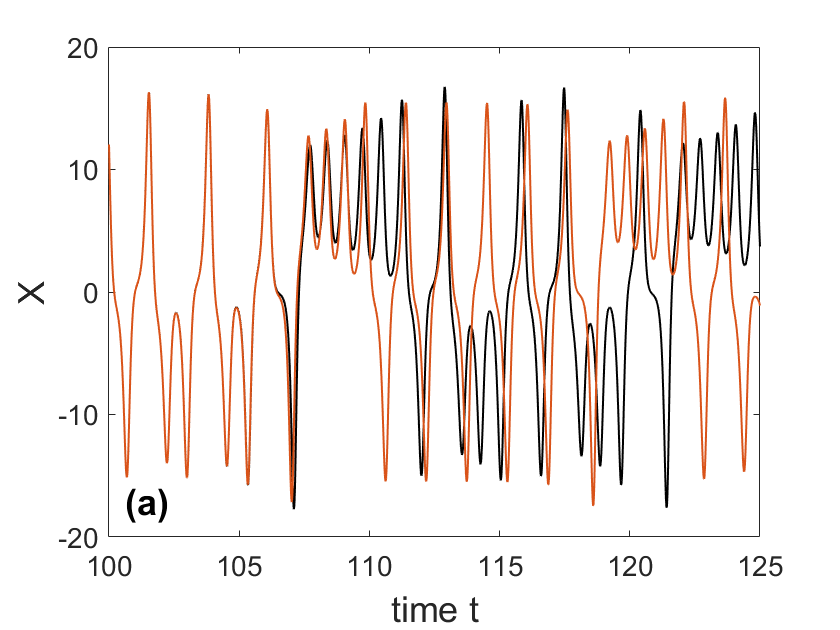}
\includegraphics[width=0.49\linewidth]{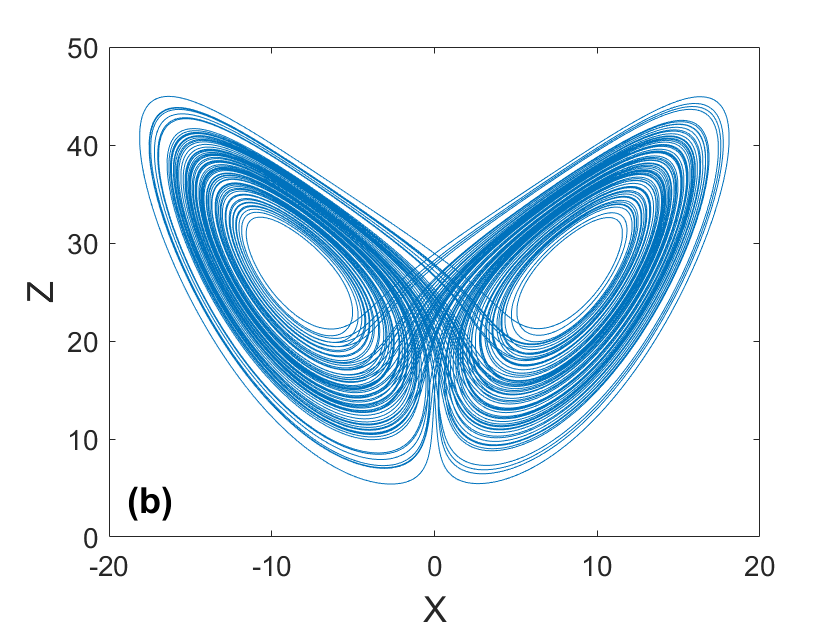}
\caption{(Color online) (a) Time evolution of the $X$ variable of the Lorenz equations Eqs. (\ref{Lorenz}) for $\sigma=10$, $\rho=28$, and $\beta=8/3$ for two trajectories initially displaced from each other by less than $5x 10^{-4}$. (b) Chaotic oscillations of Eqs. (\ref{Lorenz}) in a $(X,Z)$ projection plane.}
\label{fig:Lorenz_Chaos}
\end{figure}
The numerical integration of Eqs. (\ref{Lorenz}) here has been done by using a Runge-Kutta method of the 4th order as described in Appendix I. The results are shown in Fig. \ref{fig:Lorenz_Chaos} for the paradigmatic values of the parameters $\sigma=10$, $\rho=28$, and $\beta=8/3$. In Fig. \ref{fig:Lorenz_Chaos}(a) we show the variable $X$ versus time for two trajectories during a chaotic evolution and initially displaced from each other by less than $5 \times 10^{-4}$. At the beginning the two trajectories overlap with each other but after around $15$ time units they separate from each other leading to very different dynamics. If the initial separation was smaller than $5 \times 10^{-4}$, the overlap time would increase but eventually the two trajectories will separate from each other thus demonstrating high sensitivity from initial conditions and absence of asymptotic periodicity. In Fig. \ref{fig:Lorenz_Chaos}(b) the full beauty of the Lorenz attractor in a projection on the $(X,Z)$ plane is presented (transients have been discarded). In Section \ref{sec:TDlaser} we will see that single mode, perfectly tuned, mean-field laser equations are equivalent to the Lorenz model (\ref{Lorenz}) although the difference in parameter values makes chaos more difficult to achieve in photonic devices than in the Lorenz model. 

There are several dynamical routes to deterministic chaos including period doubling, intermittency, quasiperiodic and attractor crisis. Here we provide a simple description of the period doubling cascade to chaos while we refer the reader to specialised literature for the others \cite{StrogatzBook}. In Section \ref{subsec:Bifurc} we have introduced the period doubling bifurcation where an oscillatory state of a system of nonlinear ODEs can suddenly see its period to double upon the variation of a control parameter. In particular in Fig. \ref{fig:PD} we showed the period doubling bifurcation for a forced, damped pendulum with torque. An intriguing characteristic of period doubling bifurcations in nonlinear systems is that often, but not always, the first period doubling bifurcation is followed when changing the control parameter by a second period doubling bifurcation and then a third one and so on in what is known as a period doubling cascade. The interesting feature is that the parameter values of successive period doubling bifurcations build up a convergent series with intervals between consecutive period doublings progressively reducing with the length of the periodic orbit and at a rate known as the Feigenbaum number $4.6692...$ \cite{StrogatzBook}. Being a convergent series, there exist an accumulation point in the control parameter values where an orbit of infinite period is finally reached. This is the threshold of deterministic chaos in the period doubling route to chaos. After such a threshold value, the three conditions of deterministic chaos, non-periodicity, high sensitivity to initial conditions and a continuous background in the power spectrum, are satisfied. We note that in real experiments in photonics it is sometime difficult to measure an accurate Feigenbaum rate because only few bifurcations are available and also because there are more than one control parameters. When the line in a two parameter space corresponding to the threshold of chaos is approached close to tangency, the bifurcation rate changes to the square root of the Feigenbaum number $2.1608...$ and period doubling cascades are followed by reverse period doubling cascades as demonstrated in 1984 \cite{Oppo84}.

\begin{figure}[h]
\centering
\includegraphics[width=0.49\linewidth]{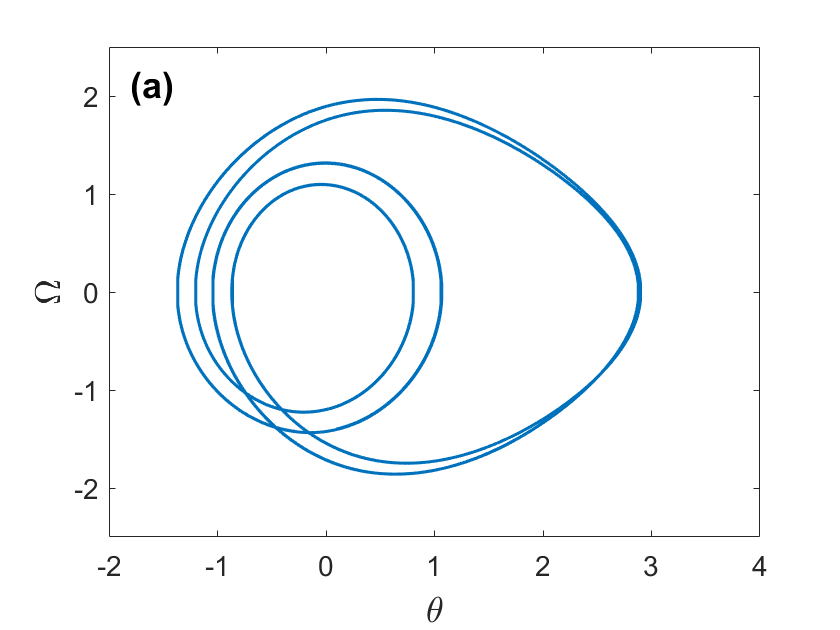}
\includegraphics[width=0.49\linewidth]{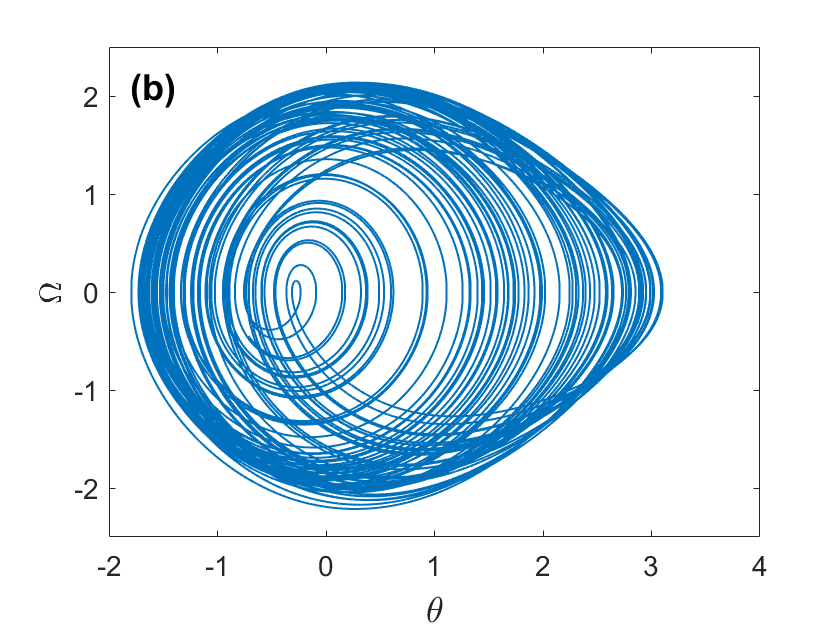}
\includegraphics[width=0.49\linewidth]{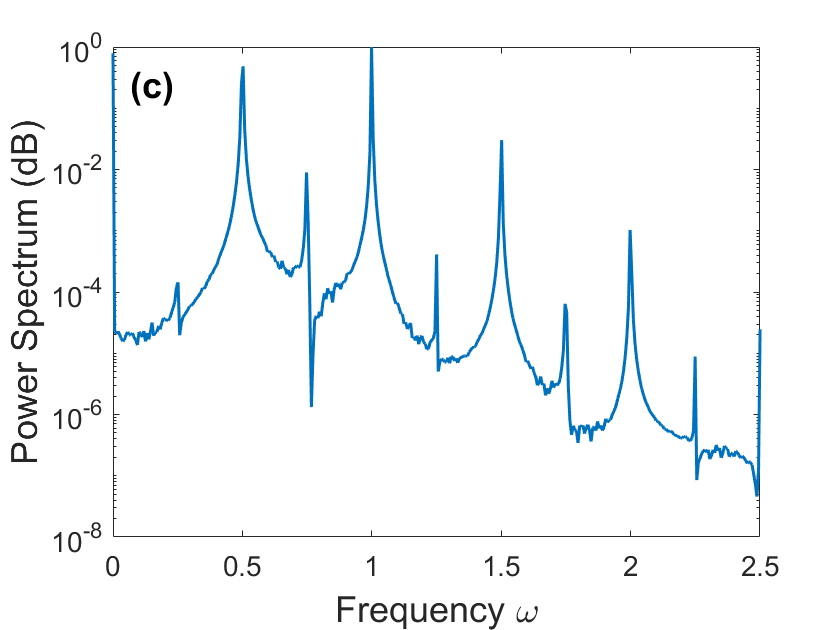}
\includegraphics[width=0.49\linewidth]{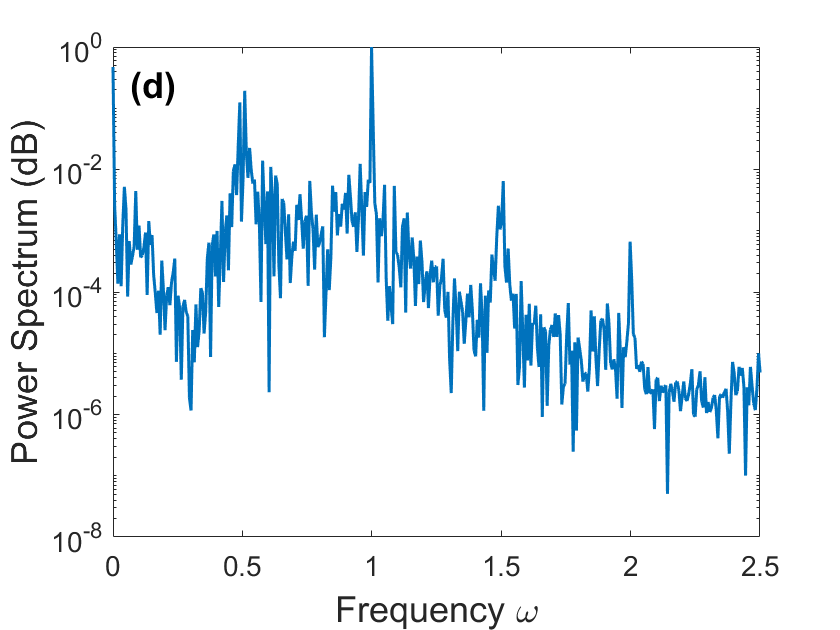}
\caption{(Color online) (a) Limit cycle oscillation of Eqs. (\ref{FPDT}) for $\gamma=0.801$, and $\alpha=\beta=0.1$. Period four orbit. (b) Chaotic oscillations of Eqs. (\ref{FPDT}) for $\gamma=0.815$, and $\alpha=\beta=0.1$. (c) Power spectrum of the period four orbit shown in (a). (d) Power spectrum of the chaotic state shown in (b).}
\label{fig:P4_Chaos}
\end{figure}
An example of a chaotic attractor after a period doubling cascade from Eqs. (\ref{FPDT}) is shown in Fig. \ref{fig:P4_Chaos}(b). The first period doubling bifurcation of this cascade was shown in Fig. \ref{fig:PD} while the solution trajectory after the second period doubling bifurcation is shown in Fig. \ref{fig:P4_Chaos}(a). The power spectrum for the period four orbit obtained numerically by using fast Fourier Transforms is presented in Fig. \ref{fig:P4_Chaos}(c) and that for the chaotic evolution in Fig. \ref{fig:P4_Chaos}(d). In Fig. \ref{fig:P4_Chaos}(c), the highest peak is at the fundamental frequency $\omega=1$ with sub-harmonic peaks at $\omega=1/2$ (period two) and $\omega=1/4$ and $\omega=3/4$ (period four). In Fig. \ref{fig:P4_Chaos}(d), instead, there is a large number of excited frequencies (a continuum in the case of extremely long temporal trajectories) corresponding to lack of periodicity and high sensitivity to initial conditions, i.e. deterministic chaos. We will see period doubling cascades to deterministic chaos in laser equations with modulated losses in Section \ref{sec:TDlaser}.\\

\noi {\it {\bf Spatial disorder, Spatio-Temporal Chaos and Turbulence.}}
When moving from ODEs to PDEs to describe the dynamics of nonlinear systems in both time and space, the ideas of disorder, chaos and turbulence are often mixed together with non-unique definitions. Far from attempting to resolve this issue mathematically or physically, we introduce here pragmatic definitions that will turn out to be useful to distinguish different behaviours seen in the theory of spatially extended photonic devices.\\

\begin{figure}[b]
\centering
\includegraphics[width=0.49\linewidth]{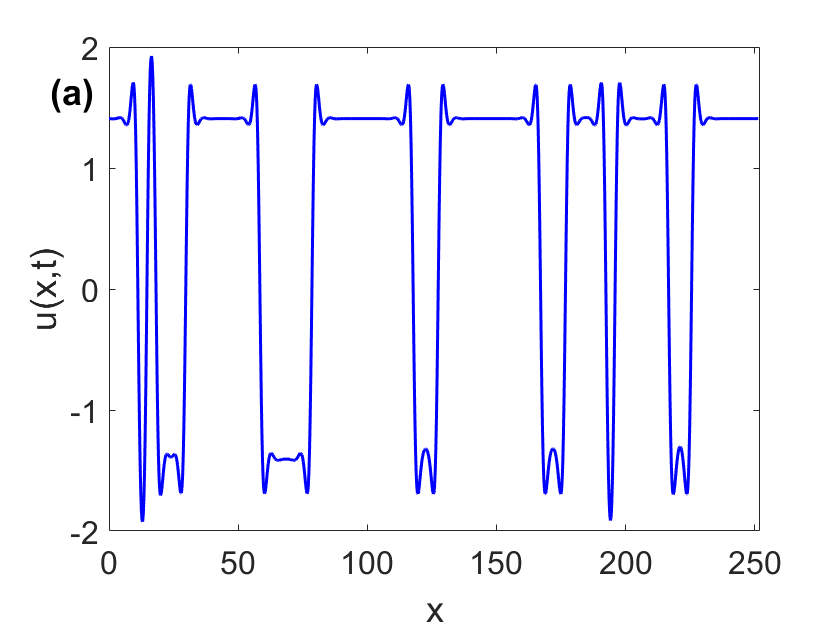}
\includegraphics[width=0.49\linewidth]{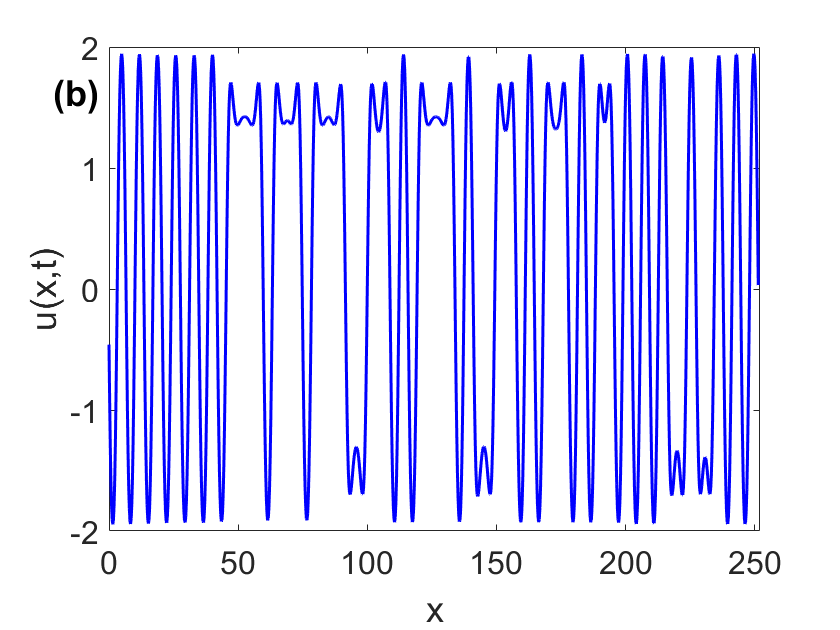}
\caption{(Color online) (a)-(b) Final (asymptotic) spatial distribution of the variable $u(x)$ of the SHE (\ref{SHE}) for two different numerical simulations with initial conditions given by the unstable state $u_0$ plus random spatial noise. }
\label{fig:SpatDisorder}
\end{figure}
{\it Spatial disorder.} We have seen that Turing instabilities can lead to the formation of spatially modulated spatial structures (patterns). In particular we have seen the onset of stable Turing patterns in the SHE. This prototypical real PDE is useful for the introduction of spatial disorder too. The time coordinate is intrinsically different in nature from the spatial coordinate. Apart from science fiction and some Feynman diagrams, the direction of the time in physical experiments is always forward without a possibility of moving backward. Space coordinates instead allow for simultaneous realizations of, for example, travelling waves in a given and/or opposite direction. There is an intrinsic symmetry for space coordinates that is absent from the time coordinate. As a consequence of this symmetry, space coordinates are bound on both sides reflecting the finite size of optical media or laser beams. This has important consequences on the generalization of transitions to temporal chaos to the spatio-temporal domain. For example, the transition to chaos via period doubling cascades requires the existence of orbits with periods tending to infinity which cannot be obtained in a finite spatial domain. This does not mean that stationary spatially disordered structures cannot be found in nonlinear systems described by PDEs. For example in \cite{Coullet87}, the SHE (\ref{SHE}) is shown to generate spatially disordered stationary states when increasing the $\epsilon$ parameter above the Turing and pitchfork bifurcations of the state $u_0=0$. In Figs. \ref{fig:SpatDisorder}(a) and (b), the asymptotic states of numerical simulations of the SHE (\ref{SHE}) for $\epsilon=3$ are shown. Spatially oscillating structures around the two pitchfork branches can lock with each other at random distances leading to a stable spatially disordered configurations. Different random initial conditions around the unstable state $u_0$ lead to different final distributions of peaks, troughs, horns and U-shaped structures as demonstrated in the two panels of Fig. \ref{fig:SpatDisorder}. \\

{\it Spatio-temporal chaos.}
Stationary Turing patterns in PDEs can undergo temporal instabilities similar to those experienced by stationary states in ODEs or homogeneous steady states in PDEs. Although Turing patterns were originally described in reaction-diffusion equations in models of chemical reactions with multiple reagents, temporal oscillations and chaotic behaviour of Turing patterns were first described in nonlinear optics for Kerr resonators \cite{Haelterman92,Gomila03} by using the Lugiato-Lefever equation \cite{Lugiato87}, counter-propagating laser beams \cite{Geddes94} and degenerate optical parametric oscillators \cite{Tlidi97}. These works started at almost the same time of experiments and modelling of oscillating chemical reactions \cite{DeKepper93,Yang02}. Since we will discuss the Lugiato-Lefever equation and spatio-temporal dynamics in Section \ref{sec:STDkerr} we only briefly summarise the main aspects of spatio-temporal chaos here. Once Turing patterns are formed their linear stability analysis can reveal the interaction with a AH mode, i.e the onset of temporal oscillations as described in Section \ref{subsec:Bifurc}. During these regular oscillations the Turing patterns maintain their spatial wavelengths (and wavevectors) with only small broadening and narrowing of the spatial spectral lines. Further bifurcations can then take place when changing control parameters leading to chaotic oscillations. Here we would differentiate between the cases of spatio-temporal chaos where these chaotic oscillations maintain an almost regular shape in space (i.e. corresponding to the excitation of very few spatial modes even in the case of coupled spatial patterns with different wavelengths \cite{Kuptsov12}) and turbulence where irregular temporal oscillations spread over a wide range and randomly excited spatial modes (see next subsection). 

In Section \ref{sec:STDlaser} we will provide specific examples of spatio-temporal chaos.\\

{\it Spatio-temporal turbulence.}
As mentioned above, we identify turbulence with erratic temporal behaviour spread over many (tending to a continuum) of spatial modes. This definition of turbulence is somewhat arbitrary but it helps to differentiate it from spatio-temporal chaos where only a small and finite number of spatial modes are excited during temporally chaotic dynamics. To exemplify what we mean with turbulence in this paper, we briefly describe 'Defect Mediated Turbulence' (DMT) originally introduced by P. Coullet, L. Gil and J. Lega in 1989 \cite{Coullet89}.  
\begin{figure}[h]
\centering
\includegraphics[width=0.49\linewidth]{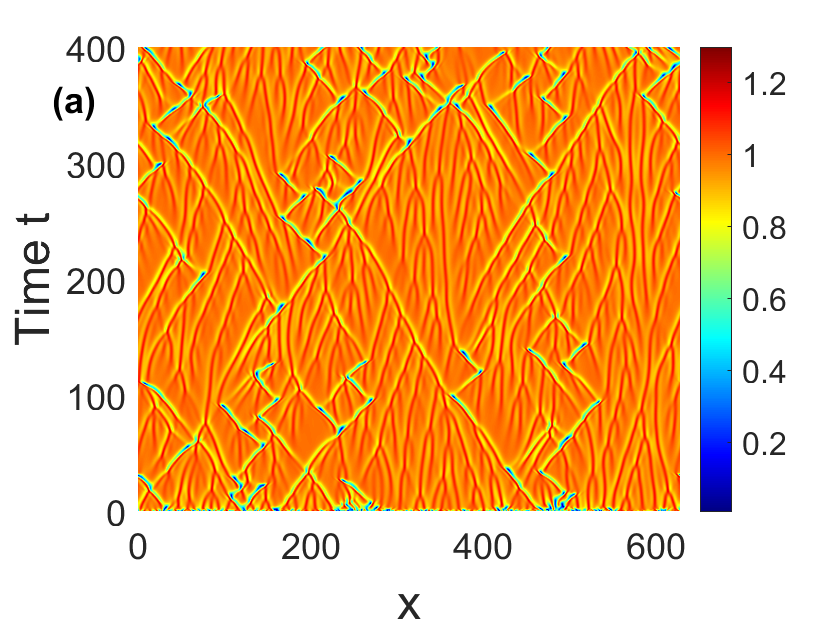}
\includegraphics[width=0.49\linewidth]{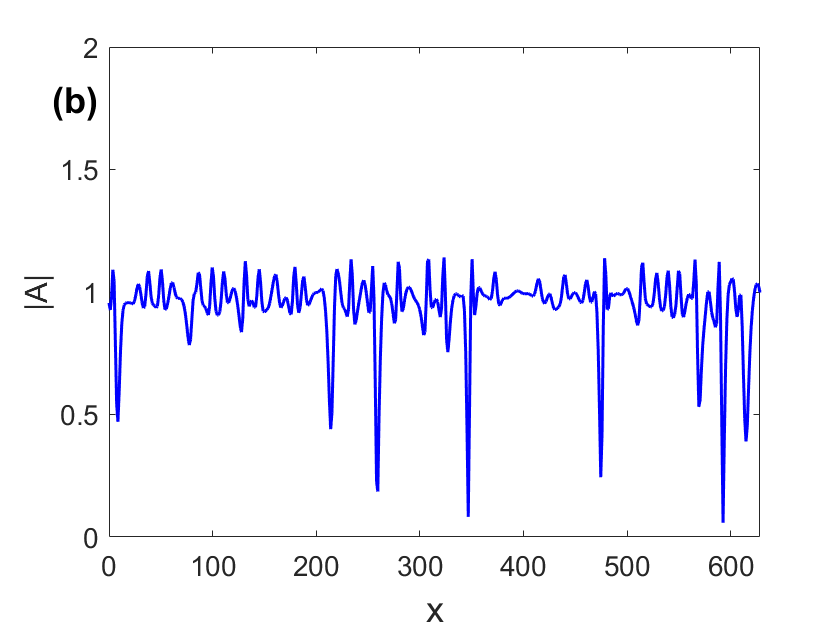}
\caption{(Color online) (a) Space-Time DMT evolution of the amplitude $|A(x,t)|$ of the CGLE (\ref{CGLE}) for $b_1=2$, and $b_3=1$. (b) Final spatial distribution of the amplitude $|A|$ displaying the presence of defects where $|A|$ is zero or very close to zero.}
\label{fig:SpatDisTurb}
\end{figure}
They considered the PDE of the Complex Ginzburg-Landau Equation (CGLE) 
\begin{equation}
\partial_t A = A - (b_3-i) |A|^2 A + (1 + i b_1) \partial_x^2 A
\label{CGLE}
\end{equation}
where $A(x,t)$ is a complex variable with $|A|^2=Re(A)^2+Im(A)^2$ being the square of amplitude of $A$, while $b_1$ and $b_3$ are control parameters. The CGLE is provided here in his one spatial dimension form and with the normalizations provided by H. Chat\'e in \cite{Chate94}. This is useful in identifying the line $b_1=b_3$ in parameter space as a Turing instability of the plane wave solutions of amplitude $a_0=1/b_3$ and phase $\omega_0=1/b_3$. Above this line, for example for $b_1=2$ and $b_3=1$, there is a region where phase turbulence extends to amplitude turbulence leading to a sudden formation and sudden disappearance of defects of zero amplitude and undefined phase. The results of numerical integrations of the CGLE (\ref{CGLE}) done by using the methods described in Appendix II are shown in Fig. \ref{fig:SpatDisTurb} where the temporal evolution of the amplitude $|A|$ is shown in (a) and the final spatial distribution of $|A|$ in (b), the very sharp troughs representing the defects. The presence of these defects at random times and in random positions strongly reduce spatial correlations leading to spatial and temporal disorder with the excitation of many frequencies and many spatial wavevectors, i.e. turbulence \cite{Coullet89}. The DMT regime can be accessed from regions of pure phase turbulence by reducing the $b_3$ parameter form, say, $b_3=1.5$ to $b_3=1$, while keeping the parameter $b_1$ fixed. In the phase turbulent regime, plane waves develop small ripples as seen in Fig. \ref{fig:SpatDisTurb}(b) around $|A|=1$ in the regions without defects. Note that all these effects survive in two spatial dimensions where the defects are topological vortices of positive or negative topological charge \cite{Coullet89}.

\begin{figure}[h]
\centering
\includegraphics[width=0.49\linewidth]{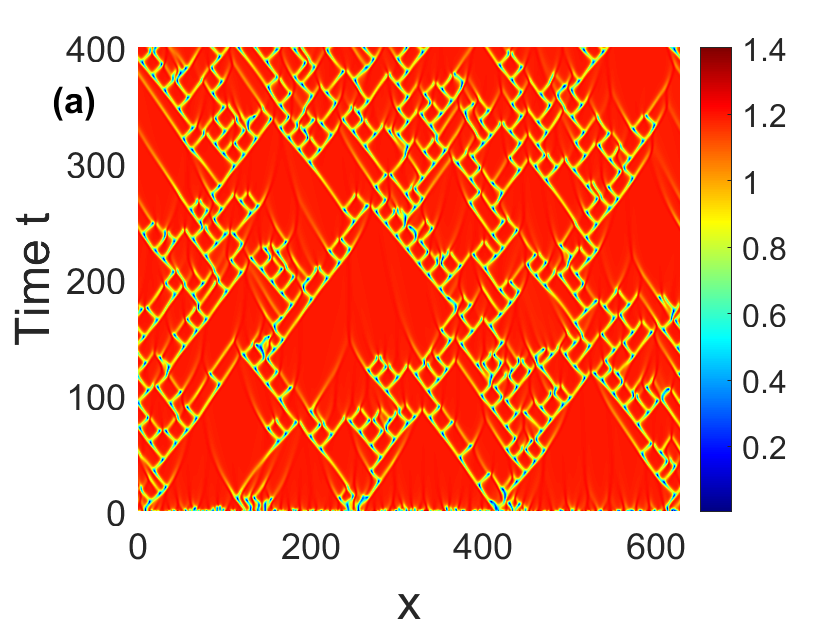}
\includegraphics[width=0.49\linewidth]{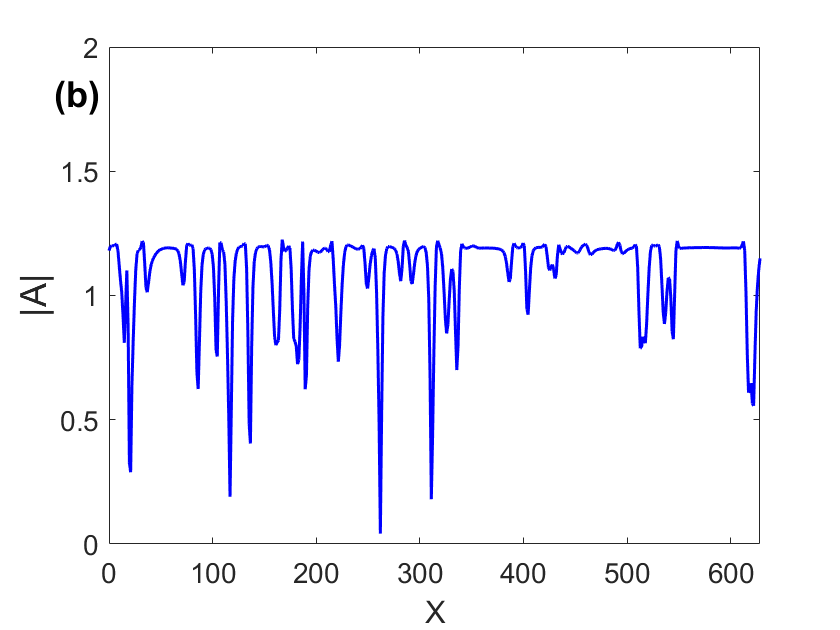}
\caption{(Color online) (a) Space-Time intermittent evolution of the amplitude $|A(x,t)|$ of the CGLE (\ref{CGLE}) for $b_1=0.5$, and $b_3=0.7$. (b) Final spatial distribution of the amplitude $|A|$ displaying the presence of defects where $|A|$ is zero or very close to zero.}
\label{fig:SpatDisInterm}
\end{figure}
Before ending this section we show that DMT also exists in the CGLE in regions without any underlying phase turbulence. Fig. \ref{fig:SpatDisInterm} presents the results of simulations of the CGLE below the Turing instability line $b_1=b_3$ for example for $b_1=0.5$ and $b_3=0.7$. Here the plane wave solution of amplitude $a_0=1/b_3$ and phase $\omega_0=1/b_3$ is stable (or long term metastable) to small perturbations but unstable to large perturbations such as those leading to the sudden formation of the amplitude defects. This regime corresponds again to a fully turbulent configuration and has been labelled as spatio-temporal intermittency in \cite{Chate94}. 

In Section \ref{sec:STDlaser}, we will see how these turbulent regimes can occur in laser configurations.

\vfill


\newpage
\section*{Part A: Temporal Description of Photonic Devices}

When neglecting space-time coupling, i.e. the effects of tranverse diffraction and/or longitudinal group velocity dispersion, the nonlinear equations describing photonic devices are ODEs. Although the reduced ODEs are approximated, they have a wide range of applications in devices with intracavity telescopes or curved mirrors to reduce diffraction or by using very short media operating close to zero-dispersion wavelenghts. Standard numerical methods for the integration and simulation of ODEs are discussed in Appendix I. Here we present numerical simulations of few but representative photonic devices. We first focus on laser cases including lasers with modulated losses and lasers with injected signals. Then we move to passive systems including Kerr media in optical cavities and saturable absorbers. Finally we investigate optical parametric oscillators where the optical nonlinearity has a different origin than the previous cases. 

\section{Temporal Dynamics of Lasers}\label{sec:TDlaser}
The exemplary derivation of standard laser equations is well detailed in \cite{LugiatoBook}. We consider the case of a single longitudinal and single transverse mode laser, i.e. no spatial effects, where longitudinal is the direction of the light propagation in the optical cavity and transverse is the plane perpendicular to the propagation direction (see Fig. \ref{fig:LaserCavities}). 
\begin{figure}[h]
\centering
\includegraphics[width=0.49\linewidth]{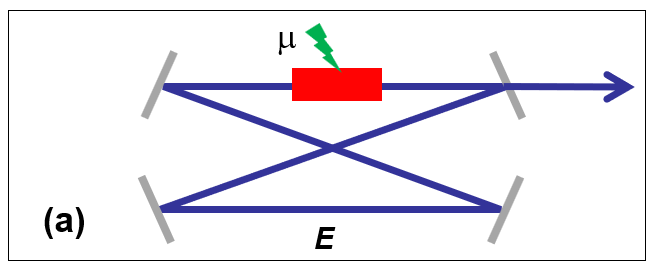}
\includegraphics[width=0.49\linewidth]{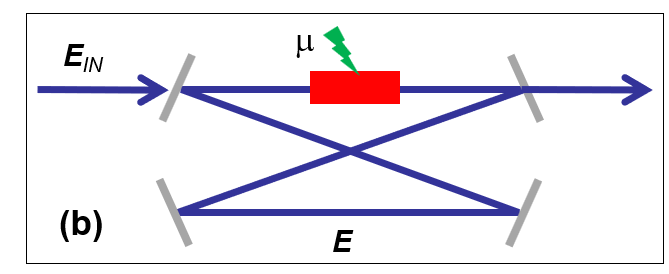}
\caption{(Color online) (a) Schematic diagram of a laser (ring) cavity with an active medium pumped by $\mu$ and an intracavity field $E$. (b) Same as (a) but with the injection of an external drive of amplitude $E_{IN}$. Note that without $\mu$ this configuration corresponds to a passive cavity.} 
\label{fig:LaserCavities}
\end{figure}
Under these approximations the laser equations are:
\begin{align}
\frac{dE}{dt} &= -k \left[ (1 + i \theta) E - P \right] \nonumber \\
\frac{dP}{dt} &= - \gamma_\perp \left[ (1 + i \Delta) P - E D \right] \label{LaserFulleq} \\
\frac{dD}{dt} &= - \gamma_\parallel \left[ D - \mu + (1/2) \, (EP^* + E^* P) \right] \nonumber
\end{align}
where $t$ is the time, $E$, $P$ and $D$ are time variables representing, respectively, the complex electric field in the cavity, the complex material polarisation and the real population inversion, i.e. the difference between the number of atoms in the excited and ground states of a two-level medium, normalized to the parameter $\mu$. The fields $E$ and $P$ are scalar fields since we are considering a single linear polarisation of the light. $k$, $\gamma_\perp$ and $\gamma_\parallel$ are the field, material polarisation and population inversion decay rates, respectively. The detuning $\theta$ is the difference between the frequency $\omega_c$ of the closest cavity resonance and the reference frequency, normalised to $k$. The detuning $\Delta$ is the difference between the atomic frequency $\omega_A$ of the two-level transition and the reference frequency, normalised to $\gamma_\perp$. Note that the two level transitions mentioned here do not correspond uniquely to discrete atomic energy levels but can include quantum features in a large variety of media, including valence and conduction bands in semiconductor media as well as different phases in liquid crystals. $\mu$ is the gain provided by an external pump if positive, while it is negative and equal to $-2C$ for a passive cavity. $C$ is positive and is known as the bistability parameter that depends on the medium absorption, its length and the transmittivity of the cavity mirrors \cite{LugiatoBook}. In this section we focus on active systems (i.e. lasers) operating in the presence of population difference $D>0$ when there is no field in the cavity, i.e. $\mu>0$, while the passive cavities are discussed in Section \ref{sec:TDkerr}. Note that in this review paper, we focus on photonic devices based on ring cavities (i.e. ring resonators). Many of these devices, however, are also realised with Fabry-Perot cavities with two mirrors facing each other. For an excellent review and extensive description of Fabry-Perot cavities in photonics see \cite{Lugiato23}. 

The physical interpretation of Eqs. (\ref{LaserFulleq}) can be done term by term. The first term in the equation for the cavity field $E$ contains losses through the cavity mirrors (and scattered by the medium), and the cavity detuning $\theta$ between the frequency of $E$ and that of the closest cavity resonance mode, the other modes being considered to be too weak for interaction and hence neglected. The second term in the first equation show that in the absence of an external drive, the source of photons in the cavity is the electric polarization field discovered by James Clerk Maxwell in 1865 \cite{Maxwell_1865}. The first term of the second equation, the one for the electric polarization field $P$, contains losses proportional to the atomic linewidth such as e.g. collisions in a gas, and the atomic detuning $\Delta$ between the frequency of $E$ and that of that of the atomic transition, i.e. the energy separation of the two atomic levels. The second term in the second of Eqs. (\ref{LaserFulleq}) is due to stimulated emission that depends on the population inversion $D$ and the electric field $E$ in the cavity. The first term in the third equation is the relaxation to equilibrium of the population inversion mainly due to spontaneous emission. The $\mu$ term is the energy provided by an external pump directly to the medium to create population inversion while the very last one describes the losses to the population inversion due to the stimulated emission of photons. The phenomena of stimulated and spontaneous emission were discovered by Albert Einstein in 1917 \cite{Einstein17}.

It is easy to see that there are two stationary states of the laser equations (\ref{LaserFulleq}). The first one, $E_s=P_s=0$ and $D_s=\mu$, corresponds to the laser being off. The second stationary state corresponds to the lasing state with $|E_s|^2=\mu - 1 - \theta^2$, $|P_s|^2=|E_s|^2 (1+\theta^2)$, $D_s= \mu - |E_s|^2 = 1+ \theta^2$ and $\Delta = - \theta$. This state only exists for $\mu > \mu_{thr} = 1 + \theta^2$ where $\mu_{thr}$ is commonly known as the laser threshold. The condition on the two detunings, $\Delta=-\theta$, determines the laser frequency $\omega_L$ during emission
\begin{equation} 
\omega_L = \frac{k \omega_A + \gamma_{\perp} \omega_C}{k + \gamma_{\perp}} \, .
\label{ModePulling}
\end{equation}
The position of the laser frequency depends on the relaxation rates of the photons in the cavity $k$ and that of the material polarisation $\gamma_{\perp}$. In the common case of $\gamma_{\perp} \gg k$, the laser frequency is very close to the cavity resonance of frequency $\omega_c$. The intensity $|E|^2$ and population $D$ for the laser stationary states are plotted in Fig. \ref{fig:laserSS}(a) for $\theta=0$, the resonant case. The laser bifurcation looks like a transcritical bifurcation for the intensity $|E_s|^2$.  
\begin{figure}[h]
\centering
\includegraphics[width=0.49\linewidth]{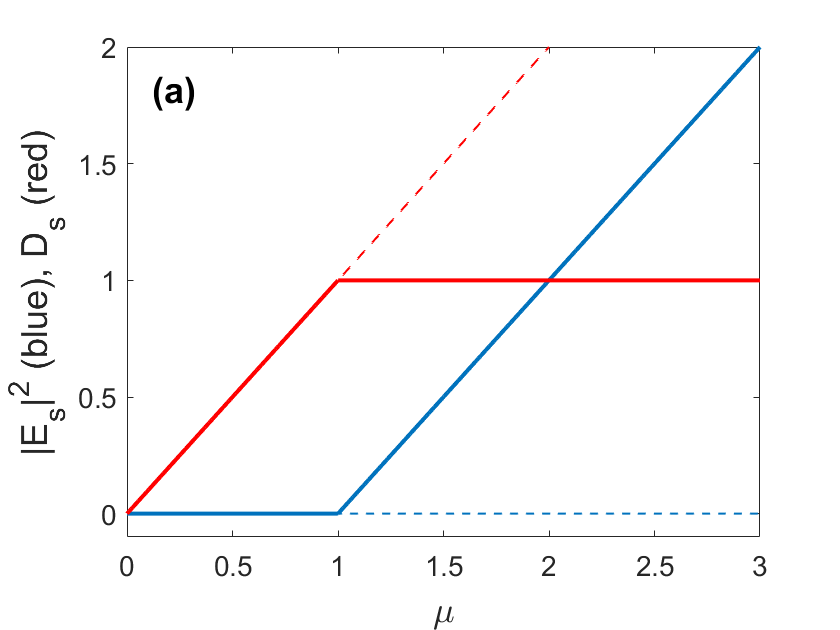}
\includegraphics[width=0.49\linewidth]{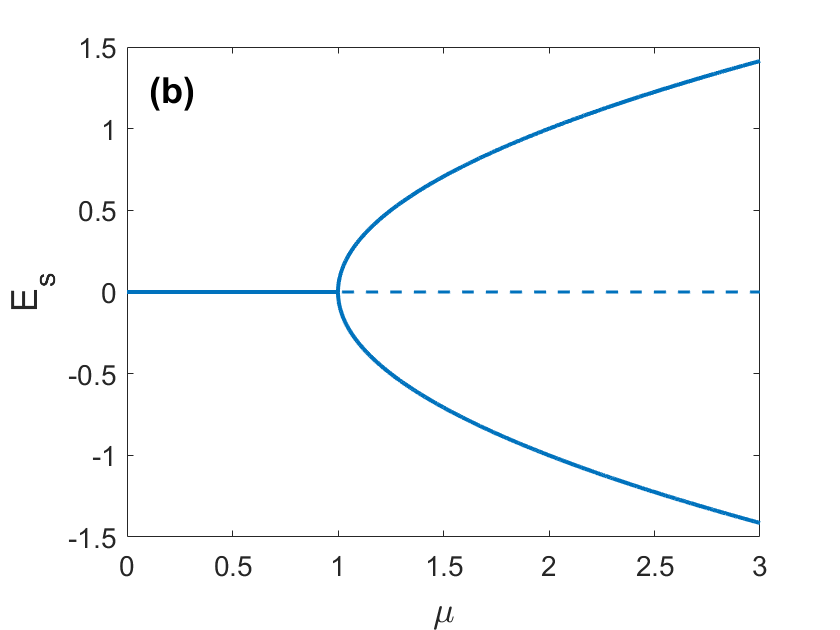}
\caption{(Color online) (a) Stationary laser intensity $|E_s|^2$ and population inversion $D_s$ of the laser bifurcation versus the gain parameter $\mu$ for $\theta=0$. (b) Same as (a) but for stationary states of the field variable $E_s$.}
\label{fig:laserSS}
\end{figure}

It should be noted, however, that above threshold the laser field $E_s$ admits two possible values of $E_s=\pm \sqrt{\mu - 1 - \theta^2}$ corresponding to a zero and a $\pi$ phase, respectively. Both of these states are possible in the laser equations and we will see their relevance when discussing spatial effects in Section \ref{sec:STDlaser}. When considering these two possible states of the laser phase, the laser threshold is a pitchfork bifurcation as shown in Fig. \ref{fig:laserSS}(b) for $\theta=0$. 

To investigate the dynamics of the laser equations, it is convenient to start from the resonant case where $\Delta=\theta=0$. In this case the imaginary part of the fields remains zero if they are initiated at zero. The electric and polarization fields are now represented by purely real variables and the laser equations are
\begin{align}
\frac{dE}{d \tau} &= -(k/\gamma_\perp) \left[ E - P \right] \nonumber \\
\frac{dP}{d \tau} &= - \left[ P - E D \right] \label{LaserReal} \\
\frac{dD}{d \tau} &= - (\gamma_\parallel/\gamma_\perp) \left[ D - \mu + EP \right] \nonumber
\end{align}
where the time has been normalised by the polarization decay rate $\gamma_\perp$ via $\tau=\gamma_\perp t$. If then we introduce the new variables $X=\sqrt{\gamma_\parallel/\gamma_\perp} E$, $Y = \sqrt{\gamma_\parallel/\gamma_\perp} P$ and $Z = \mu -D$, one obtains
\begin{align}
\frac{dX}{d\tau} &= -(k/\gamma_\perp) \left( X - Y \right) \nonumber \\
\frac{dY}{d\tau} &= \mu X - Y  - X Z \label{LaserLorenz} \\
\frac{dZ}{d\tau} &= - (\gamma_\parallel/\gamma_\perp) Z + X Y  \nonumber \, .
\end{align}
It is easy to see that if one identifies the parameters $\sigma = (k/\gamma_\perp)$, $\rho = \mu$ and $\beta = (\gamma_\parallel/\gamma_\perp)$ the resonant laser equations are equivalent of the Lorenz model (\ref{Lorenz}). This mathematical equivalence was established by H. Haken in 1975 \cite{Haken75}. As mentioned in the introduction, the Lorenz equations are prototypical for displaying deterministic chaos. One would then expect lasers to routinely display chaotic outputs since they appear to have the same underlying mathematical structure of the Lorenz model. Nothing can be further from reality. Easiness to display chaotic behaviour would be a serious blow to the thousands upon thousands of laser applications where steady or regularly pulsed laser outputs are the basic requirement. How comes that lasers are so stable and yet equivalent to the Lorenz model for deterministic chaos?

The proof is in the pudding of the parameters. We have seen that to display deterministic chaos the three parameters of the Lorenz model $\sigma$, $\rho$ and $\beta$ are positive and of order 10. For example $\sigma = (k/\gamma_\perp)$ of order 10 would require the decay time of the photons in the cavity to be much shorter that that of the material polarization, a regime often known as the 'very bad cavity' limit. In all lasers but one or two, the ratio $(k/\gamma_\perp)$ is several order of magnitudes smaller than one, i.e. in the opposite limit of the values required for Lorenz chaos. The same happens for $\beta = (\gamma_\parallel/\gamma_\perp)$ since for physical reasons associated to the decays of the diagonal and off-diagonal terms of the density matrix $\gamma_\parallel \ll 2 \gamma_\perp$ \cite{LugiatoBook}. To understand why lasers do not spontaneously emit chaotic beams of light, we need to understand better the time scales of the three dynamical variables of the laser equations (\ref{LaserFulleq}).\\

\subsection{Laser Time Scales and Relaxation Oscillations.} \label{subsec:RO}
The electric field, the atomic polarization, and the population inversion usually decay on very different time scales, which are given by the decay rates $k$, $\gamma_\perp$ and $\gamma_\parallel $, respectively. If one of these rates is much larger than the others, the corresponding variable relaxes in a fast way and consequently adjusts to and then follows (adiabatically) the other variables. As the temporal dynamics of the variable with a large relaxation rate is faster than the other variables, this variable is regarded as a dependent variable in the long term. Therefore, the variable with faster relaxation rate is considered to be dependent on the other variables with slower relaxation rates. Through the process of adiabatic elimination of fast relaxing variables, the number of equations describing a laser with variables decaying at different rates can be reduced accordingly. It has been useful to classify lasers with different time scales into three separate classes \cite{Arecchi84,Tredicce85_1}:
\begin{itemize}
\item {\bf Class-A Lasers.} $k \ll \gamma_{\parallel} \approx \gamma_{\perp}$. The decay rate of the electric field is much slower than those of the atomic polarization and the population inversion. The variables of atomic polarization and population inversion change much faster than that of the electric field in the cavity, and can be expressed as functions of the electric field (adiabatic elimination). Therefore the system can be described by means of just one dynamical equation for the field. By injecting a constant, coherent, external field, it is possible to increase the number of degrees of freedom by one (see subsection on lasers with injection \ref{subsec:LIS}). The possible results are stable, locking, or regular pulsations, depending on the detuning between the two fields and on the amplitude of the external signal. To achieve higher-order instabilities and eventually chaotic behaviour in Class-A lasers, it is necessary to also modulate one of the control parameters, namely, the external field, or the pump rate, or the cavity losses. Since many of these purely temporal features are better visible in Class-B lasers without spatial effects, we will describe complex dynamics of Class-A lasers when considering spatio-temporal effects in Section \ref{sec:STDlaser}. Examples of Class-A lasers are Quantum Dot lasers, He-Ne lasers (632.8 nm) and dye lasers.

\item {\bf Class-B Lasers.} $(k + \gamma_{\parallel}) \ll \gamma_{\perp}$ including $\gamma_{\parallel} \ll k \ll \gamma_{\perp}$. Here, the adiabatic elimination of the material polarization is feasible. The dynamic behaviour of this class of lasers is in general described by two coupled nonlinear equations: one for the electric field in the cavity and the other for the population inversion variables. The mathematical and numerical investigation of Class-B lasers is the focus of this section. We are going to see that they display relaxation oscillations that can be externally forced by modulating the cavity losses or by injection to reach sustained oscillations, pulsing and deterministic chaos. We are also going to see that the case of $\gamma_{\parallel} \ll k \ll \gamma_{\perp}$ does not lead to a single equation for the slow dynamic of the population inversion with the material polarization $P$ and the electric filed $E$ expressed as functions of $D$ as this limit leads to a conservative dynamics of at least two variables. Many commercial lasers are classified as Class-B lasers: semiconductor, ruby, solid-state, Nd:YAG and CO$_2$ lasers.

\item {\bf Class-C Lasers.} $k \approx \gamma_{\parallel} \approx \gamma_{\perp}$. The relaxation rates of the electric field, the population inversion, and the atomic polarization variables are of the same order of magnitude. Class-C lasers require the full model (\ref{LaserFulleq}) for their description. To explore interesting dynamical regimes, however, extra conditions such as 'bad cavity limits' and extremely high pumping rates are required making their investigation dull with respect to Class-B lasers. Only very few far-infrared lasers belong to this class, He-Ne (3.39 $\mu$m), He-Xe, and NH$_3$ lasers with limited applications.

\end{itemize}

Standard adiabatic elimination procedures can be applied to Eqs. (\ref{LaserFulleq}) to obtain reduced models for Class-B lasers where $k \approx \gamma_{\parallel} \ll \gamma_{\perp}$. By choosing the system the reference frequency at the laser frequency $\omega_L$ of the stationary states $\Delta=-\theta$, see Eq. (\ref{ModePulling}), and by setting the time derivative of the material polarization equal to zero, the second equation of Eqs. (\ref{LaserFulleq}) provides $P=ED/(1-i\theta)$ thus giving
\begin{align}
\frac{dE}{dt} &= -k \left[ (1 + i \theta) E (1 - D/(1+\theta^2))  \right] \nonumber \\
\frac{dD}{dt} &= - \gamma_\parallel \left[ D - \mu  + |E|^2 D/(1 + \theta^2) \right]  \, . \label{LaserClassB}
\end{align}

In the resonant case when $\theta=0$, it is convenient 
to write these equations in terms of the laser intensity $I=|E|^2$, i.e.
\begin{align}
\frac{dI}{dt} &= - 2 k \left[ I (1 - D)  \right] \nonumber \\
\frac{dD}{dt} &= - \gamma_\parallel \left[ D - \mu  + I D \right]  \, . \label{LaserRateEq}
\end{align}
These are the renowned nonlinear {\it laser rate equations} first introduced by H. Statz and G. deMars in 1960 \cite{Statz60,Shirley68}. 

Like the full model (\ref{LaserFulleq}) there are two stationary states $I=0$, $D=\mu$ below threshold and $I=\mu-1$, $D=1$ for the lasing case (see Fig. \ref{fig:laserSS}). It is easy to perform the linear stability analysis of the lasing state via small perturbations $\delta I = I-(\mu-1)$, and $\delta D= D - 1$ providing us with the characteristic polynomial and stability eigenvalues 
\begin{equation}
\lambda^2 + \mu \gamma_{\parallel} \lambda + 2 \gamma_{\parallel} k (\mu-1) = 0 \;\;\; \rightarrow \;\;\;
\lambda_\pm= \frac{1}{2} \left[ -\mu \gamma_{\parallel} \pm \sqrt{(\mu \gamma_{\parallel})^2-8 \gamma_{\parallel} k (\mu -1)} \right] \, .
\end{equation}
As the pump parameter $\mu$ is real and positive these eigenvalues are for sure real for $2k<\gamma_{\parallel}$. This means that there are no ROs unless $2k>\gamma_{\parallel}$. For many Class-B lasers, $2k$ is much larger than $\gamma_{\parallel}$ due to metastable excited energy levels and slow relaxations of the population inversion variable. In these cases, the stability eigenvalues are complex with an imaginary part well approximated by $\omega_{RO}=\sqrt{2 k \gamma_{\parallel} (\mu-1)}$, the frequency of the RO. 
\begin{figure}[h]
\centering
\includegraphics[width=0.49\linewidth]{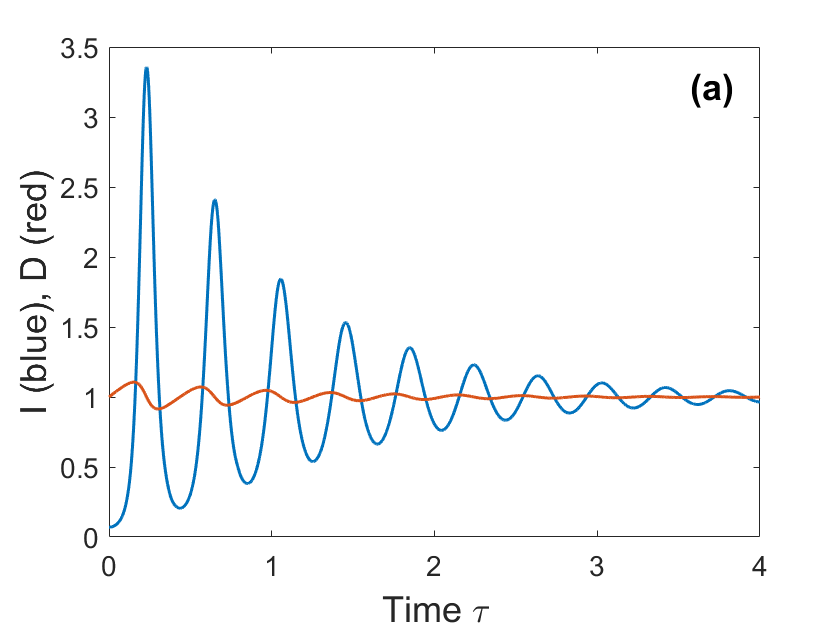}
\includegraphics[width=0.49\linewidth]{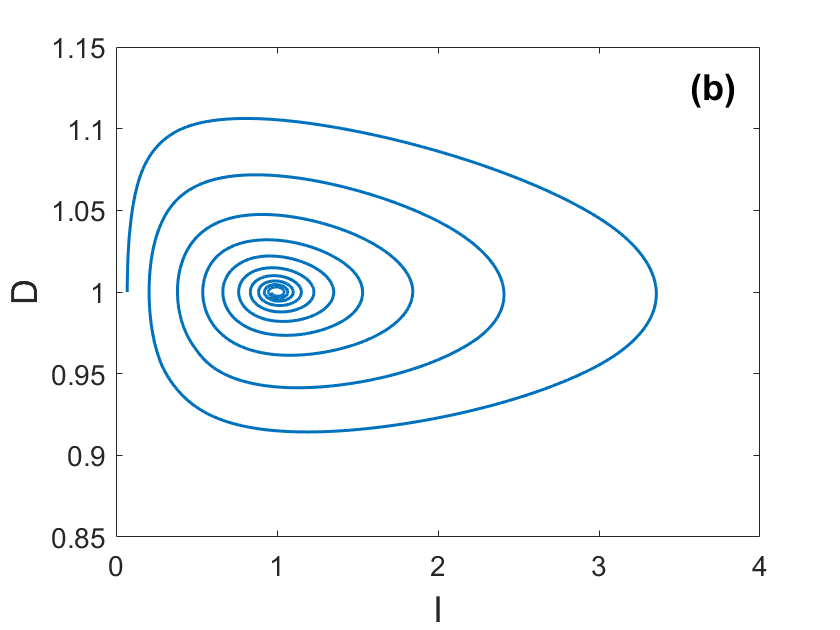}
\caption{(Color online) (a) Relaxation oscillations of the laser intensity $I$ (blue) and population inversion $D$ (red) from direct numerical integration of Eqs. (\ref{LaserRateEq}) for $2k/\gamma_{\parallel}=256$, $\mu=2$ and starting from $I=0.07$ and $D=1$ with $\tau=\gamma_{\parallel} t$ . (b) Same as (a) but in the $(I,D)$ variable space.}
\label{fig:RO}
\end{figure}

Results of the numerical simulations of Eqs. (\ref{LaserRateEq}) with the time normalized via $\gamma_{\parallel}$ and by using the ODE methods described in Appendix I, are presented in Fig. \ref{fig:RO} for $2k/\gamma_{\parallel}=256$, $\mu=2$ and initial conditions given by $I=0.07$ and $D=1$. These are the same values and time normalizations as those used by J. Shirley in 1968 \cite{Shirley68}. Fig. \ref{fig:RO}(a) shows typical RO of the intensity and population inversion variables when switching a laser from an initially small light intensity and is in excellent agreement with the computer simulations of the Statz and deMars model presented in \cite{Shirley68}. The period of these oscillations matches quite carefully the value $0.393$ that one obtains from $\omega_{RO}$ above. In Fig. \ref{fig:RO}(b) we show the RO orbit in the $(I,D)$ plane. A physical understanding of RO in laser equations is provided in the next section.\\

\noi {\it {\bf Laser Oscillations in a Toda Potential.}}
The rate equation for the dynamics of the laser intensity is mathematically interesting since its time derivative is directly proportional to the intensity itself. Following our 1985 work \cite{OpPoliti85}, it is then useful to find the appropriate time scale by introducing a logarithmic variable $s=\ln(I)$ such that $ds/dt = 2 k (D-1)$. This latter equation is even more interesting since its time derivative, i.e. the second derivative of $s$ with respect to time, is immediately provided by the dynamical equation of $D$. The population inversion variable can then be expressed in terms of $s$ via $D=(1/(2k))ds/d\tau+1$ leading to
\begin{align}
\frac{d s}{dt} &= 2 k (D-1) \nonumber \\
\frac{d D}{dt} &= \gamma_{\parallel} \left( \mu - D - D e^s \right) \nonumber \\
\frac{1}{2k \gamma_{\parallel}} \frac{d^2 s}{dt^2} &+ \frac{1}{2k} \frac{ds}{dt} \left(1+e^s \right) - (\mu -1) + e^s = 0 \, . \label{LaserTodaDamp}
\end{align}
The last of Eqs. (\ref{LaserTodaDamp}) shows that the correct time normalization for Class-B lasers with $2k \gg \gamma_{\parallel}$ is not $\gamma_{\parallel}$ as suggested by the standard theory of adiabatic elimination, but $\tau = \sqrt{2k \gamma_{\parallel} } t$ with the introduction of a useful small parameter $\epsilon = 
\sqrt{\gamma_{\parallel} / 2k}$ 
\begin{equation}
\frac{d^2 s}{d\tau^2} +  \epsilon \frac{ds}{d\tau} \left(1+e^s \right) - (\mu -1) + e^s = 0 \, .
\label{LaserTodaDamp2}
\end{equation}
This equation describes the damped motion of a particle in a Toda potential $V(s) = e^s - (\mu-1) s = I - (\mu-1) \ln(I)$ (see Fig. \ref{fig:Toda}(a)) where the damping is ruled by $\epsilon$ and becomes less and less relevant the more $\gamma_{\parallel}$ decreases with respect to $2k$ \cite{OpPoliti85}. By using the new time normalization $\tau$ and the parameter $\epsilon$, the laser rate equations (\ref{LaserRateEq}) are
\begin{align}
\frac{dI}{d\tau} &= - \frac{1}{\epsilon} \left[ I (1 - D)  \right] \nonumber \\
\frac{dD}{d\tau} &= - \epsilon \left[ D - \mu  + I D \right]  \, \label{LaserRateEqEpsi}
\end{align}
suggesting the introduction of the fluctuation of the population inversion from its stationary value via $D=1+\epsilon W$ leading to
\begin{align}
\frac{dI}{d\tau} &= I \, W  \nonumber \\
\frac{dW}{d\tau} &= \mu  - (1+\epsilon W)( 1 + I)  \, . \label{LaserRateEqW}
\end{align}
\begin{figure}[h]
\centering
\includegraphics[width=0.49\linewidth]{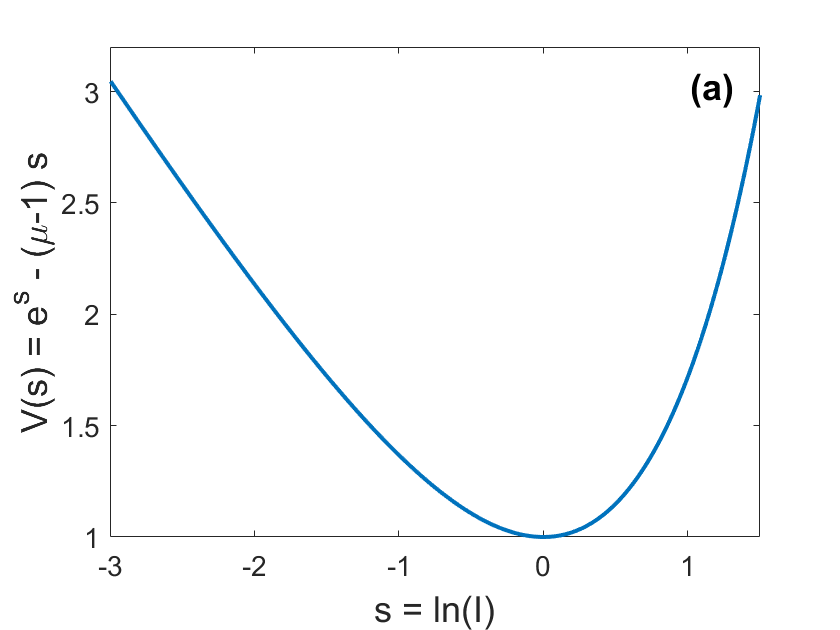}
\includegraphics[width=0.49\linewidth]{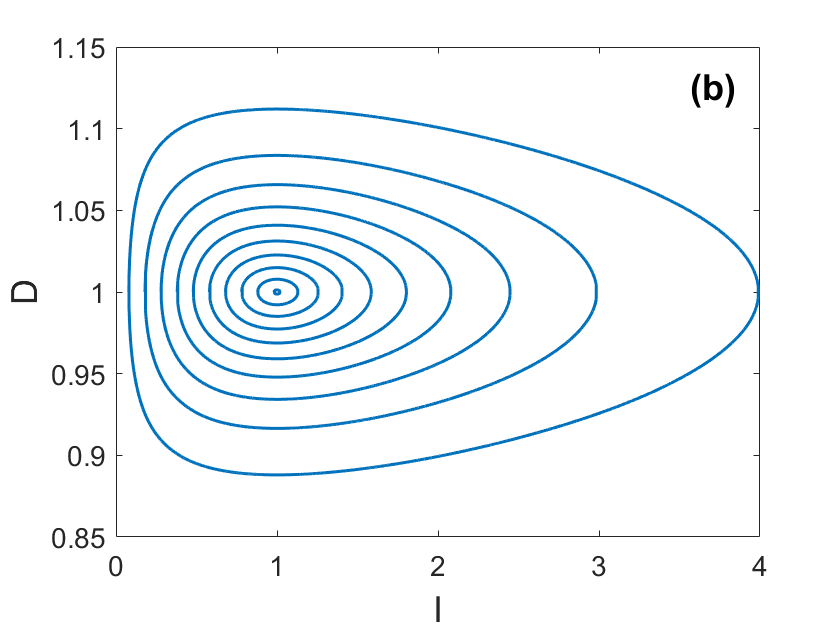}
\caption{(Color online) (a) Toda Potential. (b) Conservative laser dynamics in a Toda Potential.}
\label{fig:Toda}
\end{figure}

A conservative motion in the Toda potential is then readily obtained by neglecting the term in $\epsilon$ in the second of Eqs. (\ref{LaserRateEqW}), i.e. $d W/d\tau = \mu  - 1 - I$. The numerical integration of Eqs. (\ref{LaserRateEqW}) for $\epsilon=0$ is much faster than the numerical integration of Eqs. (\ref{LaserTodaDamp}) and is shown in the $(I,D)$ space in Fig. \ref{fig:Toda}(b) for ten different initial conditions with $I$ ranging from $0.08$ to $0.98$ and $W=0$.  The conservative dynamics clearly show that the damped orbit in Fig. \ref{fig:RO}(b) spirals down from one to the next of the conservative orbits. It may appear strange that an intrinsically dissipative system, like the laser where energy losses are present because of the partially reflecting mirrors that allow the light to exit the cavity, can be described by a Toda dynamics that is conservative in nature. The conserved 'energy-like' quantity in the Toda potential is given by $W^2/(2(\mu-1))- \ln(I) + I/(\mu-1) + \ln(\mu-1) - 1 $ and remains constant in time while the total energy of the system $I+W$ oscillates in time. The laser system alternatively stores the energy absorbed from the pump $\mu$ and re-emits it as a laser light pulse. By considering a long metastable excited level in the medium, all photon emissions are stimulated emissions (laser) and not spontaneous emissions (neglecting the term in $\epsilon$ in the population inversion equation of (\ref{LaserRateEqW})). All these theoretical and computational considerations were experimentally verified in \cite{Cialdi13} by using a Nd:YAG laser in 2013.

The common laser case of $\gamma_{\parallel} \ll k \ll \gamma_{\perp}$ shows that the standard methods of adiabatic elimination by setting the derivatives of the fast variables to zero can fail miserably in important cases \cite{OpPoliti86}. In this limit corresponding to $\epsilon \rightarrow 0$, the dynamics becomes more and more two dimensional, the Toda potential, instead of projecting onto that of a single variable, the population inversion. We will see more intriguing effects of this limit when investigating lasers with injected signals later in Subsection \ref{subsec:LIS}.\\

\noi {\it {\bf The Effect of the Linewidth Enhancement Factor.}}
Eqs. (\ref{LaserClassB}) for Class-B lasers cannot straightforwardly be applied to Semiconductor lasers as the effect of the Linewidth Enhancement Factor (LEF) is missing. In 1982, Charles H. Henry \cite{Henry82} found that in semiconductor materials the refractive index displays a dependence on the carrier density due to an increased linewidth from a coupling between intensity and phase noise. Henry introduced the LEF $\alpha$ factor (also called Henry factor) to quantify the amplitude–phase coupling mechanism and related phase changes to the modifications in the gain in reasonable agreement with experimental data \cite{Henry82}. In Semiconductor lasers under the condition $\gamma_{\parallel} \ll k \ll \gamma_{\perp}$, the following modified Class-B laser equations can be obtained \cite{Oppo09}
\begin{align}
\frac{dE}{d\tau} &= (1-i\alpha) W \, E \nonumber \\
\frac{dW}{d\tau} &= \mu - (1+\epsilon W)( 1 + |E|^2) \label{LaserClaB_LEF}
\end{align}
where $\tau=\sqrt{k \gamma_{\parallel}} \, t$ (the factor 2 not being here because this is the amplitude $E$ and not the intensity $I$), $\alpha$ is the LEF and $W$ is introduced via $D=(1+\epsilon W)$ with $\epsilon = \sqrt{\gamma_{\parallel} / k}$. In standard operational conditions of Semiconductor lasers the detuning is very small and it is neglected in (\ref{LaserClaB_LEF}) \cite{Oppo09}. By using the complex conjugate of the first of these equations, it is possible to see that the dynamics for the intensity and the population variable $W$ are the same as those of Eqs. (\ref{LaserRateEqW}). Hence the underlying dynamics of Semiconductor lasers is still described by relaxation oscillations in a Toda potential since the LEF affects the laser phase only.\\ 

\subsection{Lasers with Modulated Losses.} \label{subsec:LasModLoss}
Having studied the physical and historical origin of the RO in lasers it is possible to understand the mechanisms behind the first generation of large energy light pulses via the Q-switching technique. Q-switching was demonstrated in 1961 by R. Hellwarth and F. McClung at Hughes Research Laboratories using electrically switched Kerr cell shutters in a ruby laser \cite{Hellwarth61}. In general Q-switching is achieved by using variable attenuators inside the laser's optical cavity that is directly proportional to the cavity finesse. When the attenuator is functioning, lasing cannot begin. This attenuation inside the cavity prevents lasing action and corresponds to a decrease in the Q factor, the quality factor of the optical resonator. The variable attenuator is commonly called a "Q-switch", when used for this purpose. Initially the laser medium is pumped while the Q-switch prevents feedback by producing an optical resonator with low Q. This increases the population inversion and energy stored in the medium up to saturation, but no laser action. When the Q-switch device is quickly changed from low to high Q, the large amount of energy already stored in the gain medium is released in a short pulse of light at the output of the laser which may have very high peak intensity. The process is then repeated leading to a sequence of large energetic laser pulses. 

\begin{figure}[h]
\centering
\includegraphics[width=0.49\linewidth]{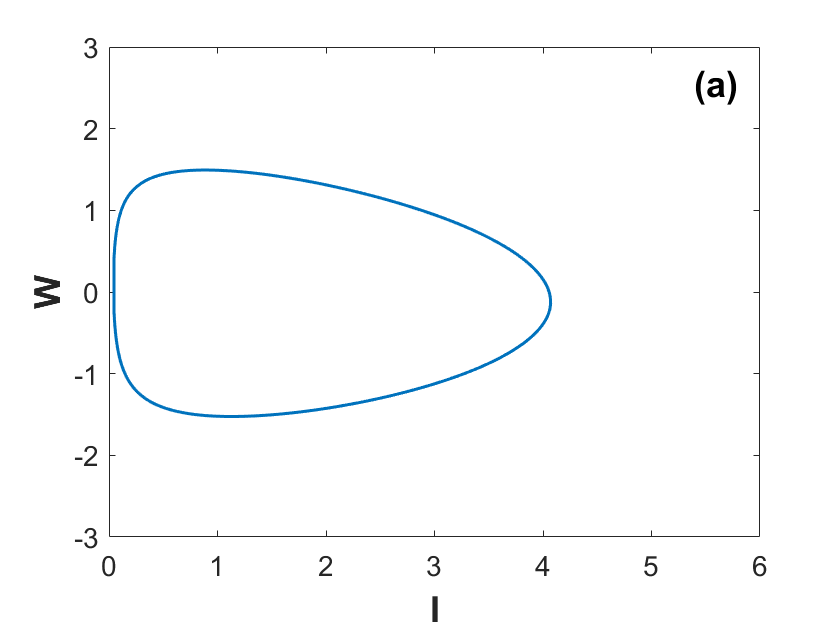}
\includegraphics[width=0.49\linewidth]{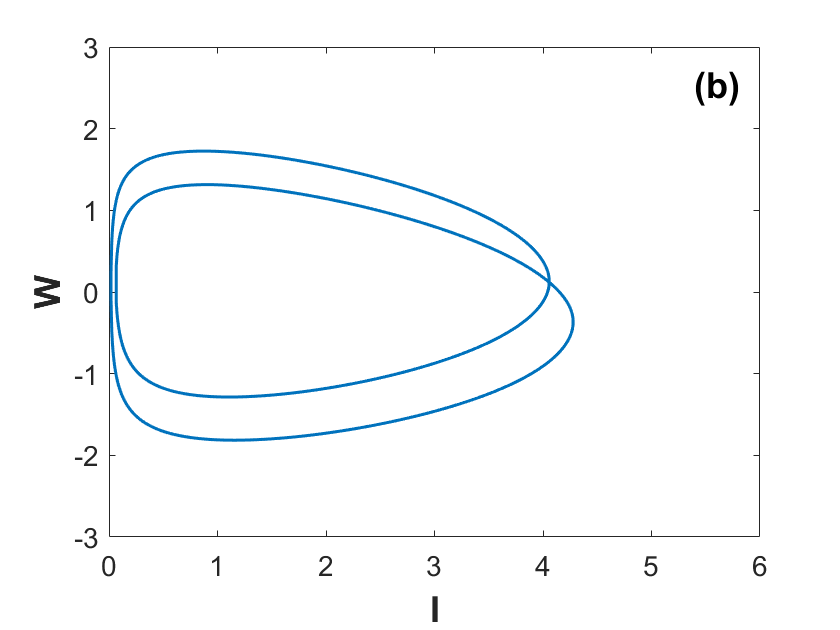}
\includegraphics[width=0.49\linewidth]{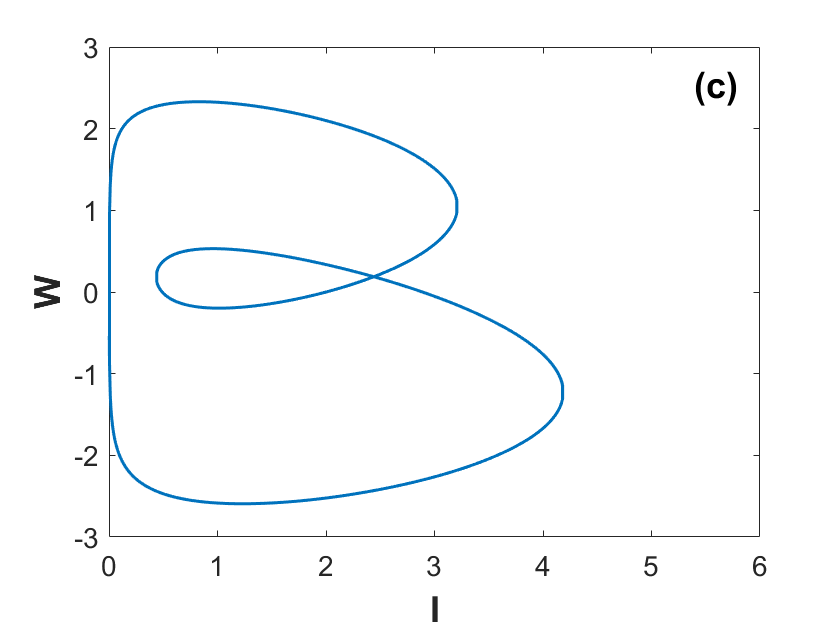}
\includegraphics[width=0.49\linewidth]{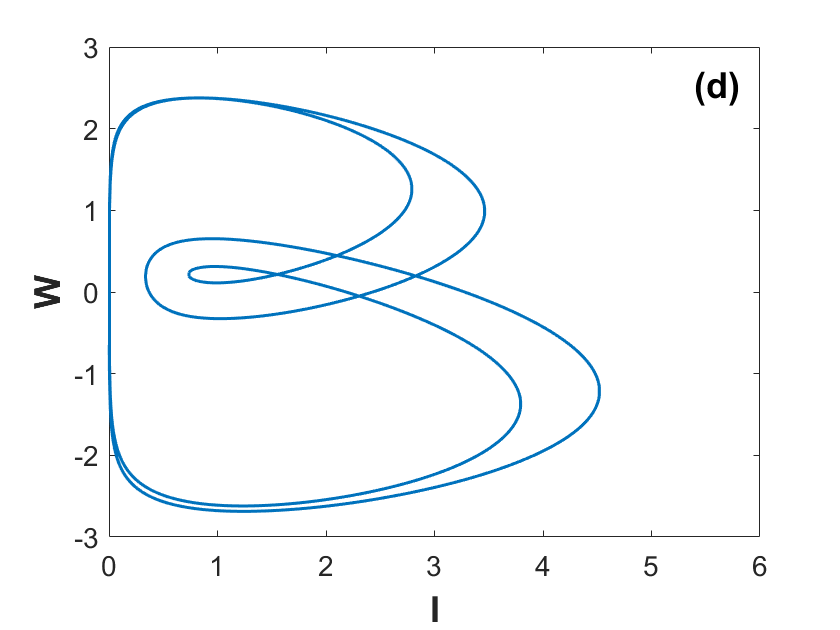}
\caption{(Color online) Asymptotic dynamics in the $(I,W)$ plane of a Class-B laser with modulated losses, Eqs. (\ref{LaserRateEqWmodu}) for $\mu=2$, $\epsilon=0.04$, $\omega = 1.1$ and (a) $m=0.0440$, (b) $m=0.0480$, (c) $m=0.0568$, and (d) $m=0.0584$. The orbits between panels (a)-(b) and (c)-(d) have undergone a period doubling bifurcation.}
\label{fig:LaserModulation}
\end{figure}
The Q-switching technique works particularly well for Class-B lasers as it makes use of the different time scales of the variables to produce the high energy pulses as shown in the first part of the evolution in Fig. \ref{fig:RO}(a). Slow modulations of the control parameters such as losses, pump rate or detuning, can indeed produce Q-switch pulses with very high peak powers. When moving closer to the RO, the natural laser pulse frequency, many nonlinear dynamical effects arise. In \cite{Tredicce85_2}, for example, a comparative analysis of the effectiveness of modulating either the cavity losses or the pump rate was presented. The final outcome is that for typical Class-B lasers such as CO$_2$ lasers, modulating the losses can easily be a hundred times more effective than modulating the pump. Three years earlier a team at the National Institute of Optics in Florence, Italy, had achieved a clear-cut transition to deterministic chaos via period doubling bifurcations in a CO$_2$ laser with modulated losses \cite{Arecchi82}. With the use of computational methods we show here how easy it is to verify transitions to deterministic chaos in laser rate equations with modulated losses given by
\begin{align}
\frac{dI}{d\tau} &= I \, \left(W  + \frac{m}{\epsilon} \cos(\omega t) \right ) \nonumber \\
\frac{dW}{d\tau} &= \mu  - (1+\epsilon W)( 1 + I)  \label{LaserRateEqWmodu}
\end{align}
where $m$ and $\omega$ are the amplitude and frequency of the modulation, with $\omega$ normalized to $\tau=\sqrt{2k \gamma_{\parallel}} \, t$ with $\epsilon = \sqrt{\gamma_{\parallel} / 2k}$ as in Eq. (\ref{LaserRateEqW}).

Fig. \ref{fig:LaserModulation} shows long-term (asymptotic) laser oscillations in the $(I,W)$ plane for $\mu=2$, $\epsilon=0.04$, $\omega = 1.1$ and $m=0.0440$ (panel (a)), $m=0.0480$ (panel (b)), $m=0.0568$ (panel (c)) and $m=0.0584$ (panel (d)). Period doubling bifurcations between panels (a)-(b) and (c)-(d) are clearly identifiable.
\begin{figure}[h]
\centering
\includegraphics[width=0.49\linewidth]{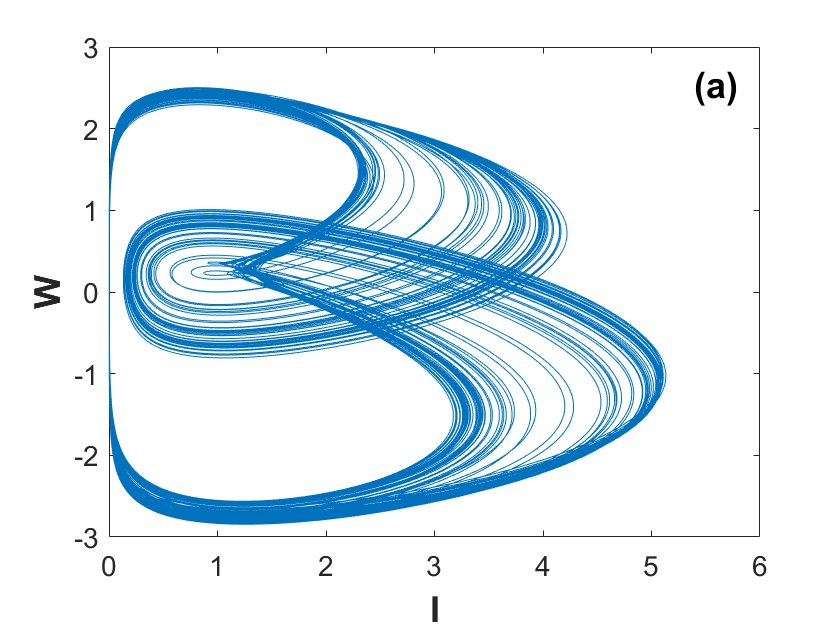}
\includegraphics[width=0.49\linewidth]{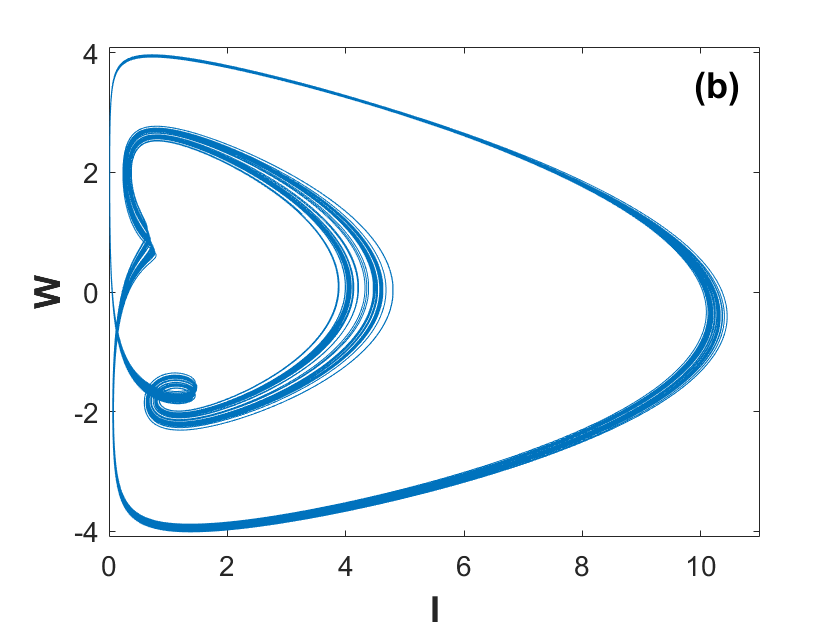}
\caption{(Color online) Asymptotic chaotic dynamics in the $(I,W)$ plane of a Class-B laser with modulated losses, Eqs. (\ref{LaserRateEqWmodu}) for $\mu=2$, $\epsilon=0.04$, and (a) $\omega = 1.1$, $m=0.0608$ and (b) $\omega = 0.47$, $m=0.080$.}
\label{fig:LaserModulationChaos}
\end{figure}
Fig. \ref{fig:LaserModulationChaos} shows long-term (asymptotic) chaotic laser oscillations in the $(I,W)$ plane for $\mu=2$, $\epsilon=0.04$, $\omega = 1.1$ and $m=0.0608$ (panel (a)), and $\omega = 0.47$ and $m=0.080$ (panel (b)) to demonstrate that deterministic chaos is ubiquitous in lasers with modulated losses.

\subsection{Lasers with Injected Signals.} \label{subsec:LIS}

A popular photonic device that bridges passive and active systems is the laser with an injected signal (or simply a laser with injection). Here a master laser of desirable frequency but perhaps limited power is injected into a subordinate laser through one of its mirrors (see Fig. \ref{fig:LaserCavities}(b)). The aim is to stabilise and/or to shift the output frequency of the subordinate laser that may operate at a higher power than the injection for useful and practical applications \cite{Lugiato83,Arecchi84,Tredicce85_1}. When the original frequencies of the master and subordinate lasers are originally not the same, an equal frequencies regime is attained at a finite value of the injection power, corresponding to a regime of locking between the two lasers. Before the locked state is reached, lasers with injected signals oscillate in time and display complex evolutions and even deterministic chaos. We start with the full laser equations (\ref{LaserFulleq}) with injection \cite{Oppo86,Solari94} 
\begin{align}
\frac{dE}{dt} &= -k \left[ (1 + i \theta) E - P \right] + i \eta' E + E'_{IN} \nonumber \\
\frac{dP}{dt} &= - \gamma_\perp \left[ (1 - i \theta) P - E D \right] + i \eta' P \label{LISFullEq} \\
\frac{dD}{dt} &= - \gamma_\parallel \left[ D - \mu + (1/2) \, (EP^* + E^* P) \right] \nonumber
\end{align}
where $E'_{IN}$ is the (real) amplitude of the injected signal and $\eta'$ is the difference between the external frequency of the injection (now the reference frequency $\omega_R$) and that of the laser frequency without injection $\omega_L$. In the general case of $(\gamma_{\perp} + k ) \gg \gamma_{\parallel}$ the adiabatic elimination of the material polarization variable leads to dynamical equations for Class-B lasers with injection given by \cite{Solari94}:
\begin{align}
\frac{dE}{d\tau} &= E \, W \left( 1 + i\theta  \right ) + i\eta E + E_{IN} \nonumber \\
\frac{dW}{d\tau} &= \mu  - 1  - \theta^2 - |E|^2 - \epsilon W \left( 1 + g |E|^2 \right)  \label{LISClassB_EW}
\end{align}
where the time has been normalized by the inverse of $\beta = [(1+\theta^2)(k^{-1}+\gamma_{\perp}^{-1})/\gamma_{\parallel}]^{1/2}$, $\epsilon=\gamma_{\parallel} \beta$, $\eta=\eta' \beta$, $E_{IN}=E'_{IN} \beta$ and $g=[(1+\theta^2)(1+k/\gamma_{\perp})]^{-1}$. These normalizations may appear cumbersome but in the case of relatively small $\theta$ and $k \ll \gamma_{\perp}$ they reduce to $\beta = [k \gamma_{\parallel}]^{-1/2}$, $\epsilon=\sqrt{\gamma_{\parallel}/k}$ and $g=1$ which are the normalizations used for Class-B lasers above and in \cite{Oppo86}.

In the following we focus on the almost tuned case of the injected laser where $|\theta| \ll 1$, $g \approx 1$ and $\eta$ is around $1$ although the results remain valid for wide ranges of these parameter values. The stationary states $(|E_S|^2, W_S)$ of Eqs. (\ref{LISClassB_EW}) are given by the implicit equations
\begin{align}
E_{IN}^2 &= |E_S|^2 \left( W_S^2 + \eta^2 \right) \nonumber \\
W_S &= \frac{\mu - 1 - |E_S|^2}{\epsilon \left( 1 + |E_S|^2 \right)} \label{LIS_SS}
\end{align}
and are plotted in Fig. \ref{fig:LIS_SS} for $\mu=2$, $\eta=1$ and $\epsilon=0.01$. 
\begin{figure}[h]
\centering
\includegraphics[width=0.49\linewidth]{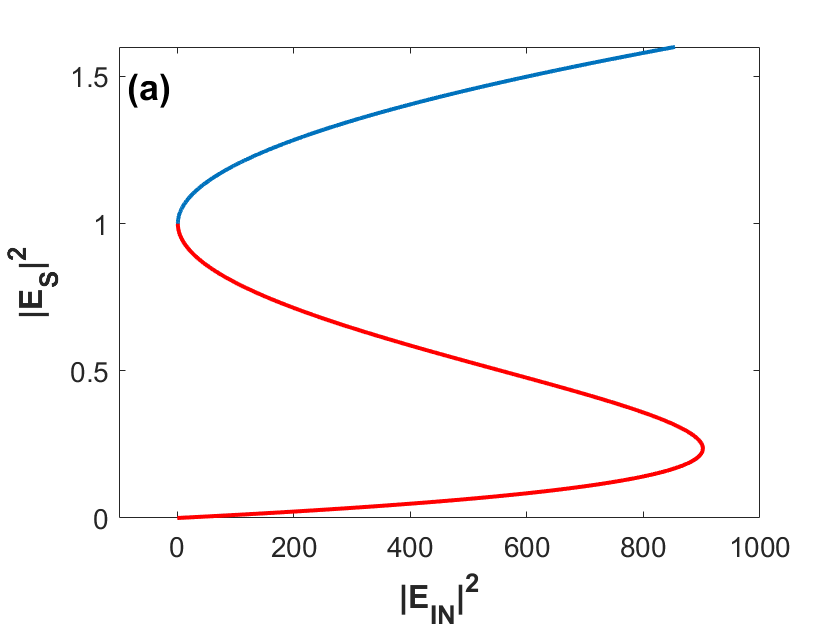}
\includegraphics[width=0.49\linewidth]{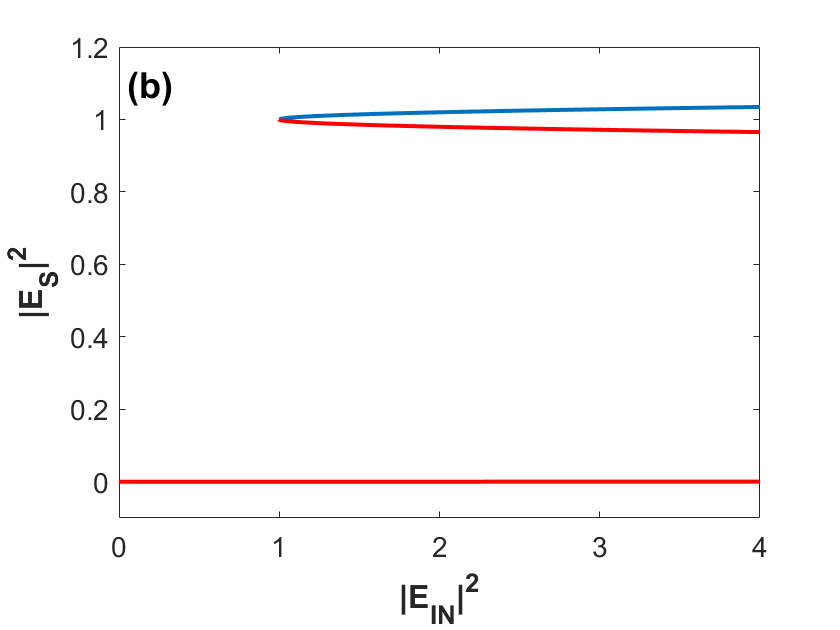}
\caption{(Color online) (a) Stationary states curve in the $(E_{IN}^2,|E_S|^2)$ plane (Eqs. (\ref{LIS_SS})) for a laser with an injected signal for $|\theta| \ll 1$, $g=1$, $\mu=2$, $\epsilon=0.01$ and $\eta=1$. The blue (red) curve denote stable (unstable) stationary states. (b) Magnification of (a) to show the region where there are no stable stationary states. The AH and saddle-node bifurcation overlap at the upper turning point of the S-shaped curve.}
\label{fig:LIS_SS}
\end{figure}
The stationary state curves are characteristically S-shaped, see Fig. \ref{fig:LIS_SS}(a), with two saddle-node bifurcations at the turning points of the S-shaped curves. It is possible to demonstrate that there is a further AH bifurcation wondering around the S-shaped stationary curves. In \cite{Oppo86} it was discovered that for certain values of the parameters, the AH bifurcation can overlap with the saddle-node bifurcation at the top turning point that is reached when increasing the intensity of the injection $E_{IN}^2$, specifically for $\mu-1=\eta^2$. This is a codimension-two bifurcation point leading to a rich dynamical unfolding \cite{Solari94}. For our parameters of choice $\mu=2$, the overlap of an AH and a saddle-node bifurcation takes place at $\eta=1$ and corresponds to a critical value of the intensity of the injected laser very close to one (see Fig. \ref{fig:LIS_SS}(b)). This means that the lower and central branches of the S-shaped curve of the stationary states are unstable while the upper branch is stable. The stable solution corresponds to a locking between the external laser and the subordinate laser where the emission frequency and phase of the two lasers are the same. This phenomenon is generally referred as an Adler frequency locking \cite{Adler46}.

\begin{figure}[h]
\centering
\includegraphics[width=0.49\linewidth]{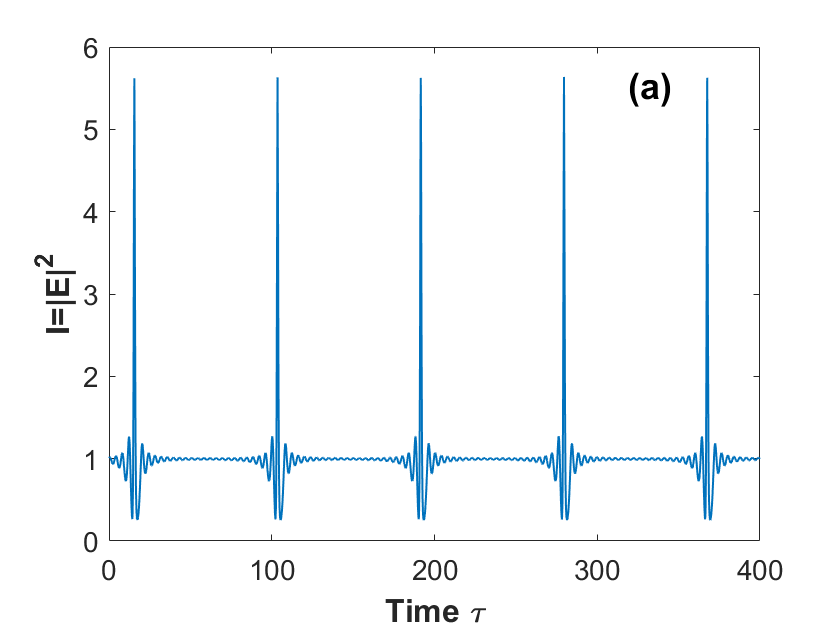}
\includegraphics[width=0.49\linewidth]{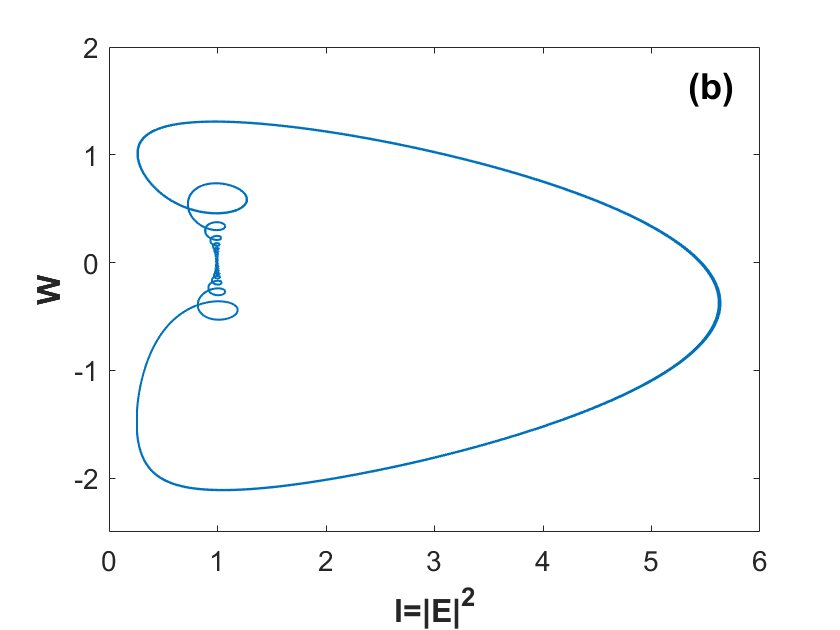}
\caption{(Color online) (a) Asymptotic intensity oscillations of the subordinate laser from direct integration of Eqs. (\ref{LISClassB_EW}) for $|\theta| \ll 1$, $g=1$, $\mu=2$, $\epsilon=0.01$, $E_{IN}^2=0.8$ and $\eta=0.9$. (b) Closed orbit in the $(|E|^2,W)$ plane.}
\label{fig:LIS_eta0_9}
\end{figure}
The presence of a codimension two bifurcation means that small variations of a control parameter can induce major changes in the behaviour of the system. For example, the dynamics close the locked stationary state drastically depend on the value of the detuning $\eta$. For $\eta<1$, the locking is reached through a saddle-node bifurcation where the frequency of the oscillation goes to zero and its amplitude displays a sudden jump to zero. For $\eta>1$, instead, the locking is reached through an AH bifurcation where the amplitude of the oscillation goes to zero while the frequency of the subordinate laser displays a sudden jump to zero. This is demonstrated by direct integration of Eqs. (\ref{LISClassB_EW}) for $\eta=0.9$ in Fig. \ref{fig:LIS_eta0_9} and for $\eta=1.3$ in Fig. \ref{fig:LIS_eta1_3}, respectively, while maintaining all the other parameters fixed. 

In Fig. \ref{fig:LIS_eta0_9}, we see that for $E_{IN}^2=0.8$ and $\eta=0.9$ the subordinate laser remains locked to the injection for long periods of time until sudden bursts of light with high intensity peaks are emitted when the phase performs a rotation of $2 \pi$. This corresponds to reaching the locking state by progressively extending the quiet regimes between pulses to infinity and decreasing the oscillation frequency to zero, which is typical of saddle-node dynamics. In this case the locking is reached through a frequency pulling phenomenon where the frequency of the subordinate laser is progressively pulled to that of the master laser when increasing the intensity of the injected signal $E_{IN}^2$ \cite{Oppo86}.

\begin{figure}[h]
\centering
\includegraphics[width=0.49\linewidth]{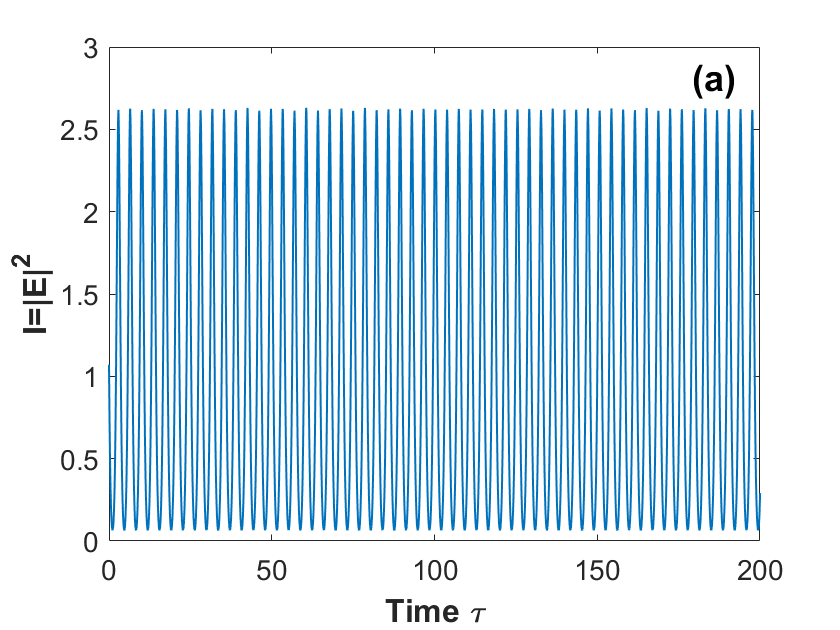}
\includegraphics[width=0.49\linewidth]{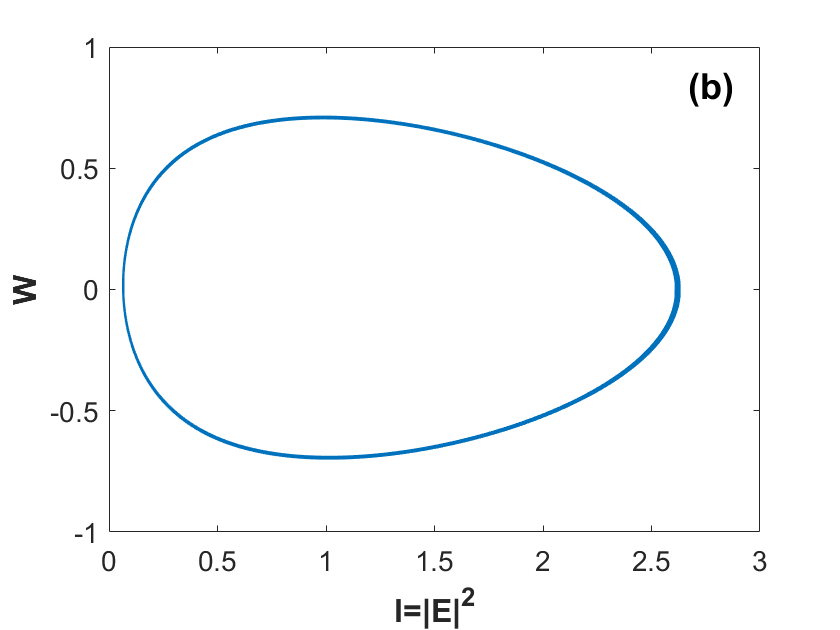}
\caption{(Color online) (a) Asymptotic intensity oscillations of the subordinate laser from direct integration of Eqs. (\ref{LISClassB_EW}) for $|\theta| \ll 1$, $g=1$, $\mu=2$, $\epsilon=0.01$, $E_{IN}^2=0.8$ and $\eta=1.3$. (b) Closed orbit in the $(|E|^2,W)$ plane.}
\label{fig:LIS_eta1_3}
\end{figure}
The phenomenology is quite different when the detuning $\eta$ takes values above the codimension-two bifurcation value of one. In Fig. \ref{fig:LIS_eta1_3}, we see that for $\eta=1.3$ and close to the final locked regime, the subordinate laser performs small-amplitude oscillations at a frequency that is much higher than that observed for $\eta=0.9$. This corresponds to reaching the locking state by progressively decreasing the oscillation amplitude to zero, which is typical of an AH bifurcation. In this case the locking is reached through a frequency pushing phenomenon where the frequency of the subordinate laser is progressively pushed away from that of the master laser when increasing the intensity of the injected signal $E_{IN}^2$ \cite{Oppo86}. This may appear at first counter-intuitive but it is in line with the features of the AH bifurcation where it is the amplitude of the oscillations that goes to zero at the locked state and not the frequency difference between the master and the subordinate lasers.

For values of the detuning $\eta$ close to the codimension-two bifurcation value of one, things can become even more intriguing. In Fig. \ref{fig:LIS_eta1_1}, we see that for $\eta=1.1$ and close to the final locked regime, the system is confused about what approach to chose to reach locking between the master and subordinate lasers; both kinds of oscillations, frequency pulled and frequency pushed, are stable and the subordinate laser approaches one or the other orbit depending only on the initial condition. 

\begin{figure}[h]
\centering
\includegraphics[width=0.49\linewidth]{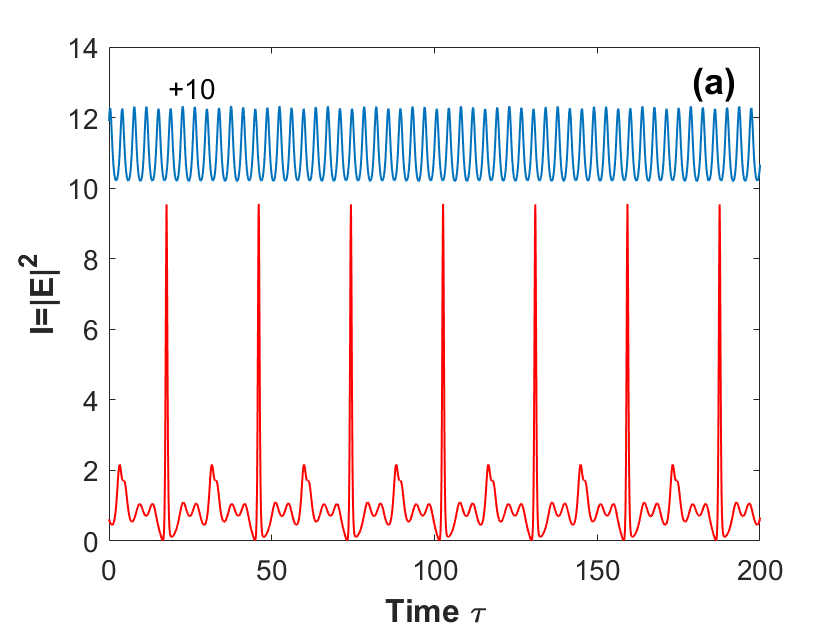}
\includegraphics[width=0.49\linewidth]{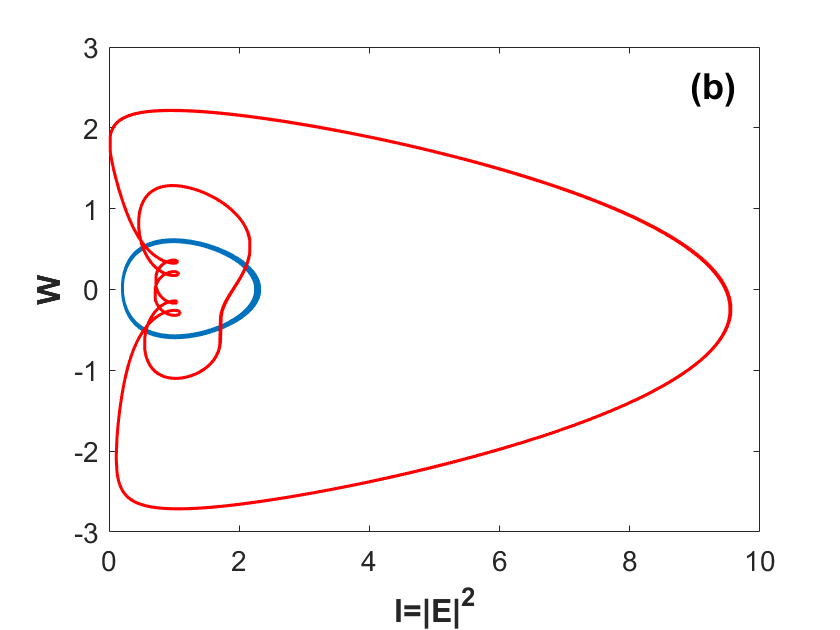}
\caption{(Color online) (a) Coexistence of two kinds of laser intensity oscillations in a laser with injected signal described by Eqs. (\ref{LISClassB_EW}) for $|\theta| \ll 1$, $g=1$, $\mu=2$, $\epsilon=0.01$, $E_{IN}^2=0.8$ and $\eta=1.1$. The blue time trace has been shifted by $+10$ to avoid overlaps with the red trace. (b) Closed orbits of (a) in the $(|E|^2,W)$ plane.}
\label{fig:LIS_eta1_1}
\end{figure}

The frequency pushing phenomenon discovered in \cite{Oppo86} and described above is not the only remarkable feature of lasers with injected signals. The dynamic implications of the codimension-two bifurcation have been exposed in their full glory in 1994 \cite{Solari94} and in 1997 \cite{Zimmermann97} and extended few years later to the full analysis of global bifurcations \cite{Zimmermann01}. We note that large and negative detunings $\theta$ in Eqs. (\ref{LISClassB_EW}) have exactly the same effect of the Linewidth Enhancement Factor making its analysis relevant to Semiconductor lasers too \cite{Solari94,Wieczorek05}.\\

\noi {\it {\bf Conservative and Dissipative features.}}
Having seen that in the limit of small $\epsilon=\sqrt{\gamma_{\parallel}/k}$ laser oscillations approach the conservative limit of the Toda potential, it is reasonable to ask what happens in this limit to a laser with an injected signal. The limit of small $\epsilon=\sqrt{\gamma_{\parallel}/k}$ in the case of negligible $|\theta|$ corresponds to neglecting the term $\epsilon W \left( 1 + g |E|^2 \right)$ in the equation for $W$ in Eqs. (\ref{LISClassB_EW}). We start with the stationary states. The equation for $W$ provides us with $|E_S|^2=\mu-1$, i.e. there are only two  stationary states (instead of the possible three states shown in the S-shaped curves of Fig. \ref{fig:LIS_SS}) of equal intensity and opposite phase. The stationary states are now given by
\begin{align}
|E_{S}|^2 &= \mu -1 \;\;\;\;\;\; E_S = - \frac{\mu-1}{E_{IN}} \left( W_S - i \eta \right) \nonumber \\
W_S &= \pm \sqrt{\frac{E_{IN}^2}{\mu-1} - \eta^2 } \label{LIStoda_SS}
\end{align}
requiring $E_{IN}^2 > (\mu-1) \eta^2$ and clearly showing the change of sign of the real part of the stationary field for the two existing states. By writing the electric field in real and imaginary part as $E=\alpha+i\beta$, Eqs. (\ref{LISClassB_EW}) in this limit can be rewritten as
\begin{align}
\frac{d \alpha}{d\tau} &= W \alpha - \eta \beta + E_{IN} \nonumber \\
\frac{d \beta}{d\tau} &= W \beta + \eta \alpha \label{LIS_CDalpha} \\ 
\frac{d W}{d\tau} &= \mu - 1 - \alpha^2 - \beta^2 \nonumber
\end{align}
These equations are equivalent to those investigated in \cite{Politi86} apart from a change of sign of $\eta$. Of course for $E_{IN}=0$ the reduced laser equations display a conservative motion in a Toda potential as shown in Fig. \ref{fig:Toda}. It is interesting to see that the linear stability analysis of the two stationary states given in Eqs. (\ref{LIStoda_SS}) show that one is stable, i.e. attractive, and the other is unstable, i.e. repulsive \cite{Politi86}. One would then expect that when changing the intensity of the injection from zero to the locking threshold $(\mu-1) \eta^2$, the system would switch suddenly from conservative, as for $E_{IN}^2=0$, to dissipative, as for $E_{IN}^2 > (\mu-1) \eta^2$. 

For $\mu=7/3$ and $\eta=-1$, this seems to be confirmed for, for example, $E_{IN}=0.20$ as shown in Fig. \ref{fig:LIS_ConsDiss}(a) where the Poincare' sections at the maxima of the oscillations of the laser intensity (i.e. for $W_{PS}= - E_{IN} \alpha / (\alpha^2+\beta^2)$) are displayed for four different initial conditions (colours black, blue, red and green) representative of the relevant dynamics in the $(\alpha,\beta)$ plane. 
\begin{figure}[h]
\centering
\includegraphics[width=0.49\linewidth]{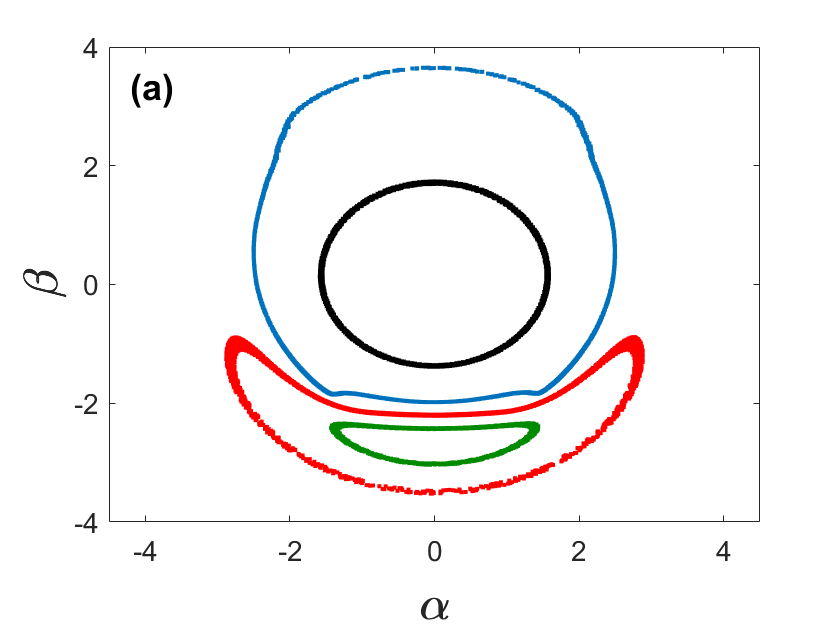}
\includegraphics[width=0.49\linewidth]{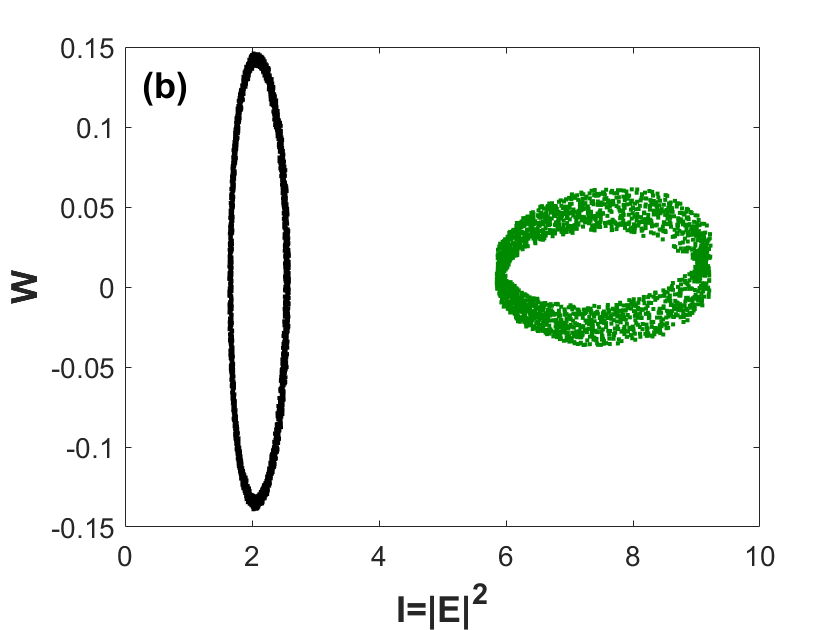}
\includegraphics[width=0.49\linewidth]{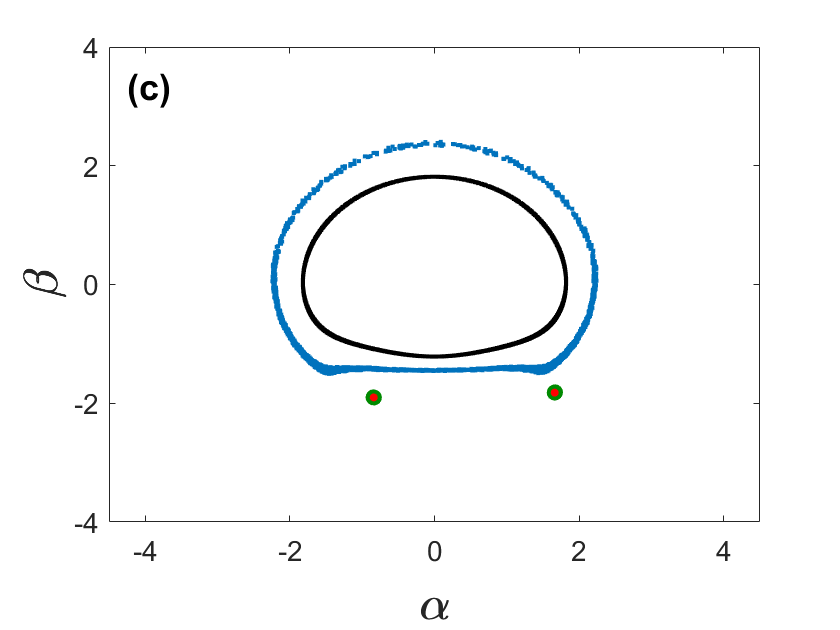}
\includegraphics[width=0.49\linewidth]{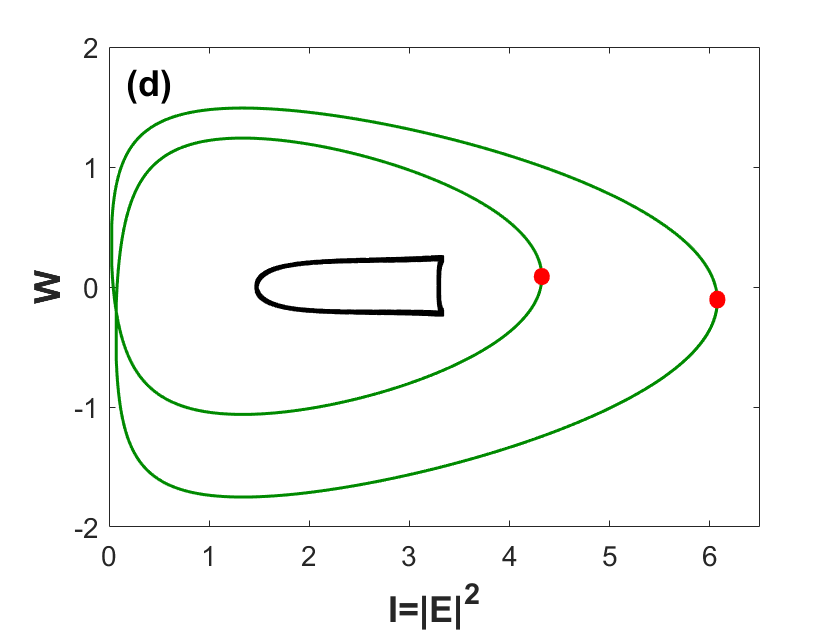}
\caption{(Color online) Poincare' sections at each maxima of the intensity $|E|^2=\alpha^2+\beta^2$ of the orbits obtained from numerical integration of Eqs. (\ref{LIS_CDalpha}) for $\mu=7/3$ and $\eta=-1$ and $E_{IN}=0.20$, (a)-(b), and $E_{IN}=0.42$, (c)-(d). (a) and (c) display the Poincare's sections in the $(\alpha,\beta)$ plane for four initial conditions while (b) and (d) in the $(|E|^2,W)$ plane for only two initial condition. In (d) the period two orbit corresponding to the two point Poincare' section (red points) has been added in green.}
\label{fig:LIS_ConsDiss}
\end{figure}
Each point of each of these curves represents the intersection of the temporal trajectory after one period of the oscillation when the next maximum of the laser intensity is detected on the $(\alpha,\beta)$ plane. The fact that all the four curves close onto themselves is a signature that the dynamics is now quasi-periodic and that the trajectories evolve on a torus. There are two families of tori in the phase space. The top one shows trajectories where the phase of the intensity peaks rotates around the centre of the plane $\alpha=\beta=0$. The second family of tori of banana shape and in colours of the Italian flag shows trajectories where the phase of the intensity peaks is bound and does not rotate around the centre of the plane. Any other initial condition generates a torus belonging to either the circular or the banana-shaped families. For $E_{IN}=0.20$ the phase space is foliated by these two families of tori of quasiperiodic motion. There exist a critical initial condition evolving on the separatrix between these two families of tori. The blue curve in Fig. \ref{fig:LIS_ConsDiss}(a) is indeed quite close to the separatrix torus and shows that motion slows down close to two cusps at the bottom of the curve. The separatrix torus is formed by an unstable period two orbit with Poincare' points that are close to the two cusps of the blue trajectory of Fig. \ref{fig:LIS_ConsDiss}(a) and are symmetric with respect to the axis $\alpha=0$, plus two heteroclinic orbits connecting the period-two points in the upper and lower parts of the curve \cite{Politi86}. The important message is that for $E_{IN}=0.20$ (and values below it) the overall dynamics of the laser with an injected signal is conservative in a way similar to what we have seen for the Toda potential in Subsection \ref{subsec:RO}. In Fig. \ref{fig:LIS_ConsDiss}(b) we show the Poincare' points of the black and green curves of panel (a) in the $(I,W)$ plane. The black curve is formed by points that appear on oval-shaped curves close to each other period after period during the evolution, typical of the circular shaped tori of panel (a). The green points instead appear on the plane with alternating positive and negative values of $W$ showing an underlying period-two dynamics typical of the banana shaped tori of panel (a).   

The behaviour of the system changes when we increase the injection intensity to, for example, $E_{IN}=0.42$ as shown in Fig. \ref{fig:LIS_ConsDiss}(c). While the family of circular tori with conservative features is still clearly visible, the family of the banana-shaped tori has collapsed into a single period two orbit (the two red and green points in Fig. \ref{fig:LIS_ConsDiss}(c)). This means that all initial conditions outside the separatrix orbit that contains the circular tori converge towards the two highlighted points in Fig. \ref{fig:LIS_ConsDiss}(c) or, in other words, dissipative dynamics towards an attractor is observed for all initial conditions outside the separatrix orbit. This is extraordinary. The dynamics of a laser with injected signal in the limit of negligible $\epsilon$ is either conservative or dissipative depending only on the initial condition: conservative for all initial conditions inside the separatrix orbit and dissipative for all others. This remarkable phenomenon was first discovered via numerical simulations in models of lasers with injection in \cite{Politi86} but is universal in dynamical systems with reversibility such as the invariance of Eqs. (\ref{LIS_CDalpha}) under the transformation $(\alpha,\beta,W,\tau) \rightarrow (-\alpha,\beta,-W,-\tau)$  as demonstrated, for example, in lattices of oscillators \cite{Topaj02} and for heat conduction in molecular dynamics \cite{Sprott14}.

Before concluding this section about the temporal dynamics of laser models, it is important to note that many other laser configurations benefit from the application of methods of computational physics such as lasers with two polarizations \cite{Puccioni87}, multi-level laser media \cite{Oppo89,Ciofini93}, lasers with saturable absorbers \cite{LugiatoBook}, coupled lasers \cite{Erzgraber08} and the investigation of noise and quantum fluctuations in laser devices \cite{Pashotta09}. And, of course, the generalizations to spatially dependent laser systems that we will see in Section \ref{sec:STDlaser}.

\section{Temporal Dynamics of Passive and Kerr Resonators}\label{sec:TDkerr}

We have seen that the prototypical active photonic devices, i.e. lasers, can be described by relatively simple ordinary differential equations. These equations, however, are intrinsically nonlinear and require numerical and computational methods for their investigation and for comparison with experimental configurations. The same applies to passive systems in the absence of a pump that creates population inversion and consequently amplification, i.e. not a laser. The nonlinear optics equations for passive resonators are remarkably similar to those seen for lasers but require some tuning of the parameters to guarantee physical accuracy. For convenience, we rewrite the full equations for the electric field in the cavity $E$, the electric polarization $P$ and the population difference $D$ in the single longitudinal and single transverse mode limit for a passive cavity \cite{LugiatoBook}   
\begin{align}
\frac{dE}{dt} &= -k \left[ (1 + i \theta) E - P \right] + E_{IN} \nonumber \\
\frac{dP}{dt} &= - \gamma_\perp \left[ (1 + i \Delta) P - E D \right] \label{PassiveFulleq} \\
\frac{d D}{dt} &= - \gamma_\parallel \left[ D + 2C + (1/2) \, (EP^* + E^* P) \right] \nonumber
\end{align}
Note the presence of the external drive term $E_{IN}$ like in the laser with injection, the replacement of the pump parameter $\mu$ with $-2C$ where $C$, the bistability parameter \cite{LugiatoBook}, is positive and describes medium absorption. When compared with the laser equations (\ref{LaserFulleq}) the definitions of the detuning $\theta$ ($\Delta$) is the difference between the cavity (atomic) frequency $\omega_c$ ($\omega_A$) and the reference frequency of the external drive, normalised to $k$ ($\gamma_\perp$). There is no laser frequency $\omega_L$ in the case of passive cavities. The passive cavity configuration is the same as that of Fig. \ref{fig:LaserCavities}(b) when the external pump $\mu$ is removed. 

At difference with the laser case where we focused on the Class-B case with $(k + \gamma_{\parallel}) \ll \gamma_{\perp}$, for the passive case we investigate the case of $k \ll \gamma_{\parallel} \approx \gamma_{\perp}$ that would correspond to Class-A dynamics for lasers. By applying a straightforward adiabatic elimination of the fast atomic variables $P$ and $D$ we obtain a single differential equation for the cavity field $E$ given by
\begin{equation}
\frac{dE}{d(kt)} = - (1+i\theta) E - \frac{2C (1-i\Delta)}{1+ \Delta^2 + |E|^2} \, E + y
\label{PassCav}
\end{equation}
where $y=E_{IN}/k$ is a normalised amplitude of the external drive. From the dynamical point of view, Eq. (\ref{PassCav}) is less exciting than the equations of Class-B lasers investigated in the previous sections. It is the nature and stability of the steady states that is interesting here and where computational tools can be of help in the interpretation of the physical phenomena. We are going to analyse two important limits.

\subsection{Bistability in Absorptive Media} \label{subsec:TDabsorb}
The first case of interest is the atom-cavity resonance when $\omega_A$ is very close to $\omega_C$. In this case $\Delta=k \theta/\gamma_{\perp}$ and is always much smaller than $\theta$ since we are in the limit of $k\ll\gamma_{\perp}$. By introducing $\kappa = k/\gamma_{\perp} \ll 1$, we can then rewrite Eq. (\ref{PassCav}) as
a function of the detuning $\theta$ only
\begin{equation}
\frac{dE}{d(kt)} = - (1+i\theta) E - \frac{2C (1 - i \kappa \theta)}{1 + |E|^2} \, E + y
\label{PassCavAbs}
\end{equation}
where we have neglected a term in $\Delta^2$ in the denominator of the nonlinear terms. At full resonance with the external drive, i.e. when $\theta=\Delta=0$, the medium in the cavity behaves as a pure saturable absorber whose transmission properties are modified in a nonlinear way by the intensity of the electric field in the cavity,  
\begin{equation}
\frac{dE}{d(kt)} = - E - \frac{2C}{1 + |E|^2} \, E + y
\label{PassCavAbsRes}
\end{equation}
\begin{figure}[h]
\centering
\includegraphics[width=0.49\linewidth]{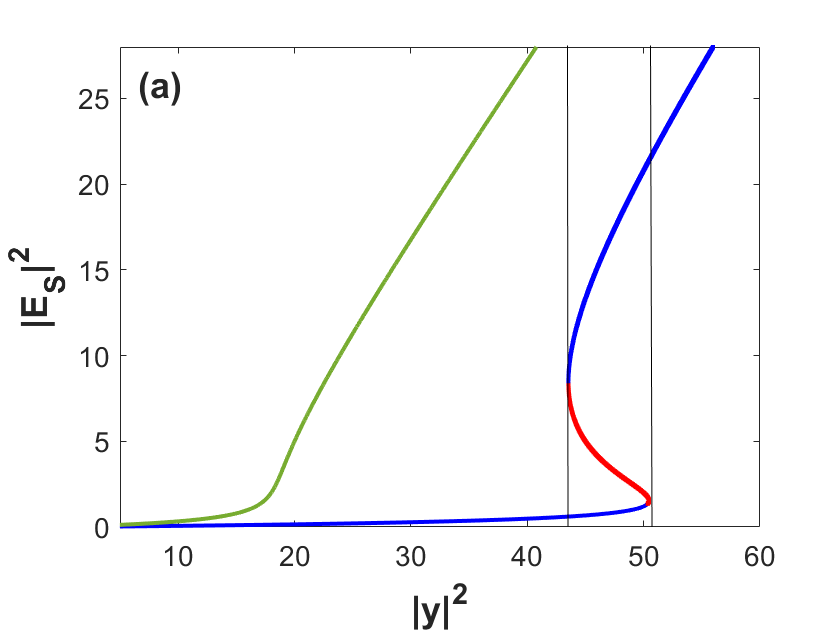}
\includegraphics[width=0.49\linewidth]{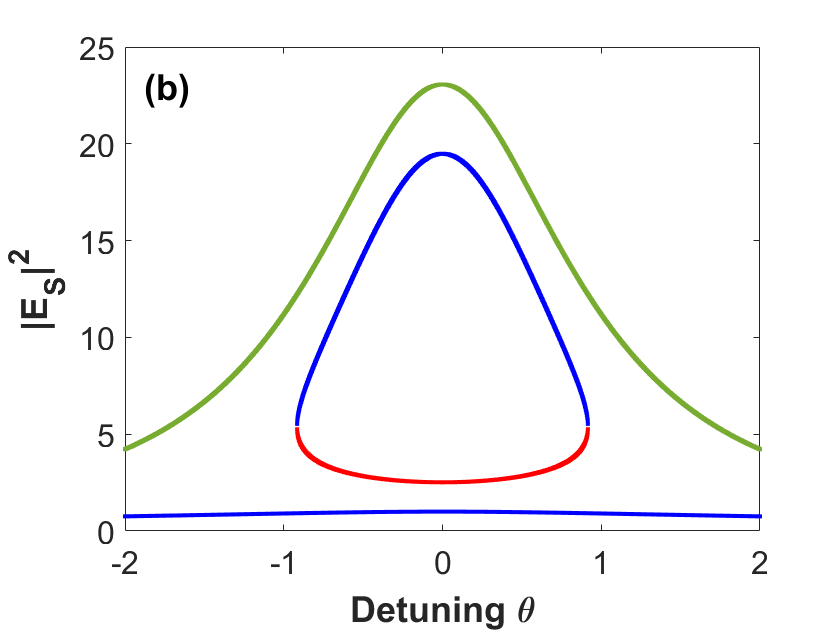}
\caption{(Color online) (a) Stationary state intensity $|E_S|^2$ of the fully resonant case of Eq. (\ref{PassCavAbsRes}) versus $y^2$ for $C=3$, green curve, and $C=6$, blue and red curves. (b) Stationary state intensity $|E_S|^2$ of Eq. (\ref{PassCavAbs}) versus the detuning $\theta$ for $\kappa=0.01$, $C=3$, $y=6$ (green curve) and for $C=6$, $y=7$ (blue and red curves). Red curves correspond to unstable stationary states.}
\label{fig:Bist_SS}
\end{figure}
The stationary state intensity $|E_S|^2$ of Eq. (\ref{PassCavAbsRes}) is shown in Fig. \ref{fig:Bist_SS}(a) when changing the power of the external driver $y^2$ for two values of $C$, $C=3$, green curve, and $C=6$, blue and red curves. The stationary state intensity $|E_S|^2$ of Eq. (\ref{PassCavAbs}) is plotted in Fig. \ref{fig:Bist_SS}(b) but when changing the input frequency in a scan of the detuning $\theta$ for $\kappa=0.01$, $C=3$, $y=6$ (green curve) and for $C=6$, $y=7$ (blue and red curves). From the point of view of numerical simulations, the curves of Fig. \ref{fig:Bist_SS}(a) (as well as those of Fig. \ref{fig:LIS_SS}) are easy to obtain since they are single valued in the $|E_S|^2$ variable. Things are different for the curves in Fig. \ref{fig:Bist_SS}(b) where standard numerical methods for the evaluation of these multivalued curves (see for example 'sp.solve' in Python or 'solve' in MATLAB) are routinely used, further demonstrating the usefulness of computational physics for photonic devices. 

From Fig. \ref{fig:Bist_SS} it is clear that the effects of the nonlinear terms are at their highest at cavity resonance and that there is a threshold value of $C$ after which multivalued stationary curves are obtained. A linear stability analysis can easily establish the stability of these stationary states to small perturbations. This can also be done by brute force by numerically integrating Eq. (\ref{PassCavAbs}) or Eq. (\ref{PassCav}) and by using initial conditions close to the stationary states, preferably provided in their real and imaginary components. The blue (red) curves in Fig. \ref{fig:Bist_SS} corresponds to stable (unstable) states. This means that above the threshold value of $C$, there are ranges in both $y$ and $\theta$ parameters where two stationary states are simultaneously stable, a phenomenon known as optical bistability \cite{LugiatoBook}. This phenomenon is due to a balance between the absorption of the medium that is a nonlinear function of the light intensity and the maximum intensity of the light that can be stored in the cavity. Starting from zero and increasing the amplitude of the input field, the absorption term proportional to $2C$ increases resulting in a rate of increase of the intracavity field intensity smaller than that in the absence of the medium. However for large enough field intensities, the absorption term proportional to $2C$ decreases to zero resulting in an almost transparent medium and a hysteresis cycle. Note that the absorptive bistability is symmetric with respect to the sign of the detuning $\theta$ and does not show any tilting of the resonance peak unlike the case of Kerr media studied in the next subsection.

\subsection{Bistability in Kerr Media} \label{subsec:TDkerr}
The second case of interest for optical bistability is the Kerr case when $|\Delta| \gg 1$. Photonic devices based on the Kerr effect take advantage of an optical refractive index that becomes a function of the electric field intensity. This phenomenon was discovered in Scotland by the reverend John Kerr 150 years ago in 1875 \cite{Kerr1875} further demonstrating that electromagnetism, electromagnetic waves and nonlinear optics are Scottish inventions of the nineteen century and born out of the fertile environment created by the Scottish Enlightenment.

For the largest majority of cases of two level media, the Kerr limit corresponds to the so-called red-detuned case when the frequency of the external laser drive is much smaller than that of the two-level medium, i.e. positive values of $\Delta$. By introducing the normalizations $E'=E/\Delta$, $C'=C/\Delta$, $y'=y/\Delta$ and $\Theta=(2C'-\theta)$ and then omitting the prime sign, we obtain
\begin{equation}
\frac{dE}{d(kt)} = - (1 - i\Theta) E - i \frac{2C \, |E|^2}{1 + |E|^2} \, E + y \, .
\label{KerrCav}
\end{equation}
This is the Kerr cavity equation where the detuning $\Delta$ has been absorbed into the normalization of the variables and parameters. In the further case of $|E|^2 \ll 1$ the so-called 'cubic' limit is recovered. For consistency with the literature \cite{LugiatoBook,Coen16} and without loss of generality, we write the 'cubic' limit equation for the complex conjugate field $F=E^*$ 
\begin{equation}
\frac{dF}{d(kt)} = - (1 + i\Theta) F + i 2C \, |F|^2 \, F + y \, .
\label{KerrCavCubic}
\end{equation}
The stationary states of this equation are given by
\begin{equation}
y^2 = |F_S|^2 \left( 1 + (2C \, |F_S|^2 - \Theta)^2 \right) \, .
\label{SSKerrCavCubic}
\end{equation}
\begin{figure}[h]
\centering
\includegraphics[width=0.49\linewidth]{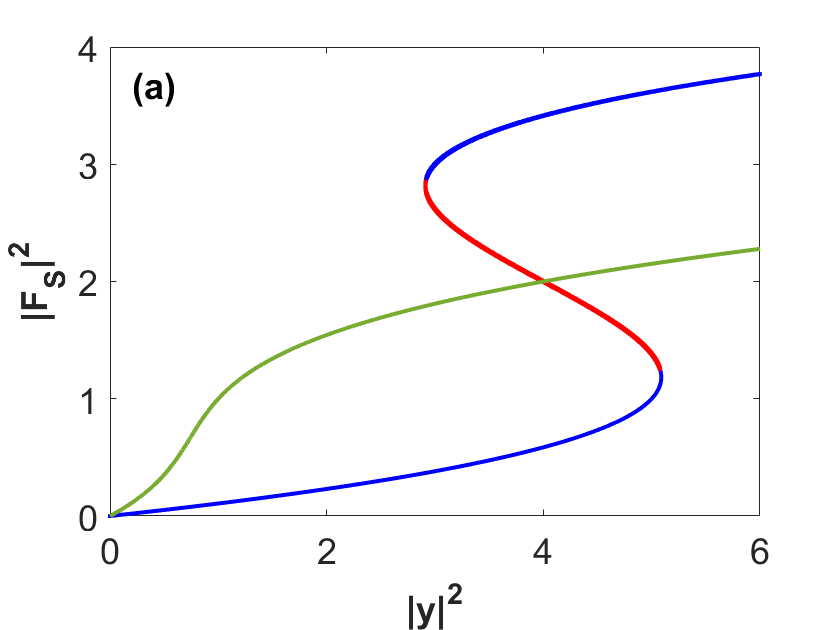}
\includegraphics[width=0.49\linewidth]{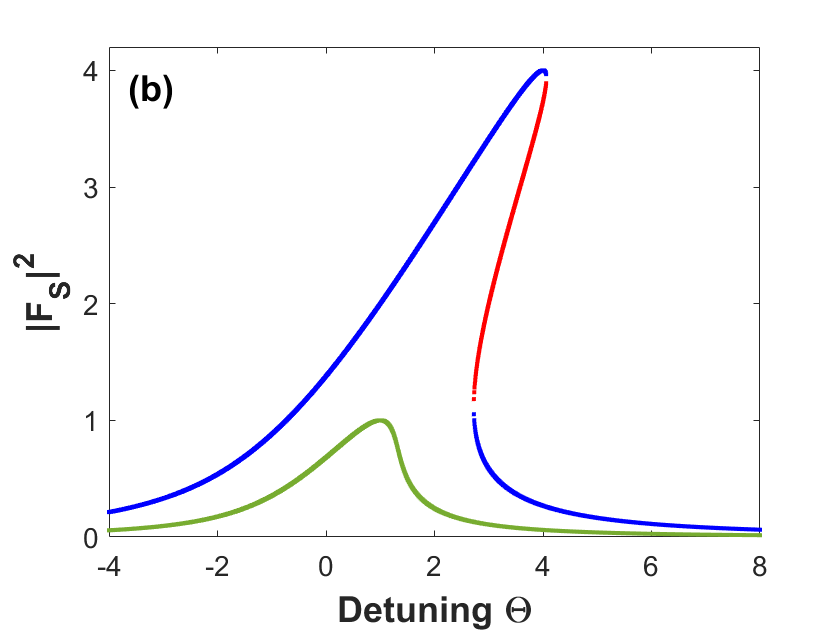}
\caption{(Color online) (a) Stationary state intensity $|F_S|^2$ of the cubic Kerr Eq. (\ref{KerrCavCubic}) versus $y^2$ for $C=0.5$, and for $\Theta=1$, green curve, and $\Theta=2$, blue and red curves. (b) Stationary state intensity $|F_S|^2$ of Eq. (\ref{KerrCavCubic}) versus the detuning $\Theta$ for $C=0.5$, and $y=1$ (green curve) and for $y=2$ (blue and red curves). Red curves correspond to unstable stationary states.}
\label{fig:Bist_KerrSS}
\end{figure}
The stationary state intensity $|F_S|^2$ of the cubic Kerr Eq. (\ref{KerrCavCubic}) is shown in Fig. \ref{fig:Bist_KerrSS}(a) when changing the power of the external drive $y^2$ for $C=0.5$ and two values of $\Theta$,  $\Theta=1$, green curve, and $\Theta=2$, blue and red curves. The same stationary state intensity $|F_S|^2$ is plotted in Fig. \ref{fig:Bist_KerrSS}(b) but when changing the input frequency in a scan of the detuning $\Theta$ for $C=0.5$ and two values of the external drive amplitude $y$, $y=1$, green curve, and $y=2$, blue and red curves. Again, from the point of view of numerical simulations, the curves of Fig. \ref{fig:Bist_KerrSS}(b) require standard numerical routines such as 'sp.solve' in Python or 'solve' in MATLAB. 

From Fig. \ref{fig:Bist_KerrSS} it is clear that the effects of the nonlinear Kerr terms are not the highest at resonance ($\Theta=0$) but for positive values of $\Theta$. When changing the input drive intensity $|y|^2$, there is a critical value of $\Theta$ that depends on the bistability parameter $C$, above which there are two stable branches of stationary states (blue curves in Fig. \ref{fig:Bist_KerrSS}) separated by an unstable branch of stationary states (red curves in Fig. \ref{fig:Bist_KerrSS}). A linear stability analysis can easily establish the stability of these stationary states to small perturbations especially for the cubic case of Eq. (\ref{KerrCavCubic}). In more complex cases of nonlinearity such as that of Eq.(\ref{KerrCav}), the stability of the different branches can be found by brute force by numerically integrating the dynamical equations and by using initial conditions close to the stationary states. In the Kerr case, the bistability phenomenon is due to the refractive index of the medium that is progressively modified by the light intensity in the cavity, exactly the Kerr effect mentioned above. In contrast to absorptive bistability that is symmetric with respect to the sign of the detuning, Kerr bistability is due to the 
tilting of the resonance peak as shown in Fig. \ref{fig:Bist_KerrSS}(b). When comparing these two kinds of optical bistability, special care need to be taken about the actual values of the parameters $C$, and $y$. These have been normalised by $\Delta$ in the Kerr case. For a typical value of $\Delta=50$, bistable values of $C$ and $y$ for the Kerr case are $50$ times larger than the corresponding values for the absorptive case. We will see the relevance of all these considerations when describing cavity solitons in passive cavity systems in Section \ref{sec:STDkerr}. 

Before concluding this section about the temporal dynamics of passive cavities, it is important to note that many other configurations benefit from the application of methods of computational physics such as Fabry-Perot cavities \cite{Lugiato23,Cole18}, passive cavities with two polarizations or counterpropagation \cite{Hill18}, cavity optomechanics \cite{Zhu23}, coupled passive cavities \cite{Biancalana11} and the investigation of noise and quantum fluctuations in passive photonic devices \cite{Reynaud89}. And, of course, the generalizations to spatially dependent passive cavities that we will see in Section \ref{sec:STDkerr}.

\section{Temporal Dynamics of Optical Parametric Oscillators}\label{sec:TDopo}

Lasers and bistable media in optical cavities are not the only interesting photonic devices that can be described effectively and efficiently by simple, although nonlinear, mathematical equations. Other important photonics devices are based on the nonlinear optical processes of second harmonic generation and parametric down conversion. Nonlinear optics through the use of laser beams was indeed discovered in 1961, just one year after the invention of the laser itself \cite{Franken61}. Here, we focus on the phenomenon of Parametric Down Conversion (PDC) in an optical cavity. The resulting photonic device is known as the Optical Parametric Oscillator or OPO with a missing 'P' \cite{OPO_Fisher77,OPO_Majid01,OPO_Melkonian21}.  

\begin{figure}[h]
\centering
\includegraphics[width=0.43\linewidth]{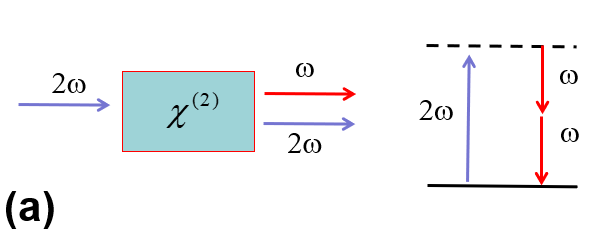}
\includegraphics[width=0.55\linewidth]{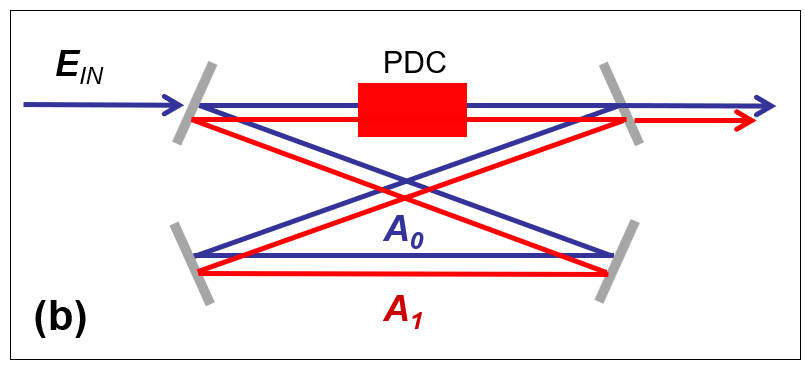}
\caption{(Color online) (a) Parametric Down Conversion (PDC) in a $\chi^{(2)}$ crystal and photon energy conservation through the generation of a virtual level (dashed line).  (b) Degenerate OPO cavity configuration. $A_0$ is the pump field at frequency $2\omega$, $A_1$ is the signal field at frequency $\omega$.}
\label{fig:DOPOcav}
\end{figure}
As shown in Fig. \ref{fig:DOPOcav}(a), in a PDC crystal an intense input laser beam with frequency $2\omega$ is capable to generate, in the degenerate case, a new output at frequency $\omega$ known as the signal. This process does not involve two energy levels from the energy quantization of the medium as studied in the previous Sections, but a 'virtual' level that exist for extremely short times that violate the energy-time Heisenberg uncertainty principle (see    Fig. \ref{fig:DOPOcav}(a)). Each photon at frequency $2\omega$ is converted into two photons at frequency $\omega$ due to energy conservation in the degenerate configuration studied here (in general there are two output fields at frequency $\omega_1$ and $\omega_2$, the signal and the idler fields such that $\omega_1+\omega_2=2\omega$). Second harmonic generation \cite{Franken61} is the opposite process where two photons at frequency $\omega$ combine to generate a photon at frequency $2\omega$. As PDC is highly inefficient being a purely quantum mechanic process, signal generation is often enhanced inside optical cavities as shown in Fig. \ref{fig:DOPOcav}(b) where an input beam $E_{IN}$ at frequency $2\omega$ is resonated in a cavity with a PDC crystal at degeneracy. The enhanced cavity field $A_0$ at frequency $2\omega$ (blue lines in Fig. \ref{fig:DOPOcav}(b)) generates the signal field $A_1$ (red lines in Fig. \ref{fig:DOPOcav}(b)) that is resonated in its own cavity too. The dynamical equations for these two fields in a Degenerate OPO (DOPO) are given by \cite{Lugiato88,LugiatoBook,Santagiustina02}
\begin{align}
\frac{dA_0}{dt} &= k_0 \left[ E_{IN} - (1 + i\theta_0) A_0 - A_1^2 \right] \nonumber \\
\frac{dA_1}{dt} &= k_1 \left[ - (1 + i\theta_1) A_1 + A_1^* A_0 \right] \label{DOPOequ}
\end{align}
where $E_{IN}$ is the real amplitude of the external drive at frequency $2\omega$, $k_0$ ($k_1$) is the cavity decay rate of the $A_0$ ($A_1$) field, $\theta_0$ ($\theta_1$) is the cavity detuning of the $A_0$ ($A_1$) field. For simplicity we consider the perfectly tuned case of $\theta_0=\theta_1=0$ since it is not easy to change the detunings while maintaining the degenerate operation of an OPO. A detailed derivation of Eqs. (\ref{DOPOequ}) is provided in \cite{Lugiato88} and is outside the scope of this paper. It is easy, however to physically justify all the terms in the two equations. The first three terms in the equation for $A_0$ corresponds to the external drive amplitude, the cavity losses and the cavity detuning. The nonlinear term $A_1^2$ describes the conversion of one photon of $A_0$ into two photons of $A_1$ with a frequency matching between $2 \omega$ and $(\omega+\omega)$. The first two terms in the equation for $A_1$ corresponds to the cavity losses and the detuning while the nonlinear term $A_1^* A_0$ describes the source of the signal field through the combination of one photon of $A_0$ with one photon of $A_1$ with an appropriate phase to satisfy frequency matching between $\omega$ and $(2\omega-\omega)$.
 
There are two stationary states solutions of Eqs. (\ref{DOPOequ}). The first one is trivially $A_0=E_{IN}$ and $A_1=0$ where not enough energy has been provided by the drive to compensate the losses and generate the signal field. The second stationary state solution is given in terms of field intensities by $|A_0|^2=1$ and $|A_1|^2=E_{IN} -1 $. Clearly this second stationary state can only exist if the amplitude of the drive $E_{IN}$ is larger than one. There is then a bifurcation at this value of $E_{IN}$ where the signal field is generated, the signal threshold. Mathematically such bifurcation is the same as that seen for the laser threshold in Fig. \ref{fig:laserSS} where $\mu$ is replaced by $E_{IN}$, $D_S$ by $|A_0|^2$ and $E_S$ by $A_1$. As for the laser case, we note that above threshold the signal field $A_1$ admits two possible values $A_1=\pm \sqrt{E_{IN} -1}$ corresponding to a zero and a $\pi$ phase, respectively. Both of these states are possible in the signal equation and we will see their relevance when discussing spatial effects in Section \ref{sec:STDopo}.
 
In Ref. \cite{Lugiato88}, bistability, stable temporal oscillations and even chaotic dynamics have been found when considering both detunings $\theta_0$ and $\theta_1$ different from zero with $k_0=k_1=1$. 
To avoid repetition and to show that chaotic dynamics is possible in fully resonant DOPOs with $\theta_0=\theta_1=0$, we consider Eqs. (\ref{DOPOequ}) at resonance and introduce a modulation of the signal cavity's losses
\begin{align}
\frac{dA_0}{dt} &= k_0 \left[ E_{IN} - A_0 - A_1^2 \right] \nonumber \\
\frac{dA_1}{dt} &= k_1 \left[ - (1 + m \cos(\omega t) ) A_1 + A_1^* A_0 \right] \, . \label{DOPOmod}
\end{align}
\begin{figure}[h]
\centering
\includegraphics[width=0.49\linewidth]{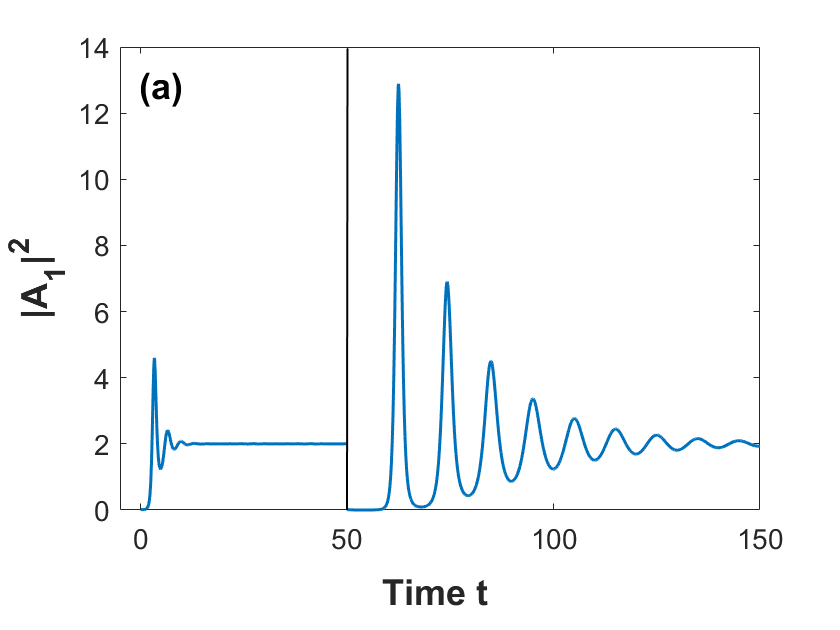}
\includegraphics[width=0.49\linewidth]{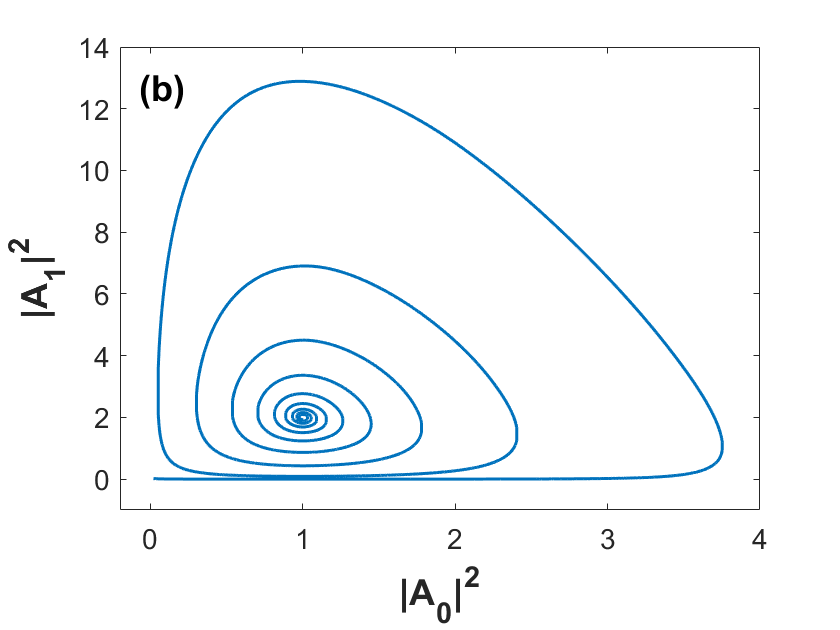}
\caption{(Color online) (a) Time evolution of the signal intensity for $E_{IN}=3$ and $m=0$ for, left hand side, $k_0=k_1=1$ and, right hand side, $k_0=0.1$ and $k_1=1$. (b) Relaxation to stationary state in the $(|A_0|^2,|A_1|^2)$ plane for $k_0=0.1$ and $k_1=1$.}
\label{fig:DOPO_RO}
\end{figure}
\begin{figure}[h]
\centering
\includegraphics[width=0.49\linewidth]{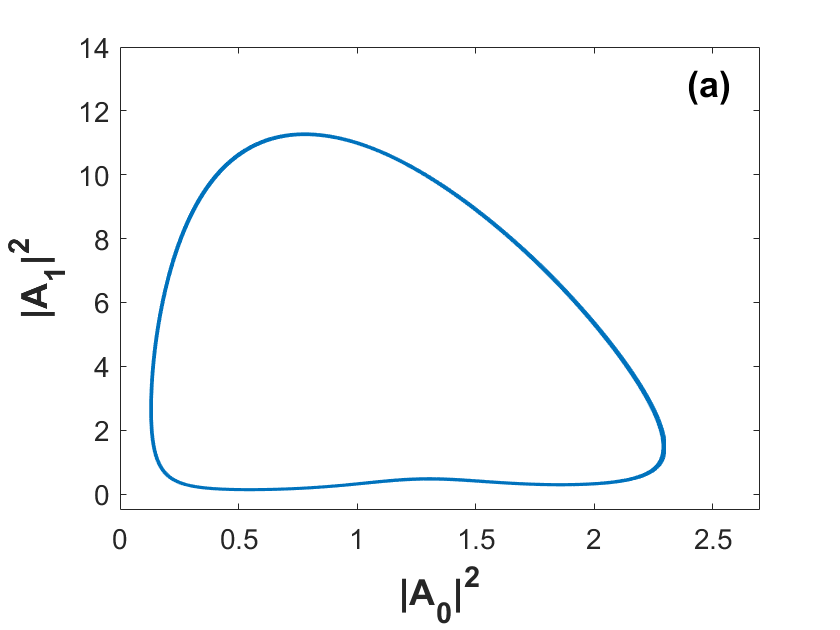}
\includegraphics[width=0.49\linewidth]{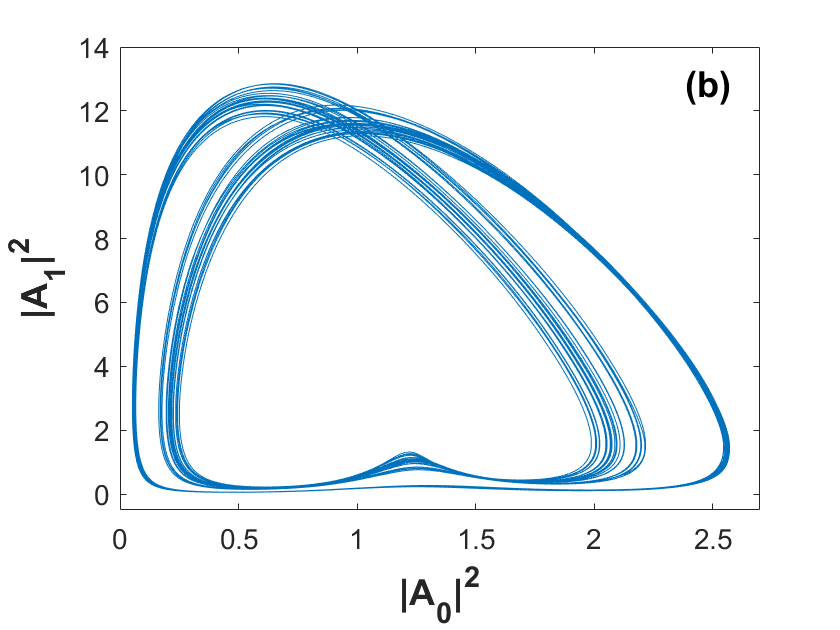}
\caption{(Color online) (a) Asymptotic limit cycle oscillations of the modulated DOPO Eqs. (\ref{DOPOequ}) for $k_0=0.1$, $k_1=1$, $E_{IN}=3$ and $m=0.5$ with $\omega=2\pi/5$ in the $(|A_0|^2,|A_1|^2)$ plane. (b) Asymptotic chaotic oscillations of the modulated DOPO Eqs. (\ref{DOPOequ}) in the $(|A_0|^2,|A_1|^2)$ plane for $m=0.62$ and the other parameters as in (a).}
\label{fig:DOPO_chaos}
\end{figure}
We have then integrated these equations numerically with the methods described in Appendix I. We start with no modulation $m=0$, $E_{IN}=3$ and see what happens when changing the decay rates of the two cavities first with $k_0=k_1=1$ and then with $k_0=0.1$ while keeping $k_1=1$, i.e. a pump cavity of higher finesse than the signal cavity.

In Fig. \ref{fig:DOPO_RO}(a) we see on the left that for $k_0=k_1=1$ there are small amplitude, fast relaxing oscillations to the stationary state $|A_0|^2=1$ and $|A_1|^2=2$ when starting with no signal field in the cavity. On the right hand side of Fig. \ref{fig:DOPO_RO}(a), the ROs to the same stationary state have hugely increased in amplitude and duration when considering $k_0=0.1$ instead of $k_0=1$. The role of changing $k_0$ while keeping $k_1=1$ on the ROs of a DOPO was investigated in details in \cite{Oppo01}. As we learned for the laser case, large amplitude and protracted ROs (i.e. small values of $k_0$) are ideal for the excitation of nonlinear chaotic dynamics by modulations of the cavity losses (see Subsection \ref{subsec:LasModLoss}).

This is exactly what is done in Fig. \ref{fig:DOPO_chaos} where the modulation amplitude $m$ is equal to $0.5$ in panel (a) and $0.62$ in panel (b) with a frequency of modulation equal to $2 \pi/5$. Sustained nonlinear oscillations are induced by the modulation of the losses of the signal cavity for $m=0.5$ as shown in  Fig. \ref{fig:DOPO_chaos}(a).   When increasing the amplitude of the modulation these oscillations undergo a full period doubling cascade to deterministic chaos (see Fig. \ref{fig:DOPO_chaos}(b) for $m=0.62$). It should then be possible to observe chaotic dynamics in DOPOs without using the detuning parameters. Since signal photons in a DOPO are quantum mechanically correlated (twin photons), DOPOs offer the possibility of coupling deterministic chaos and quantum entanglement for secure communications and cryptography.

In concluding this section, it is important to note that many other configurations of $\chi^{(2)}$ media in optical cavities benefit from the application of computational physics. These range from non-degenerate OPOs \cite{Santagiustina02} and coupled OPOs \cite{Phua98} to PDC crystals inside laser cavities \cite{OPO_Majid01}, from second harmonic generation inside optical cavities \cite{Aceves97} to the investigation of noise and quantum fluctuations in OPOs \cite{Chaturvedi02}. And, of course, the generalizations to spatially dependent OPOs that we will see in Section \ref{sec:STDopo}.

\newpage
\section*{Part B: Spatio-Temporal Description of Photonic Devices}
In Part B of this review paper, we include space-time coupling, i.e. the effects of tranverse diffraction and/or longitudinal group velocity dispersion, in the mathematical description of photonic devices such as lasers, absorbers, Kerr cavities and optical parametric oscillators. Mathematically the inclusion of spatial effects means moving from ODEs to partial differential equations (PDEs). Simple numerical methods for the integration of PDEs of relevance to photonic devices are presented in Appendix II. Building on the photonic models introduced in Part A and their dynamics, we review numerical simulations that generate important spatio-temporal structures, from Turing patterns to solitons, and their dynamics, from spatio-temporal chaos to turbulence. We first focus on the prototypical nonlinear Schr\"odinger equation  to then proceed to passive cavities (with Kerr and absorptive media) and to laser systems including lasers with injected signals, and with nested cavities. Finally we investigate OPOs where the optical nonlinearity has a different origin than the previous cases. 

\section{Light Propagation in a Kerr Medium}\label{sec:NLSE}
Building up from laser configurations that typically include optical cavities/resonators, we have seen in Part A that it is the propagation and interaction of coherent light with optically responsive media at the origin of the nonlinear terms in the model equations. Before entering into the details of cavity assisted systems, it is useful to consider the presence of spatially dependent terms during pure light propagation in a Kerr medium, typically an optical fibre \cite{AgrawalBook}. Here, we investigate a simple case and sample solutions of a nonlinear PDE system with relevance to photonic devices. We start from the standard propagation equation of coherent light $E$ in a Kerr medium, the Non-Linear Schr\"odinger Equation (NLSE)~\cite{NLSE2005}
\begin{equation}
\partial_z E + \frac {n}{c} \,\,\, \partial_t E = \frac{i}{2k} \nabla^2 E + i \sigma |E|^2 \, E
\label{NLSE}
\end{equation}
where $\partial$ denotes partial derivatives with respect to the given coordinate, $z$ is the propagation coordinate, $t$ is temporal coordinate, $n$ is the linear refractive index of the medium, $c$ is the speed of light in vacuum, $k$ is the light wave vector equal to $2\pi/\lambda$ with $\lambda$ being the wavelength, and $\sigma$ is equal to $\pm 1$ for focusing or defocusing Kerr media, respectively. The term $\nabla^2 E$ with $\nabla^2 = \partial^2_x+\partial^2_y$ describes diffraction in 2D, where $(x,y)$ are the coordinates of the plane perpendicular to the optical axis. The term $(2k)^{-1} \nabla^2 E $ can be replaced by $\gamma \partial^2_\tau E$ to account for group velocity dispersion instead of diffraction with the constant $\gamma$ being positive in the anomalous dispersion regime and negative in the normal dispersion regime with $\tau$ a retarded time.

The name Non-Linear Schr\"odinger Equation comes from the following considerations. In 1926 Erwin Schr\"odinger published his famous wave equation for quantum particles \cite{Schrod26}
\begin{equation}
i \hbar \partial_t \psi = H \psi = \left(- \frac{\hbar^2}{2m} \partial_x^2 + V(x,t) \right) \psi = - \frac{\hbar^2}{2m} \partial_x^2 \psi 
\label{SchrodEqu}
\end{equation}
where $\hbar$ is the reduced Planck constant, $\psi(x,t)$ is the complex wavefunction describing the wave properties of a quantum particle of mass $m$, $H$ is the Hamiltonian of the total energy that comprises a term for the kinetic energy and a term for the potential energy of the interactions $V(x,t)$. Note that $V(x,t)$ is zero for a free particle. 

In optics and in particular in laser physics where there is a preferential direction $z$ of the light beam, Maxwell equations for the propagation of the slowly varying amplitude $E$ of an electromagnetic wave in a medium of refractive index $n$ are provide
\begin{equation}
\partial_{\zeta} E = \left( \partial_z  + \frac {n}{c} \,\,\, \partial_t\right) E = \frac{i}{2k} \partial_x^2 E
\label{MaxProp}
\end{equation}
where $\partial_{\zeta}$ is a differential operator defined as $\partial_z + (n/c) \,\,\, \partial_t$ and we have considered a single transverse variable $x$ for convenience. This equation is mathematically equivalent to the Schr\"odinger Equation (\ref{SchrodEqu}) although the physical meaning of the terms and variables are quite different. We have already seen for passive cavities in Eq. (\ref{KerrCavCubic}) in Section \ref{sec:TDkerr} that the Kerr effect where the refractive index depends on the light intensity is mathematically described by $i |E|^2 E$. When this term is added to Eq. (\ref{MaxProp}) the NLSE is obtained with obvious meaning of the chosen name. The NLSE in nonlinear optics was first written in 1964 by Raymond Chiao, Elsa Garmire and Charles Townes \cite{Chiao64}.

Funnily enough just few year earlier, in 1961, Eugene Gross \cite{Gross61} and Lev Pitaevskii \cite{Lev61} had derived a NLSE, known as the Gross-Pitaevskii equation (GPE), when considering quantised vortices in a Bose-Einstein Condensate (BEC). For BEC with atom-atom interactions and for light propagating in a Kerr medium the GPE and NLSE are, respectively
\begin{align}
i \hbar \partial_t \psi &= \left(- \frac{\hbar^2}{2m} \partial_x^2 + g |\psi|^2 \right) \psi  \;\;\;\;\; {\rm (GPE)}\\
i \partial_{\zeta} E &= - \frac{1}{2} \partial_x^2 E - \sigma |E|^2 E \;\;\;\;\; {\rm (NLSE)} \label{NLSEv2}
\end{align}
where $g$ is proportional to the scattering length that is positive (negative) for repulsive (attractive) atomic interactions and where $\sigma$ is equal to +1 for self-focusing media and -1 for self-defocusing media. Note that the NLSE originates from a Hamiltonian $H=\int \left[(1/2) |\partial_x E|^2-|E|^4\right] \; dx $ and hence conserves the total energy \cite{NLSE2005} like, of course, the GPE. In view of the analogy above, the coordinate $\zeta$ can be seen as a temporal coordinate although it contains both time $t$ and propagation direction $z$. 

Nonlinearity used to be anathema in quantum physics and the GPE had to wait until the realization of the first BEC in 1995 \cite{Anderson95} to establish its own success. Luckily for us things have been quite different in nonlinear optics where general application of the NLSE to a variety of experimental configurations and devices started already in 1960's with major breakthroughs in 1970's and 1980's \cite{AgrawalBook} as we will see later. It is important also to mention that the NLSE finds application in the description of small-amplitude gravity waves on the surface of deep water with zero viscosity \cite{NLSE2005}.

We start the investigation of the NLSE with the simplest of the solutions, the fundamental flat solution for a generic input intensity $P$ is 
\begin{equation}
E_0=\sqrt{P} \exp(iP \zeta)=\sqrt{P} \left(\cos(P \zeta)+i\sin(P \zeta)\right)
\label{FlatNLSE}
\end{equation}
which has no dependence on the spatial coordinate $x$. The linear stability analysis of this solution is done by introducing a perturbation $\delta$ via $E=(\sqrt{P}+\delta) \exp(iP \zeta)$ so that $|E|^2=E E^* \approx P+\sqrt{P}(\delta+\delta^*)$. Then the NLSE for $E$ and $E^*$ linearised in $\delta$ provide
\begin{align}
iP(\sqrt{P}+\delta) + \partial_\zeta \delta &= i \left(P+\sqrt{P}(\delta+\delta^*)\right) (\sqrt{P}+\delta) 
+\frac{i}{2} \partial_x^2 \delta \\
-iP(\sqrt{P}+\delta^*) + \partial_\zeta \delta^* &= -i \left(P+\sqrt{P}(\delta^*+\delta)\right) (\sqrt{P}+\delta^*) 
-\frac{i}{2} \partial_x^2 \delta^* \, .
\end{align} 

We consider now $\delta = \epsilon \exp(\lambda \zeta + iKx)$, $\delta^* = \epsilon \exp(\lambda \zeta - iKx)$, where $K$ is the spatial wavevector of the perturbation of magnitude $\epsilon \ll 1$. We then replace these spatial and temporal perturbations $\delta$ in the above equations to obtain the eigenvalue matrix for $\lambda$
\begin{equation}
   \begin{vmatrix} 
   i(P-K^2/2)-\lambda & iP  \\
   -iP & -i(P-K^2/2)-\lambda \\
   \end{vmatrix} 
\end{equation}
that provides us with the characteristic polynomial whose solutions are
\begin{equation}
\lambda^2 = \frac{K^2}{2} \left(2P-\frac{K^2}{2} \right) 
\;\;\;\;\;\;\;\;\;\;\;\;\;\;
\lambda = \pm \frac{K}{\sqrt{2}} \sqrt{2P-\frac{K^2}{2}} \, .
\label{eigenNLSE}
\end{equation}
Typical of conservative systems, as well as zero eigenvalue there are eigenvalues (either real or purely imaginary) that have the same magnitude and opposite sign. In the literature this low $K$ instability is referred to as 'modulational' (or Benjamin-Feir) instability but it is nothing else than a Turing instability (see Subsection \ref{subsec:Turing}) for a conservative system. By evaluating $d (\lambda^2) / d(K^2)$ and where it is zero, it is easy to find a critical wavevector $K_{crit}=\sqrt{2P}$ corresponding to the first unstable wavevector when increasing $P$ from zero. This wavevector has the maximum value of the real part of the instability eigenvalue $\lambda$ at threshold. The band of unstable wavevectors, i.e. wavevectors $K$ with a real and positive eigenvalue, ranges from $0$ to $2\sqrt{P}$. 
\begin{figure}[h]
\centering
\includegraphics[width=0.49\linewidth]{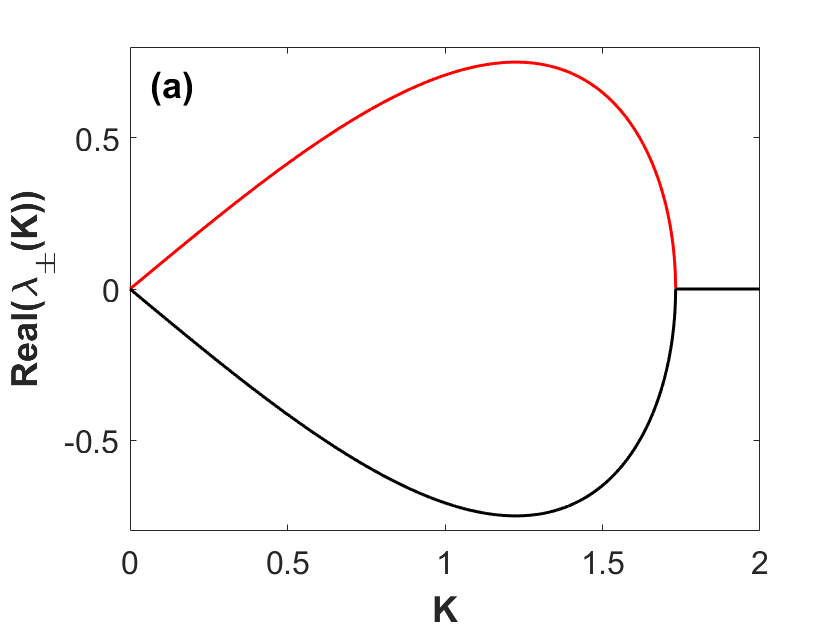}
\includegraphics[width=0.49\linewidth]{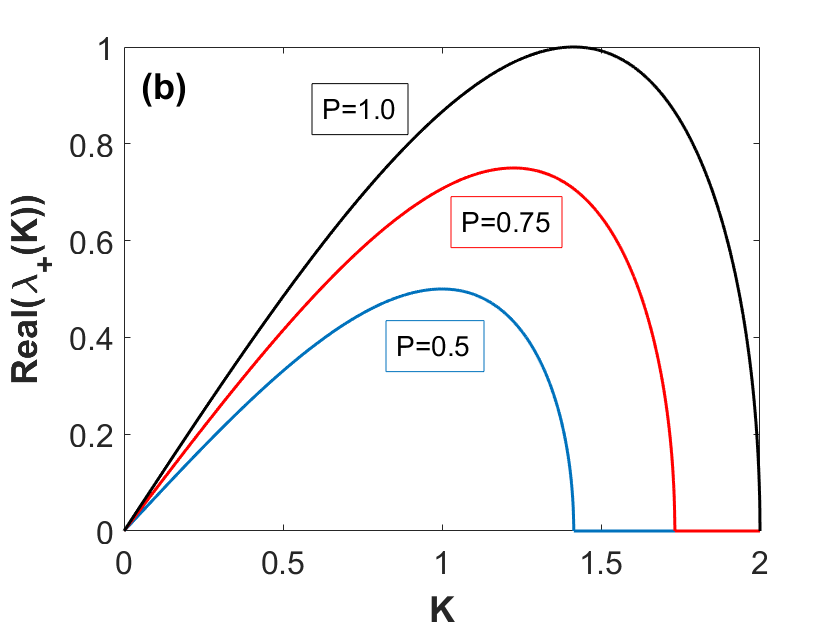}
\caption{(Color online) (a) Real parts of the eigenvalues $\lambda$ for the stability of the flat solution of Eq. (\ref{FlatNLSE}) from Eq. (\ref{eigenNLSE}) for $P=0.75$. (b) Positive real parts of the stability eigenvalues $\lambda$ from Eq. (\ref{eigenNLSE}) for $P=0.5$, blue curve, $P=0.75$, red curve, and $P=1.0$, black curve.}
\label{fig:ModInstab}
\end{figure}
%

In Fig. \ref{fig:ModInstab} we show the real part of the eigenvalues $\lambda$ for the stability of the flat solution $E_0$ (see Eq. (\ref{FlatNLSE})) from Eq. (\ref{eigenNLSE}). Fig. \ref{fig:ModInstab}(a) shows both positive and negative real parts of the eigenvalues as a function of the wavevector $K$ for $P=0.75$. The band of unstable wavevector extends from zero to $2\sqrt{0.75}$. Fig. \ref{fig:ModInstab}(b) shows the positive real parts of the eigenvalues $\lambda$ and the bands of unstable wavevectors for $P=0.5, 0.75$ and $1$, via the blue, red and black curves, respectively. These results show that any input of flat shape along $x$ at time $\zeta=0$ will develop multi-wavevector instabilities during propagation. This, however, does not exclude the existence of input profiles along the spatial variable $x$ that remain unchanged or that periodically oscillate during propagation as demonstrated in the next subsection.    

\subsection{Bright Solitons in the NLSE}
In 1834 Scottish engineer John Scott Russell observed the 'Wave of Translation' (or Soliton) in the Glasgow to Edinburgh Canal. In his own words: "I was observing the motion of a boat which was rapidly drawn along a narrow channel by a pair of horses, when the boat suddenly stopped — not so the mass of water in the channel which it had put in motion; it accumulated round the prow of the vessel in a state of violent agitation, then suddenly leaving it behind, rolled forward with great velocity, assuming the form of a large solitary elevation, a rounded, smooth and well-defined heap of water, which continued its course along the channel apparently without change of form or diminution of speed. I followed it on horseback, and overtook it still rolling on at a rate of some eight or nine miles an hour [14 km/h] preserving its original figure some thirty feet [9 m] long and a foot and a half [30-45 cm] in height. Its height gradually diminished, and after a chase of one or two miles [2–3 km] I lost it in the windings of the channel. Such, in the month of August 1834, was my first chance interview with that singular and beautiful phenomenon which I have called the Wave of Translation." 

Solitons are peculiar localised and single (sometimes multiple) peak waves that differ greatly from trains of alternating peaks and troughs typical of water surface waves after a perturbation (like a thrown stone). In 1972 V. Zakharov and A. Shabat \cite{Zakharov72} found an analytical nonlinear wave solution corresponding to a {\bf solitary} wave for the NLSE (\ref{NLSEv2}) for self-focusing media, i.e. with $\sigma=+1$ 
\begin{equation}
E(\zeta,x) = \frac{e^{i \zeta/2}}{\cosh(x)} = e^{i \zeta/2} {\rm sech}(x) \, .
\label{soliton}
\end{equation}
\begin{figure}[h]
\centering
\includegraphics[width=0.49\linewidth]{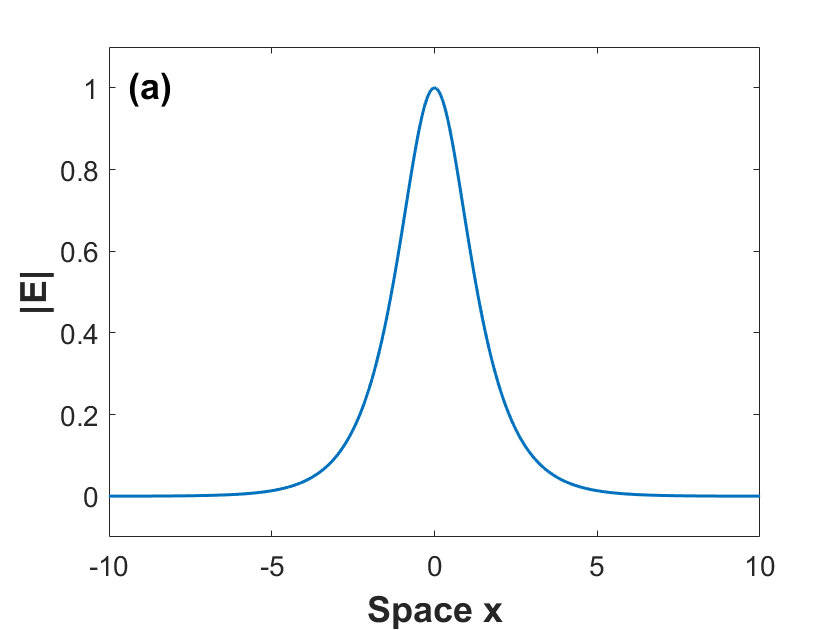}
\includegraphics[width=0.49\linewidth]{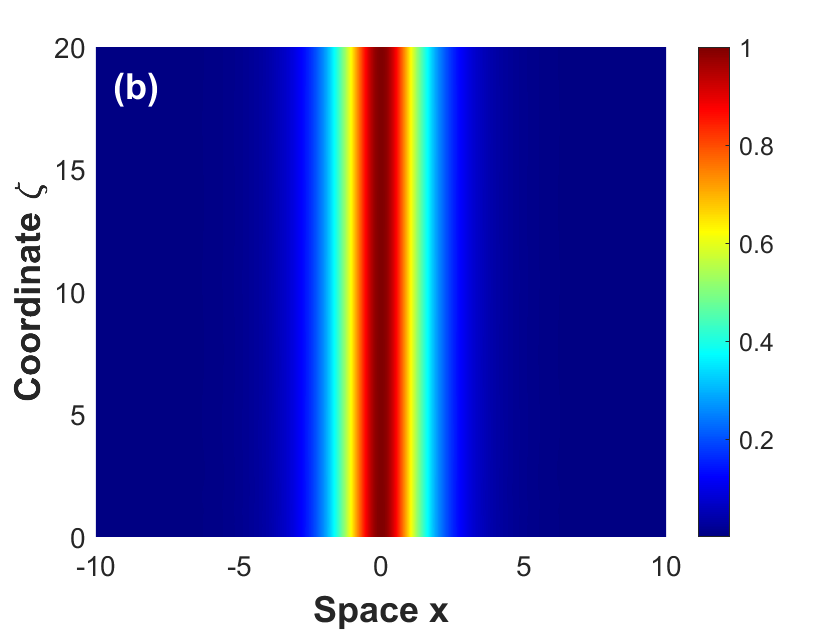}
\caption{(Color online) Numerical simulation of the NLSE (\ref{NLSEv2}) for $\sigma=+1$ and initial condition given by Eq. (\ref{soliton}) at $\zeta=0$. (a) End profile of the amplitude $|E|$ of the bright soliton after a propagation of $20$ $\zeta$ units. (b) Full spatio-temporal evolution of the bright soliton profile of Eq. (\ref{soliton}) showing no change during propagation.}
\label{fig:BrightSolitonN1}
\end{figure}

In contrast to the instabilities that affect the flat solutions investigated in the previous section, the soliton solution (\ref{soliton}) propagates unchanged along the coordinate $\zeta$ and corresponds to a balance of the self-focusing Kerr effect and the spatial spreading due to either diffraction or anomalous group velocity dispersion. In the optics literature the soliton solution (\ref{soliton}) is generally referred to  as the sech$^2$ soliton corresponding to its intensity. We demonstrate this soliton feature by direct numerical simulations of Eq. (\ref{NLSEv2}) for $\sigma=+1$ and by using the split-step method described in Appendix II. Fig. \ref{fig:BrightSolitonN1}(a) shows the final spatial profile of the amplitude $|E|$ from the numerical simulation after a propagation of $20$ units of $\zeta$. Such profile is indistinguishable from that of Eq. (\ref{soliton}). This is true not just for the amplitude profile but also for the real and imaginary part profiles of $E$ that contain the explicit dependence from $\zeta$ through the phase. Fig. \ref{fig:BrightSolitonN1}(b) displays the full spatio-temporal evolution of the amplitude of the bright soliton showing no changes and no instabilities during propagation. Bright optical solitons of sech$^2$ type via dispersion were first observed by L. Mollenauer, R. Stolen and J. P. Gordon in optical fibres in 1980 \cite{Mollenauer80} for applications in optical communications. Remarkable observations of bright optical solitons of sech$^2$ type via diffraction have been made in CS$_2$ liquids in 1985 \cite{Barthelemy85}, in glass waveguides in 1990 \cite{Aitchison90} and in nematic liquid crystals layers in 2003 \cite{Assanto03}. 

This is not the full story of solitons in the NLSE. In 1974 J. Satsuma and N. Yajima \cite{Satsuma74} showed that the bright soliton given by Eq. (\ref{soliton}) is just the first soliton solution of an entire family of localized nonlinear waves of the NLSE. They even provided an analytical expression for the second of these higher order solitary waves that turned out to be oscillating in the $\zeta$ coordinate
\begin{equation}
E(\zeta,x) =  4e^{i \zeta/2} \frac{\cosh(3x)+ 3e^{4i \zeta} \cosh(x)}{\cosh(4x)+4\cosh(2x)+3\cos(4 \zeta)} \, .
\label{solitonN2}
\end{equation}
\begin{figure}[h]
\centering
\includegraphics[width=0.49\linewidth]{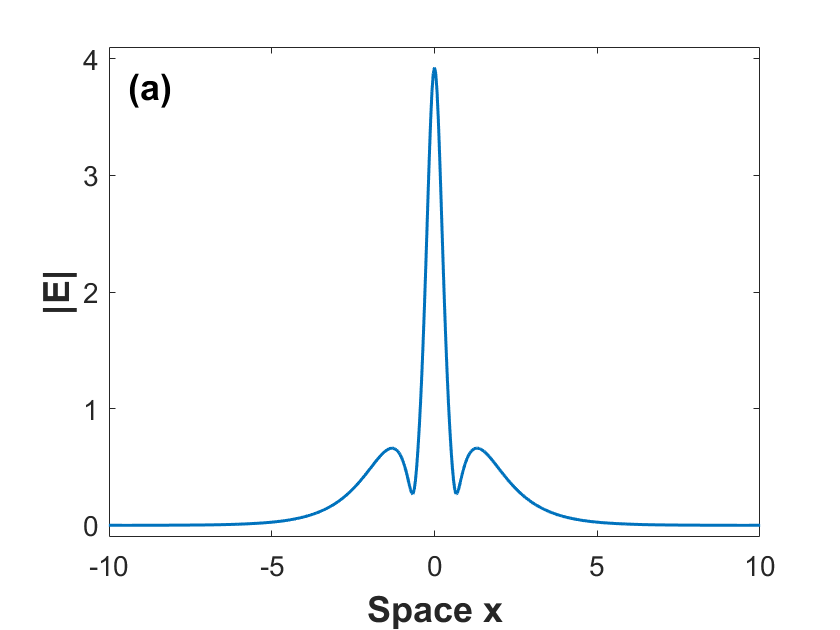}
\includegraphics[width=0.49\linewidth]{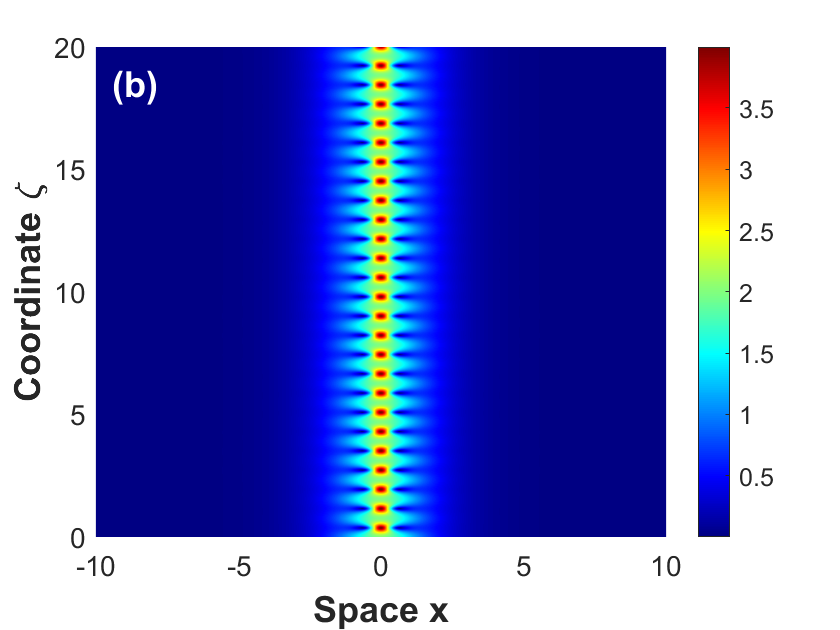}
\caption{(Color online) Numerical simulation of the NLSE (\ref{NLSEv2}) for $\sigma=+1$ and initial condition given by Eq. (\ref{solitonN2}) at $\zeta=0$. (a) End profile of the amplitude $|E|$ of the bright soliton after a propagation of $20$ $\zeta$ units. (b) Full spatio-temporal evolution of the bright soliton profile of Eq. (\ref{solitonN2}) showing the temporal oscillation during propagation.}
\label{fig:BrightSolitonN2}
\end{figure}
Again in contrast to the instabilities that affect the flat solutions investigated in the previous section, the soliton solution (\ref{solitonN2}) remains confined in space while propagating along the coordinate $\zeta$ while oscillating. We demonstrate this soliton feature by direct numerical simulations of Eq. (\ref{NLSEv2}) for self-focusing media ($\sigma=+1$) and by using the split-step method described in Appendix II. Fig. \ref{fig:BrightSolitonN2}(a) shows the final spatial profile of the amplitude $|E|$ from the numerical simulation after a propagation of $20$ units of $\zeta$ starting from Eq. (\ref{solitonN2}) at $\zeta=0$. Such final profile is indistinguishable from that obtained from Eq. (\ref{solitonN2}) for $\zeta=20$. Fig. \ref{fig:BrightSolitonN2}(b) shows the full spatio-temporal evolution of the amplitude of the higher order bright soliton clearly showing the oscillations and maintained localization during propagation. It is interesting to note that bright oscillating optical solitons of the type (\ref{solitonN2}) via dispersion were also observed experimentally by L. Mollenauer, R. Stolen and J. P. Gordon in optical fibres in 1980 together with the sech$^2$ soliton of Eq. (\ref{soliton}) \cite{Mollenauer80}. Oscillating solitons like Eq. (\ref{solitonN2}) belong to the category referred as "breathers" and have counterparts in the sine-Gordon equation and in Fermi-Pasta-Ulam-Tsingou chains.  

\subsection{Dark Solitons in the NLSE} \label{subsec:DarkNLSE}
To complete this short survey of soliton solutions in the NLSE, it is important to mention dark solitons. Dark solitons as opposed to grey (or gray) solitons possess a precise point in space where the intensity is equal to zero with a phase jump of $\pm \pi$ (hence topological). An example of a non moving dark soliton in the NLSE is
\begin{equation}
E(\zeta,x) =  E_0 \tanh (E_0 x) e^{-i E_0^2 \zeta} 
\label{DarkSoliton}
\end{equation}
where $E_0$ is the real and positive amplitude of the Continuous Wave (CW) solution $E_0  \exp(-i E_0^2 \zeta)$. It is easy to prove that both the dark solitons and the CW solution satisfy the self-defocusing NLSE
\begin{equation}
i \partial_{\zeta} E = - \frac{1}{2} \partial_x^2 E + |E|^2 E 
\label{NLSE_SD}
\end{equation}
obtained from Eq. (\ref{NLSEv2}) with $\sigma=-1$. It is important to note that Eq. (\ref{NLSE_SD}) also describes propagation of light in a focusing medium with normal group velocity dispersion when considering the complex conjugate field $F=E^*$. In this case one obtains:
\begin{equation}
i \partial_{\zeta} F = + \frac{1}{2} \partial_x^2 F - |F|^2 F \, .
\label{NLSE_SDconjg}
\end{equation}
Hence dark solitons (\ref{DarkSoliton}) are solutions of propagation in a Kerr defocusing medium with diffraction or that of a focusing Kerr medium with normal group velocity dispersion. 
\begin{figure}[h]
\centering
\includegraphics[width=0.49\linewidth]{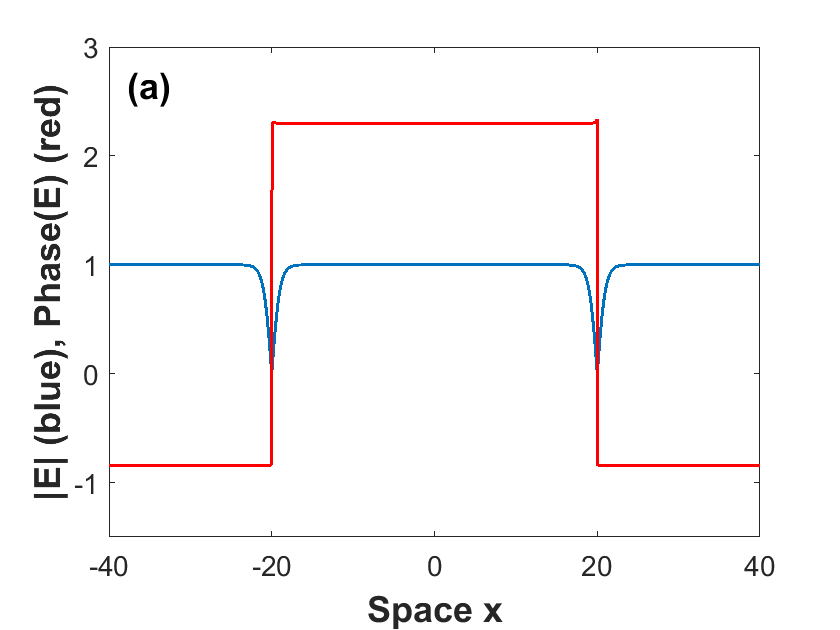}
\includegraphics[width=0.49\linewidth]{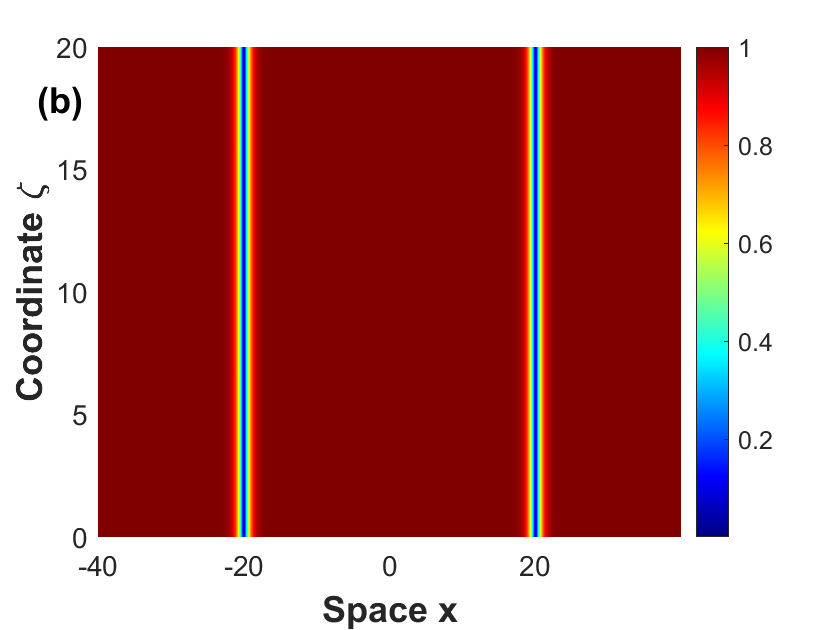}
\caption{(Color online) Numerical simulation of the self-defocusing NLSE (\ref{NLSE_SD}) with initial condition given by Eq. (\ref{DarkSoliton}) with $E_0=1$ and pinned at $x=-L/4$ and at $x=+L/4$ when $\zeta=0$, where $L$ is the size of the spatial domain. (a) End profiles of the amplitude $|E|$ (blue curve) and the phase of the field $E$ (red curve) of two dark solitons after a propagation of $20$ $\zeta$ units. (b) Full spatio-temporal evolution of the dark soliton profiles showing no change in amplitude during propagation.}
\label{fig:DarkSolitons}
\end{figure}
As already seen for the bright soltions of the NLSE, dark solitons of the form (\ref{DarkSoliton}) propagates unchanged along the coordinate $\zeta$. We demonstrate this soliton feature by direct numerical simulations of Eq. (\ref{NLSE_SD}) for self-defocusing media by using the split-step method described in Appendix II. Since the split-step method uses spatial Fourier Transforms, periodic boundary conditions are required. For example, the CW solutions on the sides of the dark soliton (\ref{DarkSoliton}) located in $x=0$ at $\zeta=0$ correspond to $E=\pm E_0$, i.e. on the left hand side we have $E=E_0$ and on the right hand side we have $E=-E_0$ with a corresponding jump of the phase of $\pi$. The hyperbolic tangent profile of the dark soliton connects two CW solutions of the same amplitude but with different phases. Hence, in order to match periodic boundary conditions required by the numerical method, we study the propagation of two dark solitons in the NLSE as shown in Fig. \ref{fig:DarkSolitons}. Fig. \ref{fig:DarkSolitons}(a) shows the final spatial profile of the amplitude $|E|$ and of the phase of the field $E$ from the numerical simulation after a propagation of $20$ units of $\zeta$. Such profile is indistinguishable from that provided by Eq. (\ref{DarkSoliton}) for two initial dark solitons. This is true not just for the amplitude profile but also for the real and imaginary part profiles of $E$ that contain the explicit dependence from $\zeta$ through the phase. Note the phase jumps of $+\pi$ and $-\pi$ when passing through the dark solitons as shown by the red curve in Fig. \ref{fig:DarkSolitons}(a). Fig. \ref{fig:DarkSolitons}(b) displays the full spatio-temporal evolution of the amplitude of the dark solitons showing no changes and no instabilities during propagation. Dark solitons in the NLSE were predicted in 1973 by A. Hasegawa and F. Tappert in the normally dispersive NLSE (\ref{NLSE_SDconjg}) with a tanh profile supported by numerical simulations \cite{Hasegawa73}. The first experimental observation was done in the group of A. Barhtelemy in 1987 \cite{Emplit87}. Odd-symmetry input pulses appropriate for launching a dark soliton were used in a 1988 experiment with good agreement with the theoretical prediction and numerical simulations \cite{Weiner88}. For further details about dark optical solitons and their applications see \cite{Kivshar98}.


\section{Spatio-Temporal Dynamics of Passive and Kerr Resonators}\label{sec:STDkerr}
As we have seen in Part A of this review article, many photonic devices use optical resonators to enhance light-matter interaction for practical purposes, the laser being the best example. Light propagation in Kerr media as described in the previous section is of course of fundamental relevance but neglects losses that are unavoidable. To compensate the losses, energy is provided from outside the cavity in the form of an incoherent pump (lasers) or a coherent injection (passive cavities). These systems are intrinsically outside the thermodynamic equilibrium.   

In this section we review passive resonators as those studied in Section \ref{sec:TDkerr} but with added spatial effects due to either diffraction or group velocity dispersion. The main aim is to show that the concept of optical soliton originally introduced in energy-conserving propagation in the NLSE (see above) extends, no problemo, to cavity configurations where energy is not conserved, the dynamics is intrinsically dissipative due to cavity losses and analytical expressions are in general not available. These soliton structures inside optical cavities are known as {\bf Cavity Solitons} (CS) \cite{Firth98,Ackemann09,Oppo24}. We start with the spatio-temporal dynamics and cavity solitons in Kerr resonators because of their enormous success in the generation and application of frequency combs since the seminal work of F. Leo and collaborators in Brussels in 2010 \cite{Leo10,Firth10}. We then follow with a section about cavity soltions in absorptive cavities.

\subsection{Cavity Solitons and Turing Patterns in Kerr Resonators}
Standard model equations for the generation and observation of CSs in passive cavities have been derived and reviewed recently in \cite{Oppo24}. We consider the optical cavities shown in Fig. \ref{fig:LaserCavities}(b) but with no external pump $\mu$ for the diffractive case and that of Fig.~\ref{Cavities}(a) for the dispersive case where an input light beam $E_{IN}$ is guided in a ring resonator (although the physics of sometimes more practical Fabry-Perot and folded resonators is very similar \cite{Firth21,Lugiato23}). 
\begin{figure}[h]
\includegraphics[width=0.98\columnwidth]{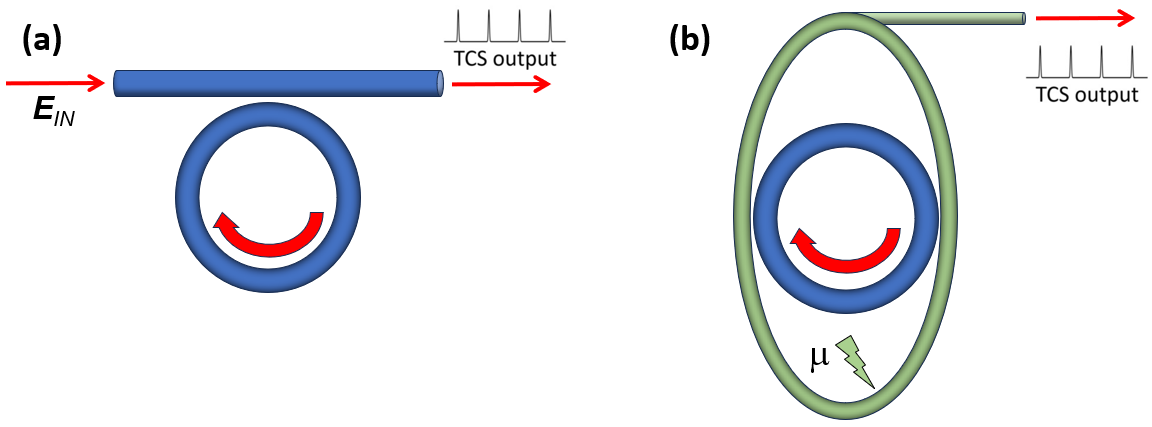}
\caption{(a) Ring cavity for the case with group velocity dispersion with a typical output formed by a train of temporal CSs. (b) Nested cavities formed by a gain loop around a Kerr cavity generating a train of laser temporal CSs. $\mu$ is the external pump.}
\label{Cavities}
\end{figure}
Starting from the NLSE (\ref{NLSE}), i.e. the standard propagation equation of coherent light $E$ in a Kerr medium, one  obtains the Lugiato-Lefever Equation (LLE) for the case with diffraction \cite{Oppo24}
\begin{equation}
\partial_{\zeta} E = E_{IN} - (1+i\theta) E + i \sigma |E|^2 E + i a \partial^2_x E \, ,
\label{LLE}
\end{equation}
in one transverse dimension $x$ (see Fig. \ref{fig:LaserCavities}(b) but with no external pump $\mu$) and the LLE for the case with group velocity dispersion \cite{Haelterman92b,Castelli17,Oppo24} 
\begin{equation}
\partial_{t} E = E_{IN} - (1+i\theta) E + i \sigma |E|^2 E + i \beta \partial^2_\tau E . 
\label{FTLLE}
\end{equation}
(see Fig.~\ref{Cavities}(a)) where $\beta$ is the group-velocity dispersion coefficient, $t$ is known as the slow time since it describes how the intra-cavity field evolves over many round trips, while $\tau$ is the fast time in a reference frame travelling at the group velocity of the driving field during a round trip of the cavity and related to the propagation direction $z$ via $\tau=nz/c$ with $c$ being the speed of light and $n$ the refractive index of the medium \cite{Coen16,Castelli17}. Mathematically, Eqs. (\ref{LLE}) and Eqs. (\ref{FTLLE}) are equivalent. They are the renowned Lugiato-Lefever model (LLE) \cite{Lugiato87}. An interesting feature of the temporal LLE is that the coefficient $\beta$ can be either positive or negative in the anomalous or normal dispersion regimes, respectively, as seen for the NLSE in the previous section. A historical review of the LLE is provided in \cite{Lugiato18}.

From the early 1990's through the work at Strathlcyde (Firth, Scroggie, Harkness), Brussels (Tlidi, Mandel) and Como (Lugiato, Brambilla, Tissoni, Prati), CSs and more generally dissipative solitons, have been identified and characterized in several models of photonic devices, starting from those describing Kerr cavities \cite{Scroggie94,Firth98}. It is incorrect and counter-productive to shift the beginning of the history of CSs in Kerr resonators to after 2013 (see for example the paper on Science Volume 361, eaan8083 (2018)). In the case of optical resonators the balance of the self-focusing Kerr effect and the spatial spreading due to either diffraction or anomalous group velocity dispersion shown by NLSE solitons, is supplemented by the balance between the input energy  and the cavity losses \cite{Firth98,Ackemann09}. By using the LLE that is basically the NLSE with an input drive, cavity losses and a detuning, CSs where predicted and observed numerically first in \cite{Scroggie94} and later generalised in \cite{Firth96_PS} and \cite{Firth02}. Such structures can be natural ‘bits’ for parallel processing of optical information, especially if they exist in semiconductor micro-resonators. In the longitudinal direction (along the cavity axis), CSs represent optimal pulses with pyramidally shaped frequency spectra. These spectra can easily span more than one octave in the frequency (or wavelength) domain while being formed by thousand of components separated by the free spectral range of the optical cavity round trip time. These kinds of spectra are known as frequency combs and CS generated frequency combs have taken the optics community by storm. CS generated frequency combs have found applications in frequency standards, optical clocks, optical communications, future GPS, astronomy and quantum technologies \cite{Pasquazi18,Oppo24}.

The CSs described here typically occur when stationary, i.e. $\partial_{\eta} E = \partial_{t} E =0$, homogeneous, i.e. $\partial^2_x E= \partial^2_\tau E= 0$, solutions coexist with spatially modulated structures. The bifurcations where homogeneous stationary states of Kerr resonators are unstable to spatial wave vectors $K$ are nothing else than Turing instabilities (see Subsection \ref{subsec:Turing}) \cite{Lugiato87}. CSs exist where a localised perturbation does not spread transversely, enabling a spatially localised stationary state. Since there is no transverse spreading, other localised CSs can be created nearby and remain independent, forming an array of independent "bits". Thus an array of $n$ CSs can support $2^n$ different states, leading to a huge information capacity. For the bistability of Turing patterns and homogeneous stationary states to exist, the instability of the homogeneous states has to be subcritical when increasing the input amplitude $E_{IN}$ while keeping the detuning $\theta$ fixed \cite{Harkness02,McSloy02}. Hence, the "cavity soliton" region is in general found for input amplitudes $E_{IN}$ below the Turing instability threshold of the HSSs.

For self-focusing Kerr nonlinearities ($\sigma=+1$), Eqs.~(\ref{LLE}) and (\ref{FTLLE}) admit homogeneous stationary solutions $E_s$ for anomalous dispersion ($\beta=+1$) obeying the implicit equation
\begin{equation}
E_{IN}^2=|E_s|^2 \left[ 1 + \left( \theta- |E_s|^2 \right)^2 \right] .
\label{HSS}
\end{equation} 
The steady-state curve of $|E_s|^2$ as a function of $E_{IN}^2$ is single-valued for $\theta < \sqrt{3}$ and 
S-shaped with possible optical bistability (see Section \ref{sec:TDkerr}), for $\theta > \sqrt{3}$. We introduce perturbations proportional to $\exp(\lambda \zeta) \exp(i K x)$ for Eq. (\ref{LLE}) with $a=1$, or proportional to $\exp(\lambda t) \exp(i K \tau)$ for Eq. (\ref{FTLLE}) with $\beta=1$ (anomalous dispersion) where $K$ is the wavevector, to perform the linear stability analysis of the HSSs. Following \cite{Scroggie94}, one finds that these solutions are unstable to the growth of modulations in the wave vector interval of
\begin{equation}
\left( 2 |E_s|^2 - \theta \right) - \sqrt{ |E_s|^4 - 1 } \;\;\;\; < K^2 < \;\;\;\; 
\left( 2 |E_s|^2 - \theta \right) + \sqrt{ |E_s|^4 - 1 } \, .
\end{equation}
When plotting these curves in a $(K^2, |E_s|^2)$ diagram one can find that given the input amplitude $E_{IN}$ and the detuning $\theta$, there are critical values $(K_c^2=2-\theta, |E_s|_c^2=1)$ corresponding to minima of these curves for $\theta<2$. For $\sqrt{3} < \theta < 2$ the entire upper branch of the hysteresis cycle of the homogeneous stationary solutions is unstable to Turing patterns as well as a segment of the lower branch, whereas for $\theta> 2$ the upper branch is still unstable but the lower branch is stable. Moreover, the Turing instability leading to patterns is supercritical for $\theta<41/30$ and subcritical for $\theta>41/30=1.3666..$ \cite{Lugiato87}. 
The subcritical condition is ideal to obtain simultaneously stable patterns and HSSs. 

Following \cite{Scroggie94}, we select $E_{IN}=1.2$ and $\theta=1.7$ where there is no bistability of homogeneous states but where stable homogeneous solutions can coexist with a stable branch of periodically modulated Turing patterns. By starting from a homogeneous input beam $E_{IN}$ with a strong perturbation in its middle, we can numerically simulate the LLE model by using the methods described in Appendix II and observe the formation of a CS. After a short transient, the perturbation is removed and the flat input beam restored while the CS solution survives indefinitely as shown in Fig.~\ref{fig:CS}. The CS intensity peak, its real and imaginary parts are shown Fig.~\ref{fig:CS}(a), and a coexisting Turing pattern solution in Fig.~\ref{fig:CS}(b). Note that the plot of the real part of $E$ faithfully and accurately reproduces Fig. 11(c) of \cite{Scroggie94} in spite of the enormous progress made by computers in the last thirty years. 
\begin{figure}[h]
	\includegraphics[width=0.49\columnwidth]{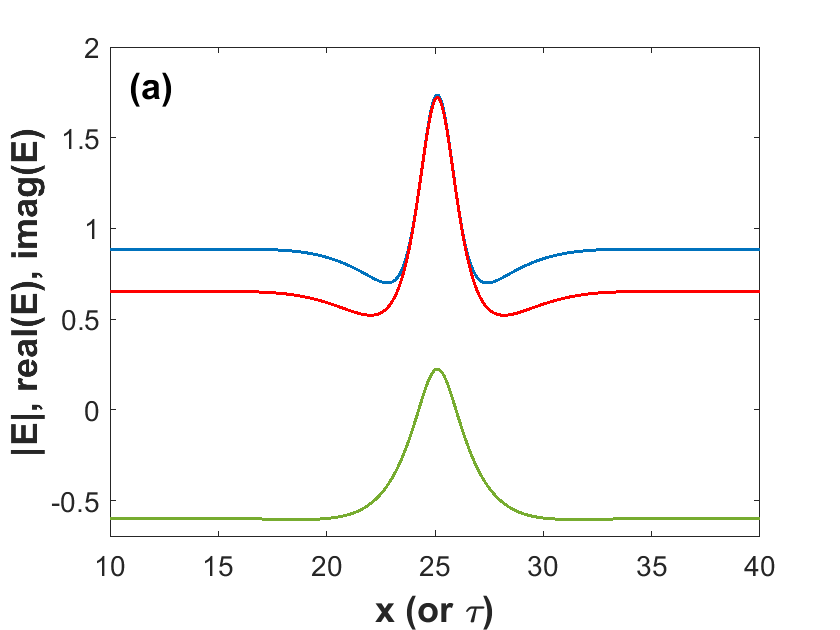}
	\includegraphics[width=0.49\columnwidth]{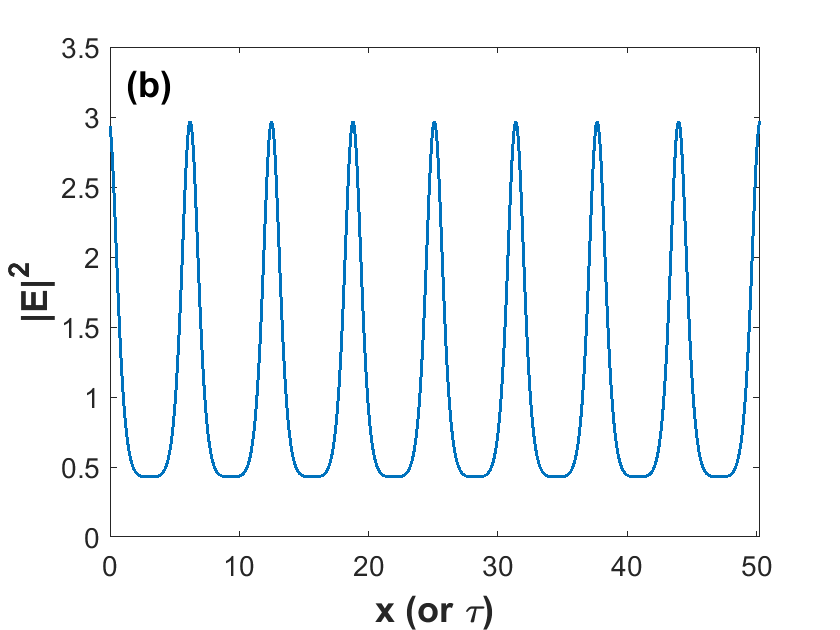}
	\caption{(a) Amplitude (blue curve), Real Part (red curve), and Imaginary Part (green curve) of the field $E$ of a CS for both of the LLE models (\ref{LLE}-\ref{FTLLE}) for $E_{IN}=1.2$ and $\theta=1.7$. (b) Stable Turing pattern at the same parameter values.}
	\label{fig:CS}
\end{figure}

It is important to note that the horizontal scale of Fig.~\ref{fig:CS} can either be the transverse coordinate $x$ with the LLE in the presence of diffraction or the fast time $\tau=nz/c$ of the LLE in the presence of anomalous dispersion. The two solutions are of course identical and for this reason a large part of the theoretical/numerical work done from the early 1990s in the diffractive case has an immediate application to the group velocity dispersion cases. Note however that unless one uses intra-cavity telescopes to drastically modify the diffraction length, the diffraction coefficient $a$ in front of the second derivative in space in Eq.~(\ref{LLE}) is always positive. 

Things are different for the dispersive case where, depending on the material used, the group velocity dispersion coefficient can change from positive (anomalous dispersion) to negative (normal dispersion). We have seen that the NLSE solitons can survive this change of sign but instead of bright solitons, one observes dark solitons in the normal dispersion regime. The situation is analogous for the LLE equations apart from the fact that instead of dark solitons one observes grey solitons (i.e. with finite intensity at the bottom of the trough) \cite{ParraRivas16,Note2}. Fig.~\ref{fig:DarkCS} shows two examples of grey CSs when the coefficient $\beta$ is negative and for parameter values of $E_{IN}=1.3515$, $\theta=1.95$ (panel (a)), and $E_{IN}=2.2$, $\theta=4$ (panel (b)) obtained from numerical simulations of Eq.(\ref{FTLLE}) by using the numerical methods described in Appendix II. When increasing the input amplitude and the detuning, grey CSs develop local oscillations around the trough.  \\

\begin{figure}[h]
	\includegraphics[width=0.49\columnwidth]{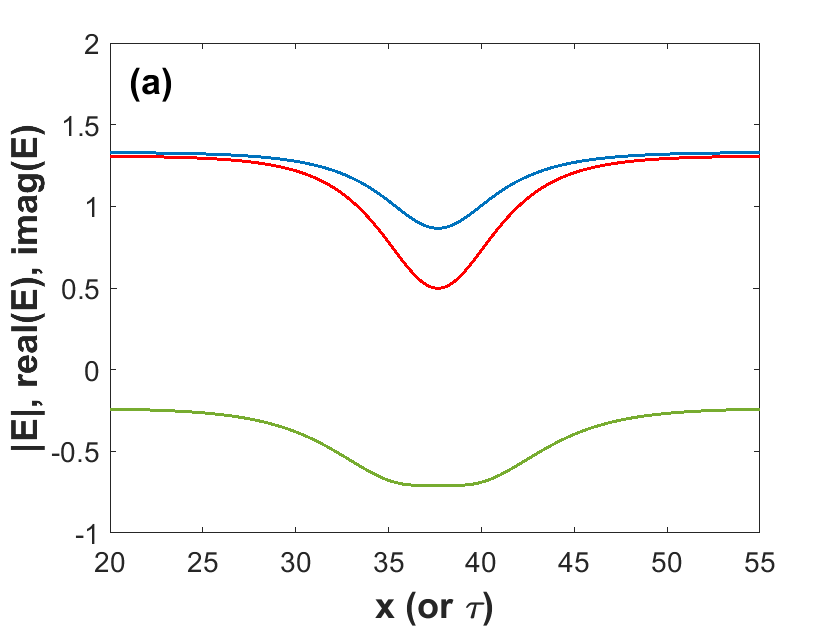}
	\includegraphics[width=0.49\columnwidth]{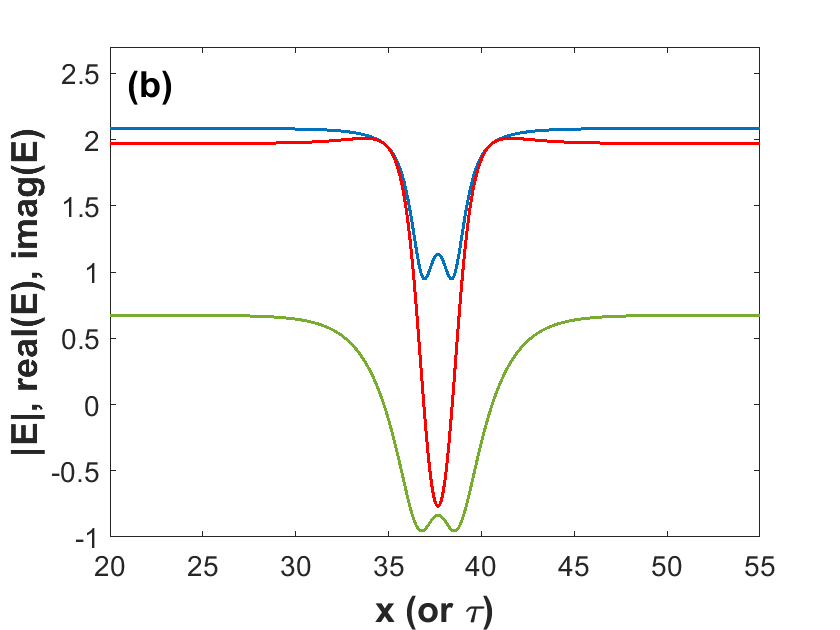}
	\caption{(a) Amplitude (blue curve), Real Part (red curve), and Imaginary Part (green curve) of the field $E$ of a grey CS for both of the LLE model (\ref{FTLLE}) for normal dispersion, $E_{IN}=1.3515$ and $\theta=1.95$. (b) Same as (a) but for $E_{IN}=2.2$ and $\theta=4$.}
	\label{fig:DarkCS}
\end{figure}

\noi {\it {\bf Frequency Combs due to CSs in Kerr Resonators.}}
Femtosecond pulses in a photonic crystal fibre led to the advent of optical frequency combs in which a broadband  optical spectrum (often spanning more than an octave) is formed by a series of finely spaced and narrow lines with stabilized absolute frequencies. Within a few years, such combs had revolutionized the precision and accuracy with which different optical transition frequencies can be measured leading to the 2005 Nobel Prize in Physics to T. H\"ansch and J. Hall \cite{HanschHall06}. In 2007, Pascal Del'Haye and his group showed that light propagating in a monolithic ring microresonator was capable to generate an output spectrum with a huge number of discrete lines and a span of over 500 nm ($\approx$ 70 THz) around 1550 nm, a frequency comb, without relying on any external spectral broadening \cite{Pascal07}. It was later realised first theoretically/numerically \cite{Coen13,Chembo13} and then experimentally by Tobias Herr and co-workers \cite{Herr14} that this new method of generation of frequency combs was due to bright CSs circulating in the Kerr ring resonator. Since then there has been an explosion of theoretical, numerical, experimental and industrial work on frequency combs generated by temporal CSs of the LLE \cite{Pasquazi18} of exactly the same kind, shape and stability of those discovered in the early 1990s at Strathclyde \cite{Scroggie94}. On the theoretical side, it is worthwhile to mention a recent model that unifies temporal CSs and frequency combs in active and passive cavities \cite{Columbo21}. 

Temporal CSs along the longitudinal direction of a resonator represent optimal optical pulses with broad pyramidally shaped frequency spectra for frequency combs. Here we consider the spectral properties of temporal CSs in photonic devices of the kind shown schematically in Fig.~\ref{Cavities}(a) where a CW input laser produces an output of regularly spaced optical pulses, the CSs. In Fig.~\ref{fig:Spectra}(a) we show the power spectrum in decibels of a train of bright LLE CSs of Fig.~\ref{fig:CS}(a). In Fig.~\ref{fig:Spectra}(b) of a train of bright purely absorptive CSs of Fig.~\ref{fig:CSPA}(a) (see next Subsection). In Fig.~\ref{fig:Spectra}(c) and (d) of a train of the grey LLE CSs corresponding to Fig.~\ref{fig:DarkCS}(a) and (b), respectively. 
\begin{figure}[h]
	\includegraphics[width=0.49\columnwidth]{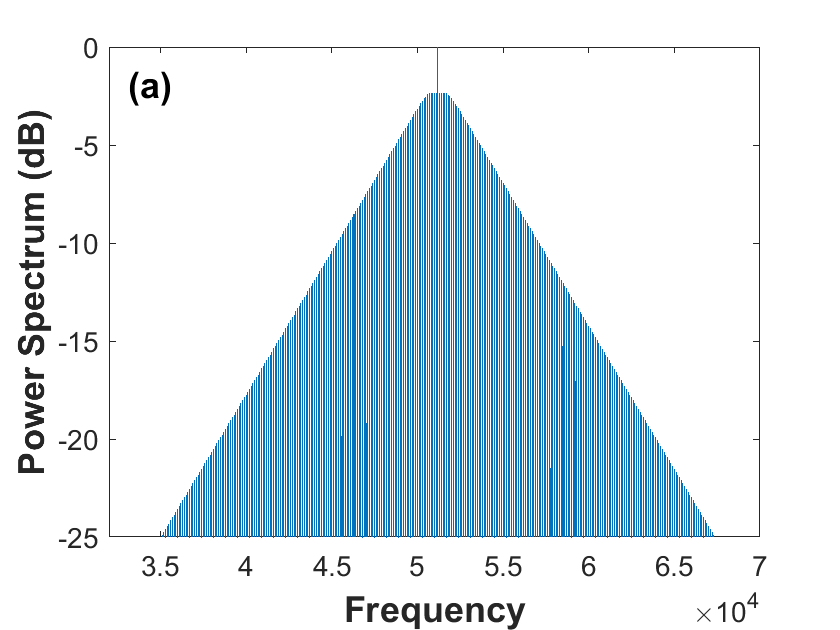}
	\includegraphics[width=0.49\columnwidth]{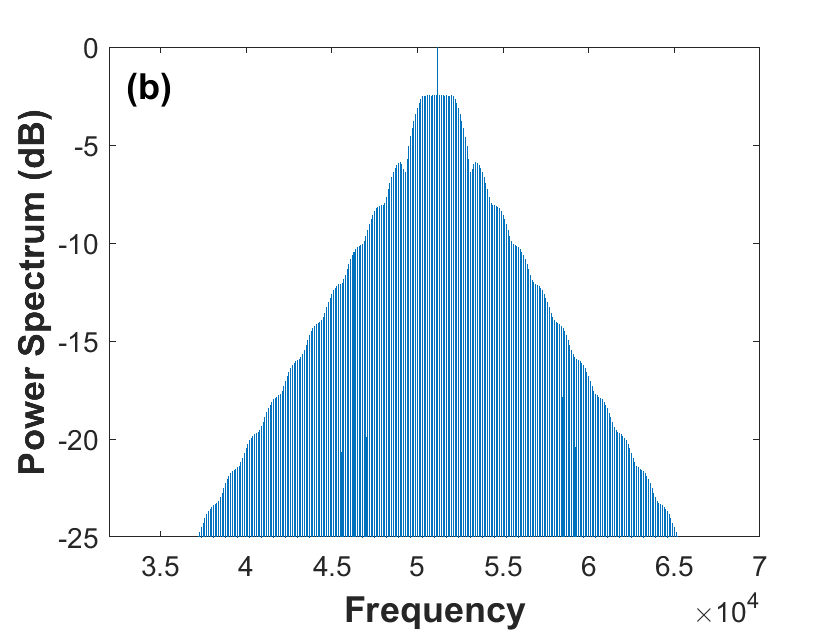}
	\includegraphics[width=0.49\columnwidth]{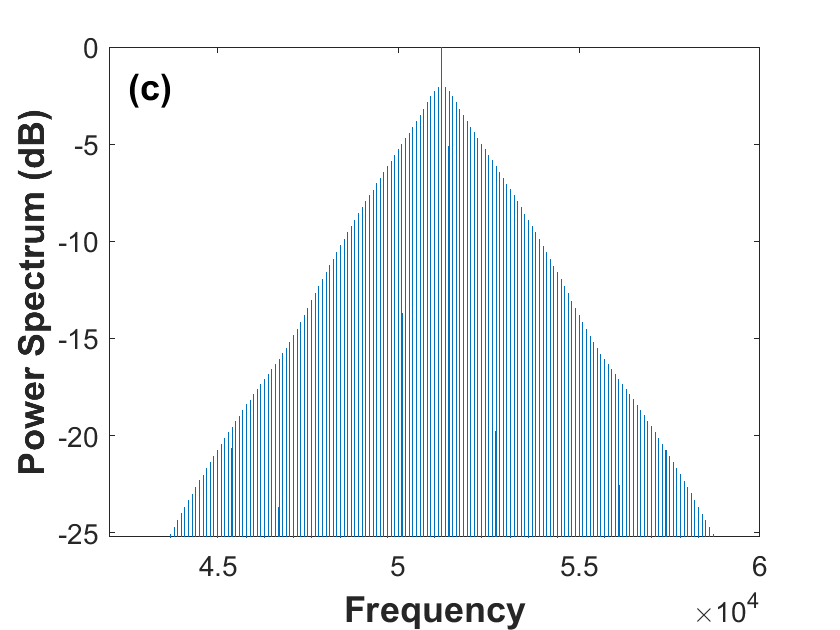}
	\includegraphics[width=0.49\columnwidth]{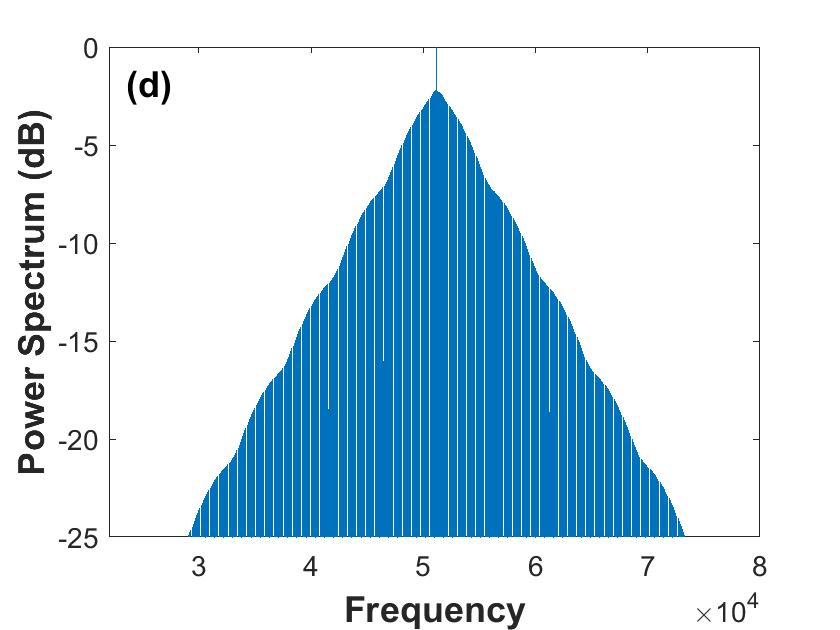}
	\caption{Frequency power spectra of CSs pulse trains for (a) the bright LLE CS of Fig.~\ref{fig:CS}(a), (b) the bright purely absorptive CS of Fig.~\ref{fig:CSPA}(a), (c) the grey LLE CSs of Fig.~\ref{fig:DarkCS}(a), (d) the grey LLE CS of Fig.~\ref{fig:DarkCS}(b).}
	\label{fig:Spectra}
\end{figure}
There are three clear common features in these four power spectra: 1) Spectral broadness that survives to very low decibel scales; 2) Finely spaced comb teeth separated by the inverse of the round trip time of the micro-resonator; 3) Almost pyramidal structure with a single central peak. These are the features that have made CSs the optimal generators of optical frequency combs and their applications \cite{Oppo24}. 

\subsection{Cavity Solitons and Turing Patterns in Absorptive Media}
If we consider propagation in a saturable absorber instead of a pure Kerr medium, the NLSE of Section \ref{sec:NLSE} changes into
\begin{equation}
\partial_\zeta E=\partial_z E + \frac {n}{c} \,\,\, \partial_t E = \frac{i}{2k} \nabla^2 E - \frac{Q (1-i\Delta) E}{1+\Delta^2+|E|^2}
\label{SA}
\end{equation}
where $Q$ is a numerical factor proportional to the atomic density in the medium and $\Delta$ is proportional to the detuning between the input laser frequency and the atomic frequency of a two energy level system as seen in Section \ref{sec:TDkerr}. By applying the mean field approach it is possible to obtain \cite{Oppo24}
\begin{equation}
\partial_{\kappa t} E = E_{IN} -(1+i\theta) E - \frac{2C (1-i\Delta) E}{1+\Delta^2+|E|^2} + i a \nabla^2 E \, .
\label{CavitySA}
\end{equation}
It is easy to see that in the limit of large $|\Delta|$ and small intra-cavity intensities $|E|^2$, Eq.~(\ref{CavitySA}) reduces to the LLE~(\ref{LLE}). As seen in Section \ref{sec:TDkerr}, another interesting limit is the atomic resonance case of $\Delta=0$ where Eq.~(\ref{CavitySA}) becomes
\begin{equation}
\partial_{\kappa t} E = E_{IN} -(1+i\theta) E - \frac{2C \, E}{1+|E|^2} + i a \nabla^2 E . 
\label{CavityPureA}
\end{equation}
This is known as the purely absorptive case. The homogeneous stationary states are given by \cite{Firth96PRL}:
\begin{equation}
E_{IN}^2=|E_s|^2 \left[ \left( 1 + \frac{2C}{1+|E_s|^2} \right)^2 + \theta^2 \right] ,
\label{HSSPA}
\end{equation} 
and, depending on the values of $\theta$ and $C$, the plane-wave input-output characteristic may be either monostable or bistable. We consider CSs in the monostable regime, demonstrating again that they are a phenomenon independent of bistability of homogeneous states. There is a Turing instability for $(I\,S) > (S+1)$ where $I=|E_s|^2$ and $S=2C/(1+I)^2$ is a saturation parameter. At threshold, the critical wave vector is $aK_c^2= -\theta$ which is real only if $\theta$ is negative. As for the LLE case, a subcritical condition for Turing patterns is ideal to obtain simultaneously stable patterns and homogeneous stationary states. Following \cite{Firth96PRL}, we select $\theta=-1.2$, $C=5.4$ and $E_I=6.65$ where the lower branch of the homogeneous solutions is stable and coexists with a stable branch of periodically modulated patterns. Again, we start from a homogeneous input beam $E_{IN}$ with a strong perturbation in its middle. We numerically simulate the Eq.~(\ref{CavityPureA}) by using the numerical methods of Appendix II and observe the formation of a CS. After a short transient, the perturbation is removed and the flat input beam restored while the CS solution survives indefinitely as shown in Fig.~\ref{fig:CSPA}(a). The CS amplitude, real and imaginary parts are shown Fig.~\ref{fig:CSPA}(a), and a coexisting Turing pattern solution in Fig.~\ref{fig:CSPA}(b). In the two-dimensional case with diffraction, CSs in the purely absorptive case have been labelled as Optical Bullet Holes \cite{Firth96PRL}. The frequency comb spectrum generated by the CSs of Fig. \ref{fig:CSPA}(a) obtained via the numerical integration of the pure absorptive model (\ref{CavityPureA}) is shown in Fig. \ref{fig:Spectra}(b). 

\begin{figure}[h]
	\includegraphics[width=0.49\columnwidth]{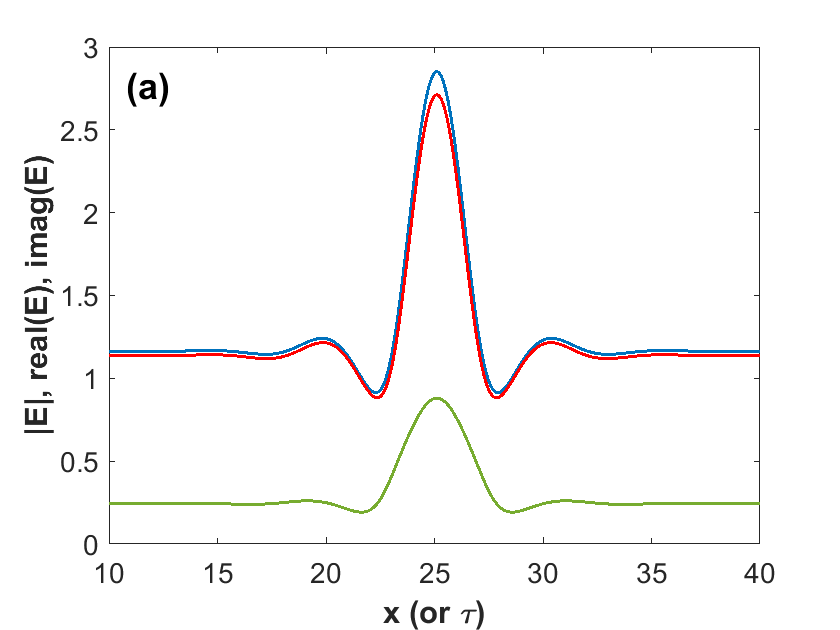}
	\includegraphics[width=0.49\columnwidth]{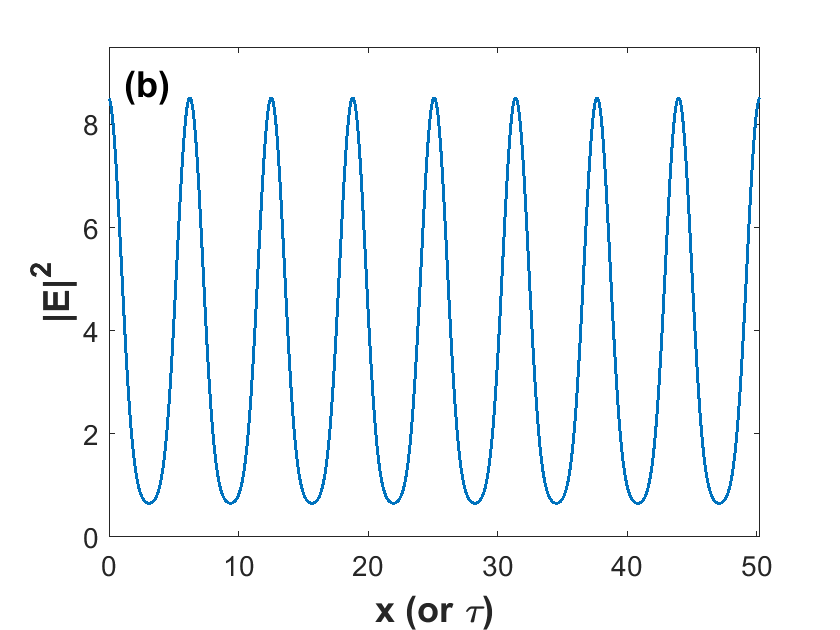}
	\caption{(a) Amplitude (blue curve), Real Part (red curve), and Imaginary Part (green curve) of the field $E$ of a CS of the pure absorptive model (\ref{CavityPureA}) for $\theta=-1.2$, $C=5.4$ and $E_{IN}=6.65$. (b) Stable Turing pattern at the same parameter values.}
	\label{fig:CSPA}
\end{figure}

Before concluding this section about the spatio-temporal dynamics of passive cavities, we note that many other configurations of passive cavities benefit from the application of methods of computational physics for PDEs such as ring resonators with two polarizations \cite{Garbin17,Garbin20,Garbin21,Xu21,Lucas25} or counterpropagation \cite{Hill18,Fan20,Woodley21,Campbell22}, Fabry-Perot cavities \cite{Lugiato23,Cole18,Campbell23,Hill24}, coupled passive cavities \cite{Miller15,Komagata21,Pal24}, cavities with three level media \cite{Oppo22,Esalmi23} and the investigation of CS interactions \cite{Jang16} as well as noise and quantum fluctuations of CS in passive photonic devices \cite{Seibold22}.

\section{Spatio-Temporal Dynamics of Lasers}\label{sec:STDlaser}
Having investigated spatio-temporal features and in particular cavity solitons in passive resonators, we turn our attention to active systems, i.e. lasers. First we review the generation of laser temporal cavity solitons in a laser system that includes a nested Kerr ring resonator. We then investigate the generalised spatio-temporal model for a laser with an injected signal previously studied in Subsection \ref{subsec:LIS} for its intriguing dynamical properties. In particular we will see the onset of spatio-temporal oscillations and turbulence in this standard photonic device.

\subsection{Spatio-temporal Dynamics of Lasers with a Nested Kerr Cavity} \label{subsec:LNested}
In the previous section we have seen that temporal CSs that balance group velocity dispersion with the nonlinear phase shift in driven and lossy cavities can have many practical applications through the generation of frequency combs. A recent successful alternative has been the generation of temporal Laser Cavity Solitons (LCS) by nesting a Kerr microresonator in a fibre loop with gain \cite{Bao19} (see Fig. \ref{Cavities}(b)). Historically LCS have been first generated in spatial configuration with semiconductor materials and by using diffraction \cite{Tanguy08,Genevet08,Scroggie09,Marconi14,Gustave17}. Their modelling is however complex and often requires time delays that are not easy to describe in this review. Hence we focus here on the simulations of temporal LCS in lasers with a nested Kerr cavity. These LCS fundamentally differ from passive Kerr CSs because they receive energy directly from the gain of the lasing medium and exist without any background light. Temporal LCS are intrinsically a very energy-efficient class of CSs using average powers less than $6\%$ of equivalent LLE CSs, a demonstrated mode efficiency of $75\%$, with a theoretical maximum predicted to be $96\%$, and output frequency combs with a bandwidth of more than $50$nm \cite{Bao19}. 

With reference to Fig. \ref{Cavities}(b), a simplified version of the experiments in \cite{Bao19,Rowley22}, the spatio-temporal equations describing the dynamics of the field $a$ in the Kerr microring and $b$ in the external loop with gain $\mu$ are given by
\begin{align}
\partial_{t} a &= - \kappa a + i |a|^2 a + i (\beta_a/2) \partial^2_\tau a + \sqrt{\kappa} b \nonumber \\
\partial_{t} b &= \mu b - (1 - 2 \pi i \Delta) b + i (\beta_b/2) \partial^2_x b + \sigma \partial^2_\tau b + \sqrt{\kappa} a \label{LaserLoop}
\end{align}
where $\kappa$ is the coupling between the two resonators, $\beta_a$ and $\beta_b$ are the normalised anomalous dispersion coefficients, $\tau=nz/c$ is the fast time coordinate along the propagation direction $z$, $\mu$ is the saturated gain, $\Delta$ is the detuning, and $\sigma$ is the spectral filtering bandwidth. Note that the equation for $b$ is linear and the nonlinearity only originates from the Kerr term in the equation for $a$ \cite{Bao19,Cutrona21}. 

Temporal LCS and broad frequency combs have been experimentally engineered and successfully compared with computer simulations of extended version of Eqs. (\ref{LaserLoop}) \cite{Bao19,Bao20,Cutrona21,Rowley22}. Here we demonstrate again the power of numerical simulations for photonic devices by showing temporal LCS of Eqs. (\ref{LaserLoop}) in Fig. \ref{fig:CSNestLaser} and Turing patterns and spatio-temporal disorder in Fig. \ref{fig:PatternNestLaser}. 
Guided by the work in \cite{Bao19,Cutrona21,Rowley22} we have selected the following parameters for Eqs. (\ref{LaserLoop}): $\kappa=2 \pi$, $\beta_a=1.25 \times 10^{-4}$, $\mu=0.044$, $\Delta=0.25$, $\beta_b=3.5 \times 10^{-4}$, and $\sigma=1.5 \times 10^{-4}$. Eqs. (\ref{LaserLoop}) with these parameters have then been integrated numerically by using the methods described in Appendix II.
\begin{figure}[h]
	\includegraphics[width=0.49\columnwidth]{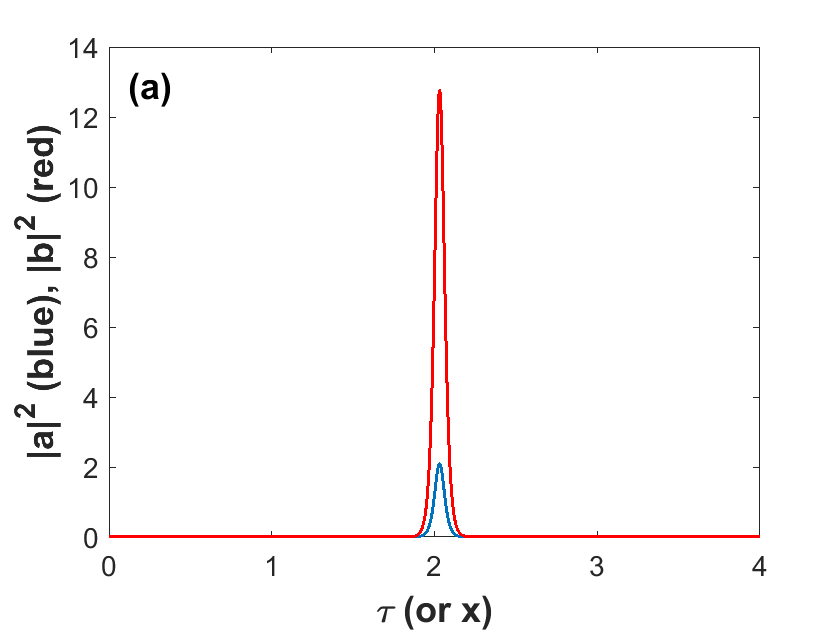}
	\includegraphics[width=0.49\columnwidth]{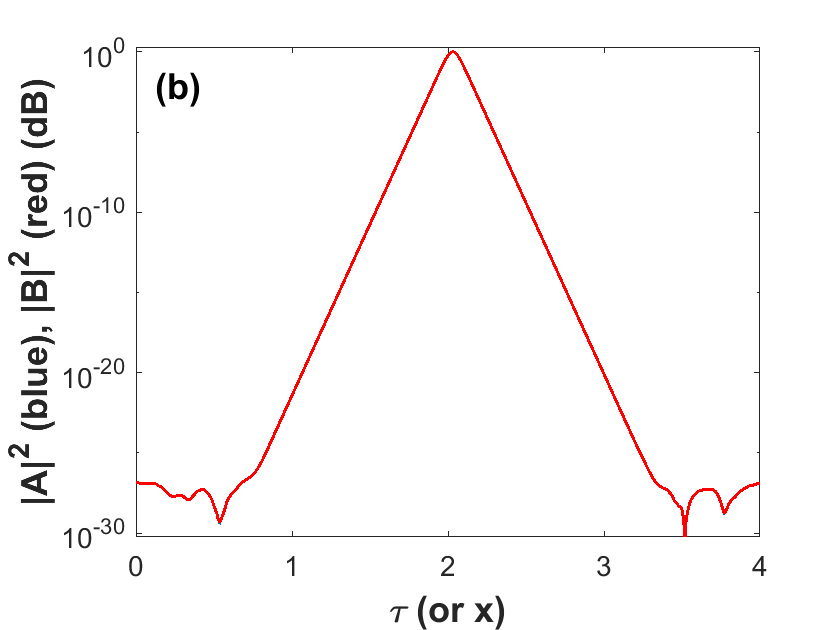}
	\caption{(a) Intensity profiles for the fields $a$ (blue line) and $b$ (red line) for a typical temporal LCS from the numerical integration of Eqs. (\ref{LaserLoop}) for $\kappa=2 \pi$, $\beta_a=1.25 \times 10^{-4}$, $\mu=0.044$, $\Delta=0.25$, $\beta_b=3.5 \times 10^{-4}$, and $\sigma=1.5 \times 10^{-4}$. (b) Log plot of the temporal LCS of (a) for the fields $A=a/\sqrt{max(|a|^2)}$ (blue line) and $B=b/\sqrt{max(|b|^2)}$ (red line). Note that the blue line is behind the red line.}
	\label{fig:CSNestLaser}
\end{figure}
The typical intensity profiles of the fields $a$ and $b$ of a temporal LCS in the laser system with a nested Kerr ring are shown in Fig. \ref{fig:CSNestLaser}. It is very interesting to see in Fig. \ref{fig:CSNestLaser}(b) where we plot the profiles for the fields $A=a/\sqrt{max(|a|^2)}$ (blue line) and $B=b/\sqrt{max(|b|^2)}$ (red line) in decibels that the LCS sit on a zero background in contrast to the CS of the LLE. There are more than $23$ orders of magnitude of difference between the LCS of Fig. \ref{fig:CSNestLaser} and typical LLE CS thus demonstrating that temporal LCS are a separate class of CS and that they are locked to nothing. It is also interesting to see that the pulse intensity in the gain loop is close to a magnification of the pulse intensity circulating in the Kerr cavity ring. The blue trace of the normalised intensity profile of the $a$ field overlaps with the red trace of the normalised intensity profile of the $b$ field, making it invisible in the plot in Fig. \ref{fig:CSNestLaser}(b).

\begin{figure}[h]
	\includegraphics[width=0.49\columnwidth]{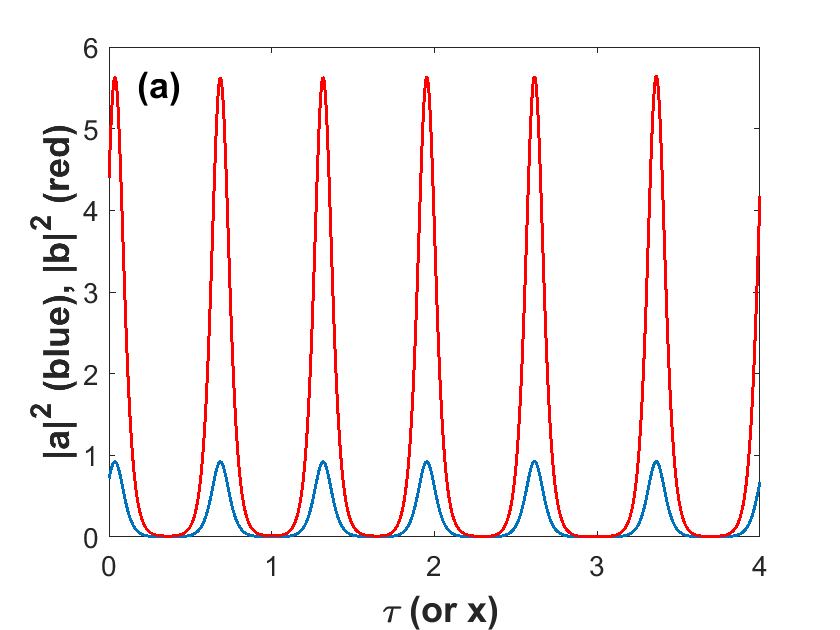}
	\includegraphics[width=0.49\columnwidth]{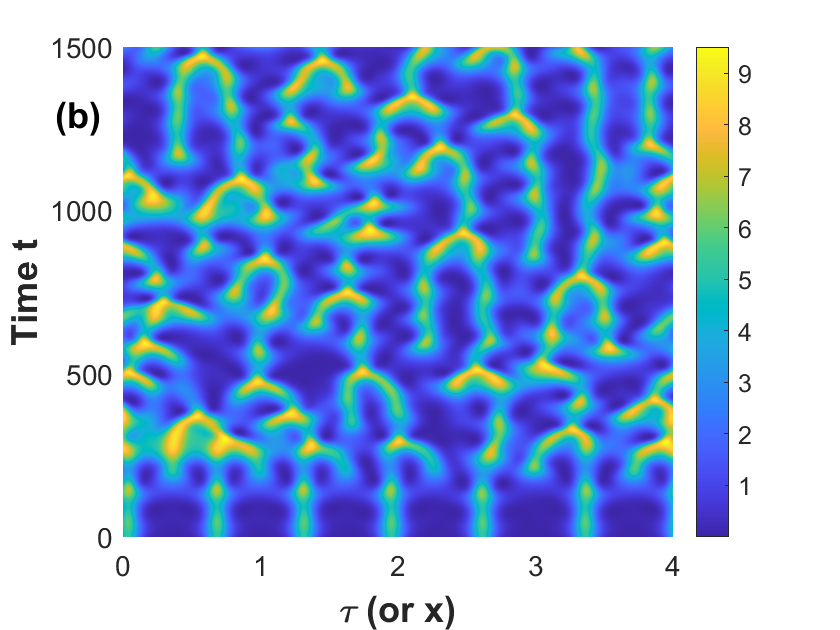}
	\caption{(a) Intensity profiles for the fields $a$ (blue line) and $b$ (red line) for a typical Turing pattern from the numerical integration of Eqs. (\ref{LaserLoop}) for $\kappa=6.25$, $\beta_a=(1.7)10^{-4}$, $\mu=0.03$, $\Delta=0.2$, $\beta_b=(3.5)10^{-4}$, and $\sigma=(2.5)10^{-4}$. (b) Spatio-temporal evolution of the amplitude of the field $a$ during a transition from a Turing pattern to spatio-temporal disorder for $\Delta=0$ and the other parameters as in (a). }
	\label{fig:PatternNestLaser}
\end{figure}
To show Turing patterns in Eqs. (\ref{LaserLoop}) we follow \cite{Bao20} and use the parameters $\kappa=6.25$, $\beta_a=1.7 \times 10^{-4}$, $\mu=0.03$, $\Delta=0.2$, $\beta_b=3.5 \times 10^{-4}$, and $\sigma=2.5 \times 10^{-4}$. The final profiles of the field intensities after discarded transients are shown in Fig. \ref{fig:PatternNestLaser}(a) and display a typical Turing pattern in this kind of laser device. 
Patterns appear in this laser system via a Turing instability in a way analogous to those described in the LLE in Subsection \ref{subsec:Turing}. Finally it is interesting to observe a transition from a stable Turing pattern to a regime of spatio-temporal disorder. By changing the detuning $\Delta$ from $2$ to $0$, i.e. moving to resonance, we observe the excitation of temporal oscillations and of multiple spatial wavevectors in the fast time coordinate $\tau$ leading to spatio-temporal disorder that persists indefinitely in the laser cavity as shown in Fig. \ref{fig:PatternNestLaser}(b).

Temporal LCS and Turing patterns in the laser system with nested Kerr resonator studied here find applications in spectroscopy, metrology \cite{Cutrona23}, optical communications \cite{Pfeifle15} and are suited to devices that require high quality repetition rates, such as microwave and terahertz generation, i.e. those necessary for low-noise ultrafast telecommunications.

\subsection{Spatio-temporal Dynamics of Lasers with an Injected Signal}

We move now to the spatio-temporal dynamics of lasers with injection. For simplicity we consider here the case of a Class-A laser in one 'spatial' dimension, i.e. either a single transverse coordinate $x$ in the presence of diffraction or a fast time coordinate $\tau=nz/c$ along a single round trip of the laser cavity in the presence of dispersion (see Fig. \ref{fig:LaserCavities}(b)). We start from Eqs. (\ref{LISClassB_EW}) and consider the case of high finesse resonators where $k \ll \gamma_{\perp}$ and $k \ll \gamma_{\parallel}$ as well, i.e. Class-A lasers with injection. Setting the time derivative of the population inversion variable $W$ to zero one obtains
\begin{equation}
\partial_{kt} E = E_{IN} - (1+i\theta) E + i \eta E  + \frac{\mu(1+i\theta)}{1+\theta^2 + |E|^2} E + (1+i\beta) \partial_x^2 E
\label{LISclassAfull}
\end{equation}
where we have normalised the amplitude of the injection and the detuning $\eta$ via $E_{IN}=E'_{IN}/k$ and $\eta = \eta'/k$, respectively, where $E'_{IN}$ and $\eta'$ where introduced in Eqs. (\ref{LISFullEq}). As for the LLE, $\beta$ represents either diffraction or group velocity dispersion in which case $x$ becomes $\tau=nz/c$ the fast time coordinate along the longitudinal cavity direction $z$. One interesting addition is the presence of a diffusion term. It has been shown that the adiabatic elimination of the polarization field variable $P$ in the presence of spatial effects such as diffraction or dispersion leads to diffusion in the field equation \cite{Oppo09}. Field diffusion can be controlled in the dispersion case by using a spectral filter as seen for example in Subsection \ref{subsec:LNested}.

To make a clear connection with the dynamical effects of the Complex Ginzburg-Landau Equation (\ref{CGLE}) (see  Subsection \ref{subsect:Chaos}), we perform the so called 'cubic' approximation \cite{LugiatoBook} by considering $|E|^2 < (1+\theta^2)$. By introducing $\mu' = \mu/ (1+\theta^2)$ and the field $F=qE$ with $q=\sqrt{\mu'/(1+\theta^2)} $ one obtains
\begin{equation}
\partial_{kt} F = y + (\mu'-1) \left( 1+i \theta \right) F + i \eta F- (1+i\theta)|F|^2 F + (1+i\beta) \partial_x^2 F
\label{LISclassA}
\end{equation}
where $y=q E_{IN}$. This is known as the driven CGLE and has been extensively studied mathematically in \cite{Chate99}.  To make the connection even closer we chose $\mu'=2$. In this case there is a one-to-one correspondence between the parameters $y$, $\theta$, and $\eta$ of Eq. (\ref{LISclassA}) for a laser with injection, and those of Ref. \cite{Chate99}, $B=y$, $\alpha=\theta$, and $\nu=\theta+\eta$. The phenomenology of this system is quite extraordinary with regimes of full stability of the lasing solution with a flat spatial profile, solitons, stable Turing patterns, oscillating regimes, spatio-temporal disorder, spatio-temporal chaos and defect mediated turbulence. The oscillating regimes can be of just the flat lasing solution or affecting both slow time $kt$ and spatial coordinate (spatio-temporal oscillations). We provide here just a couple of examples of spatio-temporal dynamics to demonstrate again the full power of simple numerical simulations in the description of important photonic devices. In Figs. \ref{fig:LIS_STchaos} and \ref{fig:LISturbo} we see examples of spatio-temporal chaos and defect mediated turbulence, respectively, displayed by Eq. (\ref{LISclassA}) by just changing its parameters.
\begin{figure}[h]
	\includegraphics[width=0.49\columnwidth]{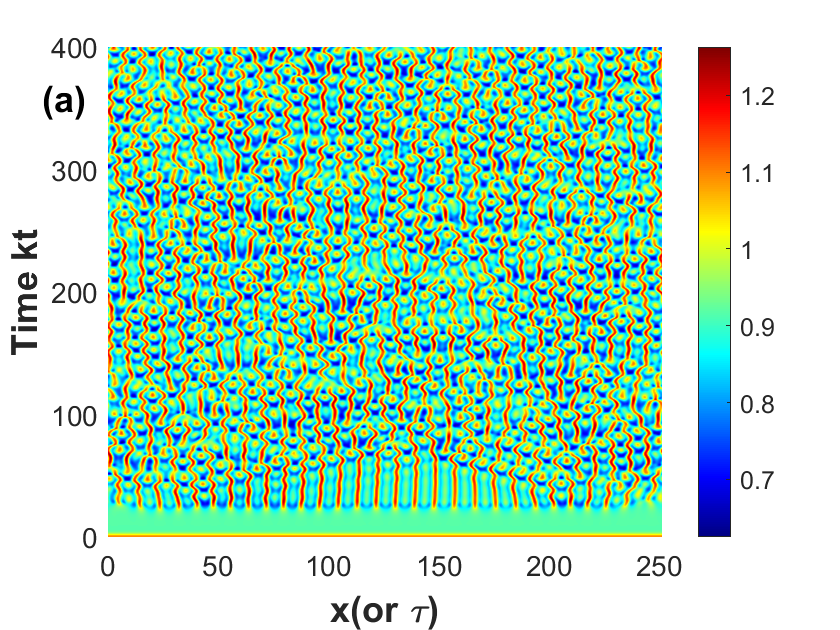}
	\includegraphics[width=0.49\columnwidth]{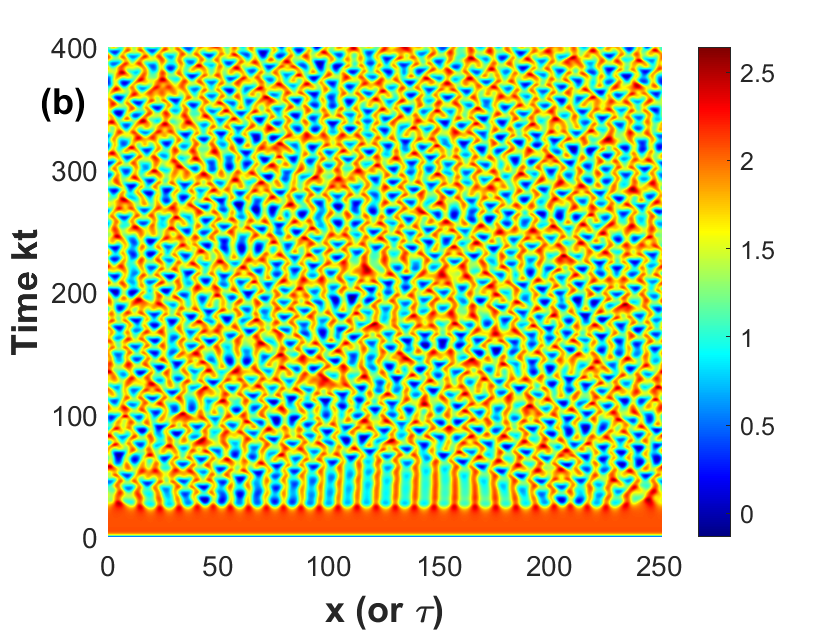}
	\caption{(a) Spatio-temporal evolution of the amplitude of the field $F$ from numerical simulations of Eq. (\ref{LISclassA}) for $\mu'=2$, $y=0.3$, $\theta=-2$, $\eta=0.6$ and $\beta=2$. (b) Same as (a) but for the phase of the field $F$.}
	\label{fig:LIS_STchaos}
\end{figure}
For example for $\mu'=2$, $y=0.3$, $\theta=-2$, $\eta=0.6$ and $\beta=2$ we observe in Fig. \ref{fig:LIS_STchaos} an instability of a Turing pattern and a transition to a spatio-temporal disordered state, somewhat similar in nature to that shown in Fig. \ref{fig:PatternNestLaser}(b) for the amplitude and phase of the field $a$ in a laser with a nested Kerr ring. Note that the excursion of the laser amplitude is limited between $0.5$ and $1.3$ (see Fig. \ref{fig:LIS_STchaos}(a)) while the phase is limited between $0$ and $0.83 \pi$ (see Fig. \ref{fig:LIS_STchaos}(b)). In comparison with a pendulum, this means that the oscillations are phase bound (no rotations) with no defects corresponding to zeros of the field amplitude. The loss of spatio-temporal coherence is present but limited in spatial and temporal scales, typical of deterministic chaos.  
\begin{figure}[h]
	\includegraphics[width=0.49\columnwidth]{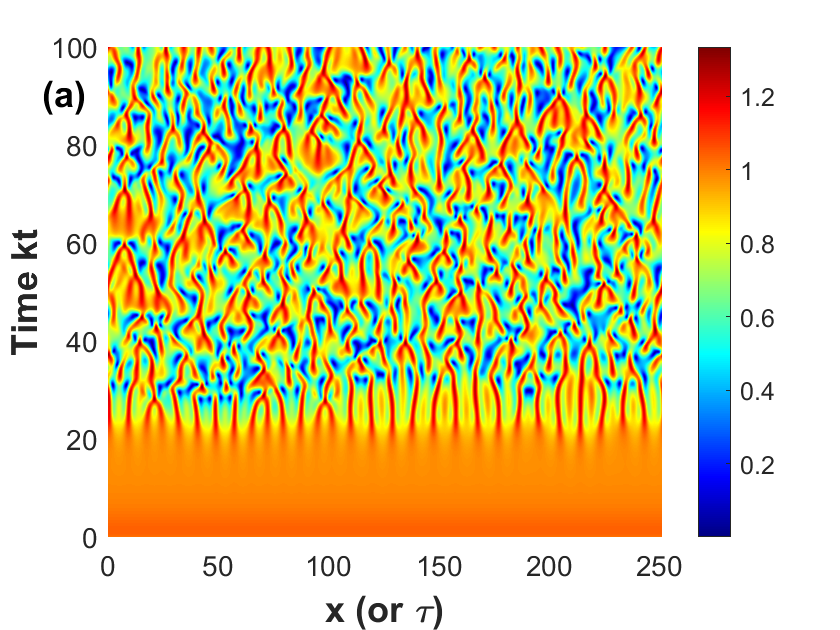}
	\includegraphics[width=0.49\columnwidth]{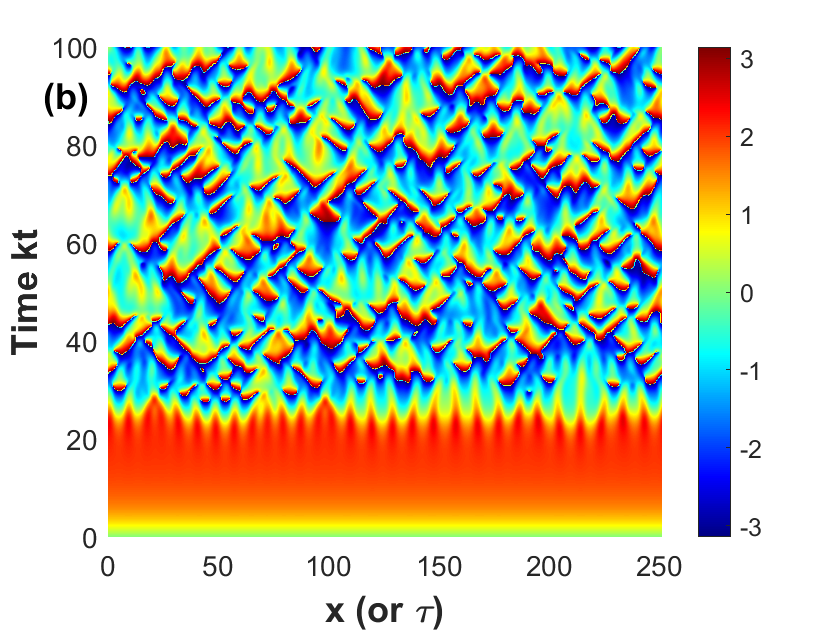}
	\caption{(a) Spatio-temporal evolution of the amplitude of the field $F$ from numerical simulations of Eq. (\ref{LISclassA}) for $\mu'=2$, $y=0.1$, $\theta=-2$, $\eta=0.2$ and $\beta=2$. (b) Same as (a) but for the phase of the field $F$.}
	\label{fig:LISturbo}
\end{figure}
By a small change of the values of the parameters $y$ and $\eta$ to the values of $0.1$ and $0.2$ respectively, numerical simulations of Eq. (\ref{LISclassA}) show a transition to a regime of fully developed defect mediated turbulence as described in Subsection \ref{subsect:Chaos}. Fig. \ref{fig:LISturbo} shows the spatio-temporal evolution of the amplitude of the field $F$ from numerical simulations of Eq. (\ref{LISclassA}) by using the numerical methods described in Appendix II. The presence of topological defects \cite{Coullet89,Chate99} during this evolution is demonstrated by the amplitude of the field touching zero values (see blue regions in Fig. \ref{fig:LISturbo}(a)) corresponding to jump in the phase (see the V-shaped regions in Fig. \ref{fig:LISturbo}(b)). During this evolution, the phase is unbound and corresponds to rotations when compared to a pendulum. The loss of spatio-temporal coherence is considerably larger in spatial and temporal scales than that of, for example, spatio-temporal chaos, thus justifying the name 'defect mediated turbulence'  \cite{Coullet89}.

These features have also been reproduced in simulations of lasers with injected signals in 2D \cite{Gibson16,Oppo22csf}. In 2D the full topological character of the defects is demonstrated via the topological charge of the vortices that appear in couples, separate and then annihilate each other if they have opposite topological charge thus decreasing progressively the spatial coherence. These structures have considerably less spatio-temporal coherence than, for example, stationary homogeneous solutions or Turing patterns. The interaction of several vortex defects in 2D simulations of Class-A lasers with injected signal can be so powerful to lead to rogue waves, i.e. rare wave event with high intensity peaks of very short lifetime and ultra-strong power on a turbulent background \cite{Gibson16}.


\section{Spatio-Temporal Dynamics of Optical Parametric Oscillators}\label{sec:STDopo}
In Section \ref{sec:TDopo} we have seen that degenerate optical parametric oscillators can display interesting temporal dynamics when changing control parameters such as the input drive amplitude $E_{IN}$ and the cavity decay rates $k_0$ and $k_1$. DOPOs can even display chaotic dynamics when the signal cavity losses are modulated. In this Section we consider partial derivatives with respect to space for transverse variable $x$, or with respect to the fast time $\tau=nz/c$ where $z$ is the propagation direction to describe diffraction or group velocity dispersion, respectively, in the two cavities. The final DOPO mean-field equations for the pump field $A_0$ and signal field $A_1$ generalise Eqs. (\ref{DOPOequ}) to give \cite{Oppo94,Oppo99,Oppo01,Parra19}
\begin{align}
\frac{dA_0}{dt} &= k_0 \left[ E_{IN} - (1 + i\theta_0) A_0 - A_1^2 \right] + i \beta_0 \partial_x^2 A_0 \nonumber \\
\frac{dA_1}{dt} &= k_1 \left[ - (1 + i\theta_1) A_1 + A_1^* A_0\right] + i \beta_1 \partial_x^2 A_1 \label{STDOPOequ}
\end{align}
where $\beta_0$ and $\beta_1$ are the diffraction (with $x$) or group velocity dispersion (with $\tau$) coefficients. In 1994, Turing patterns were found in this equations for detunings $\theta_1<0$ after a Turing instability at the threshold of the signal generation \cite{Oppo94}. We have numerically integrated Eqs. (\ref{STDOPOequ}) by using the methods described in Appendix II. 
\begin{figure}[h]
	\includegraphics[width=0.49\columnwidth]{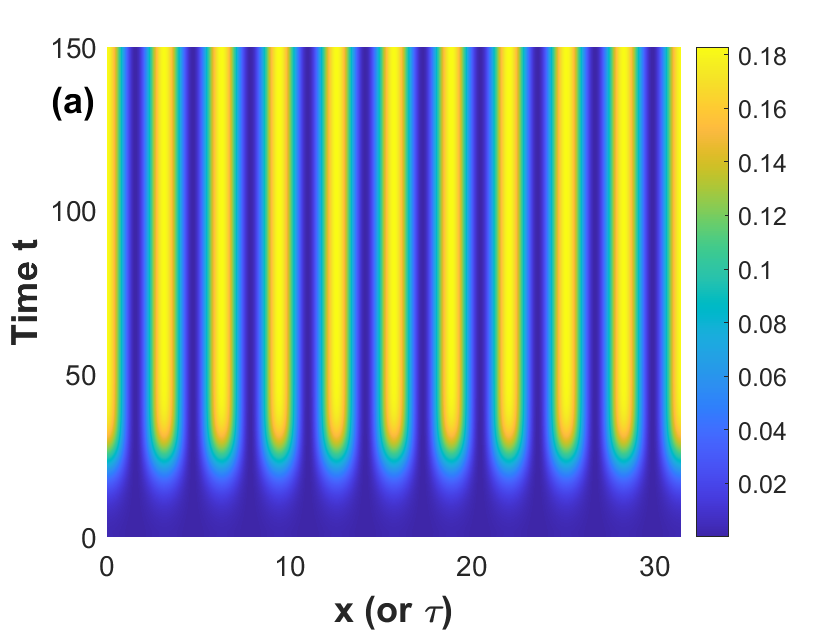}
	\includegraphics[width=0.49\columnwidth]{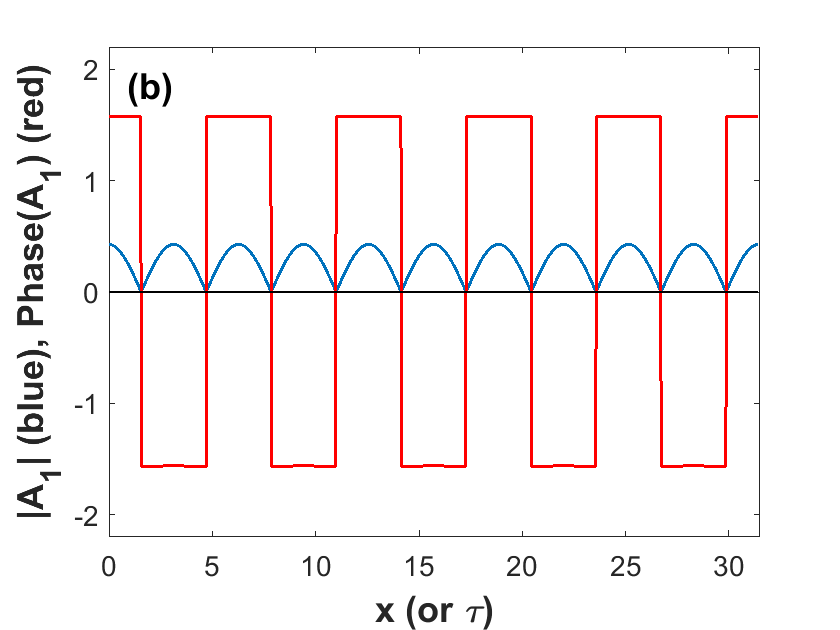}
	\caption{(a) Spatio-temporal evolution of the intensity of the signal field $|A_1|^2$ from the numerical simulations of Eqs. (\ref{STDOPOequ}) for $k_0=k_1=1$, $E_{IN}=1.1$, $\theta_0=0$, $\theta_1=-1$, $\beta_0=0.5$ and $\beta_1=1$. (b) Final spatial distributions of the signal amplitude $|A_1|$ (blue curve) and signal phase (red curve) for a Turing pattern at the end of the evolution shown in (a).}
	\label{fig:DOPO_pattern}
\end{figure}

In Fig. \ref{fig:DOPO_pattern}(a) we show the evolution of the signal field intensity for $E_{IN}=1.1$, i.e. just above the threshold of signal generation $E_{IN}^{thr}=1$ at resonance, starting from a flat distribution very close to zero amplitude and reaching a fully developed stable Turing pattern for $k_0=k_1=1$, $\theta_0=0$, $\theta_1=-1$, $\beta_0=0.5$ and $\beta_1=1$ in excellent agreement with \cite{Oppo94}. The signal field being different from zero here is surprising since in the presence of a detuning $\theta_1$ (off resonance) the signal threshold moves to $E_{IN}^{thr}=\sqrt{1+\theta_1^2}$ that in our case is equal to $\sqrt{2}$, i.e. larger than the value of $1.1$ used in Fig. \ref{fig:DOPO_pattern}. The stable Turing pattern shown in amplitude and phase of the signal field $A_1$ in Fig. \ref{fig:DOPO_pattern}(b) is even more intriguing. At each zero of the intensity of the field, the phase has a discrete jump of $\pm \pi$. We have seen in Subsection \ref{subsec:DarkNLSE} for the NLSE that localised points of zero intensity and phase jumps of $\pm \pi$ correspond to (topological) dark cavity solitons. The DOPO Turing pattern of Fig. \ref{fig:DOPO_pattern}(b) \cite{Oppo94} can then be associated to a chain of (topological) dark cavity solitons.

We continue with the topic of (topological) dark cavity solitons \cite{Note2} and move to the perfectly tuned case with both $\theta_0$ and $\theta_1$ equal to zero. Here S. Trillo, M. Haelterman, and A. Sheppard found stable dark topological cavity solitons in the DOPO equations in 1997 \cite{Trillo97}. In Fig. \ref{fig:DOPOdarksolitons} we present two examples of (topological) dark cavity solitons in the 1D DOPO for the case of equal cavity finesses with $k_0=k_1=1$ and the case of a signal cavity with higher finesse with $k_0=1$ and $k_1=0.1$. 
\begin{figure}[h]
	\includegraphics[width=0.49\columnwidth]{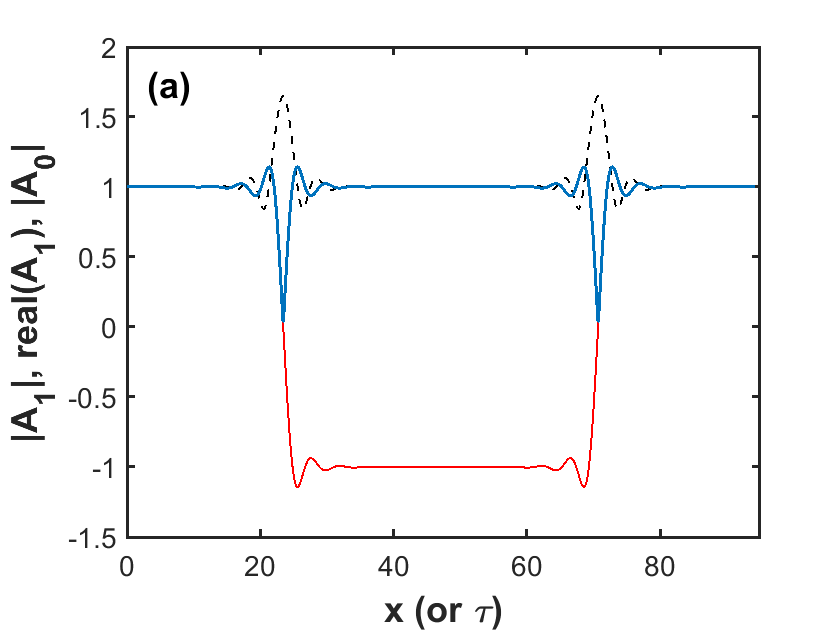}
	\includegraphics[width=0.49\columnwidth]{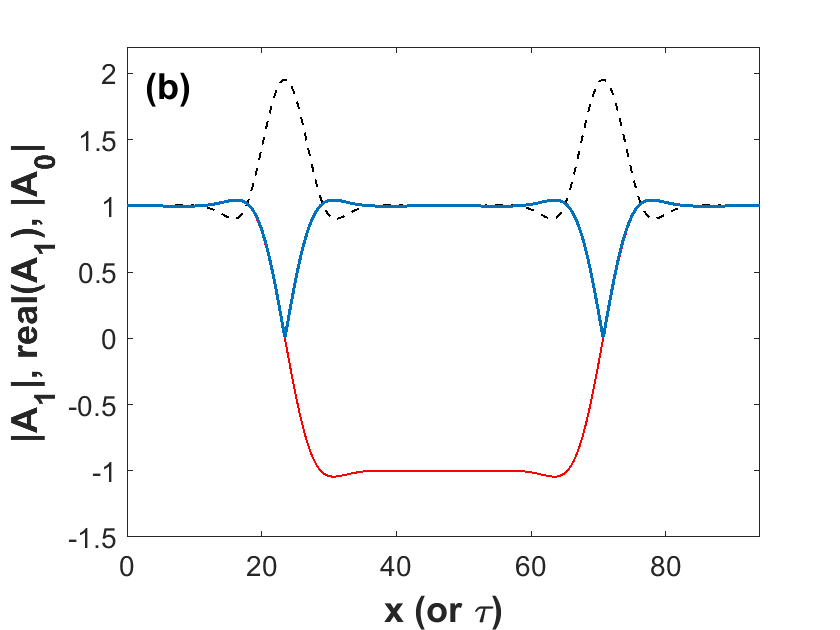}
	\caption{(a) Stable spatial distributions of the signal amplitude $|A_1|$ (blue curve), signal real part (red curve) and pump amplitude $|A_0|$ (black dashed curve) for two dark cavity solitons from the numerical simulations of Eqs. (\ref{STDOPOequ}) for $k_0=k_1=1$, $E_{IN}=2$, $\theta_0=0$, $\theta_1=0$, $\beta_0=0.5$ and $\beta_1=1$. (b) Same as (a) but for $k_1=0.1$. Note that since the imaginary part of $A_1$ is very small, the red curve of the real part of $A_1$ is hidden behind the blue curve of $|A_1|$ on the left and right sides of the plots.}
	\label{fig:DOPOdarksolitons}
\end{figure}

The dark cavity solitons displayed in Fig. \ref{fig:DOPOdarksolitons} are topological because their amplitudes are zero (see blue curves) in the exact same place where there is a phase jump of $\pm \pi$ as demonstrated by the profiles of the real part of the signal field $A_1$ (see red curves). These dark cavity solitons are also referred to as Domain Walls (DW) in analogy with Ising domain walls in magnetic systems \cite{Coullet90}. It is interesting to see that the energy drop in the signal field  $A_1$ at the center of the dark solitons is compensated by peaks in the pump field $A_0$. This is a very early example of a dark-bright soliton in an optical system \cite{Trillo97}. When changing the finesse of the signal cavity from $k_1=1$ to $k_1=0.1$ we observe that the dark cavity solitons broaden, the bright cavity solitons reach higher peaks and the local oscillations where the solitons reach the flat background solutions tend to disappear (see Fig. \ref{fig:DOPOdarksolitons}(b)). In the limit of $k_1$ going to zero, these oscillations are all but gone and the real part of the signal field approaches a hyperbolic tangent profile typical of the dark solitons of the NLSE (see \cite{Trillo97} in 1D and \cite{Oppo99,Oppo01} in 2D). The fact that the dark cavity solitons displayed in Fig. \ref{fig:DOPOdarksolitons}(a) and (b) are analytical continuations of this hyperbolic tangent profile further demonstrates that they are topological dark solitons. 

It is relevant to note that if the phase of the signal field is appropriate, the nonlinear term in the pump equation changes from a nonlinear absorption to a nonlinear gain. Indeed Fig. \ref{fig:DOPOdarksolitons} shows evidence of back conversion from the signal to the pump field at the centre of the dark cavity solitons above the flat background of $A_0$. It is a combination of nonlinear processes such as back conversion and nonlinear gain that leads to the stabilization of both signal and pump solitons within the region close to the point where the signal intensity goes to zero. It is useful to consider the energy balance in the system as a real quantity whose spatial variations gives information about transport along the spatial direction $x$ or $\tau$. For our chosen parameters, the local energy injection is $E_{IN}$Re$(A_0)$, while the local dissipation is $(|A_0|^2+|A_1|^2)$ \cite{Oppo99}. The nonlinear terms change only the energy distribution so that, in steady state, any local excess of dissipation over the external driving can only arise from energy transport. The energy balance (driving minus dissipation) across the dark DOPO cavity solitons is perfectly kept far away from the solitons while in their centres there is a large energy excess and an energy deficit in the tails around it. This implies an outward energy flow from the pump driving to intensity dissipations in the tails of the dark cavity solitons. Note that the integral of the energy balance $E_{IN}$Re$(A_0)-(|A_0|^2+|A_1|^2)$ across the full spatial area is exactly zero.
\begin{figure}[h]
	\includegraphics[width=0.49\columnwidth]{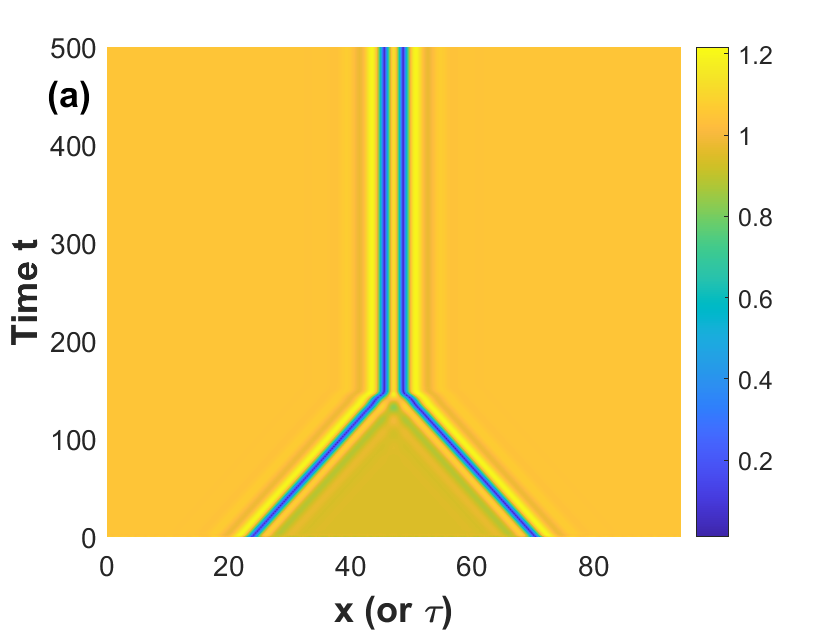}
	\includegraphics[width=0.49\columnwidth]{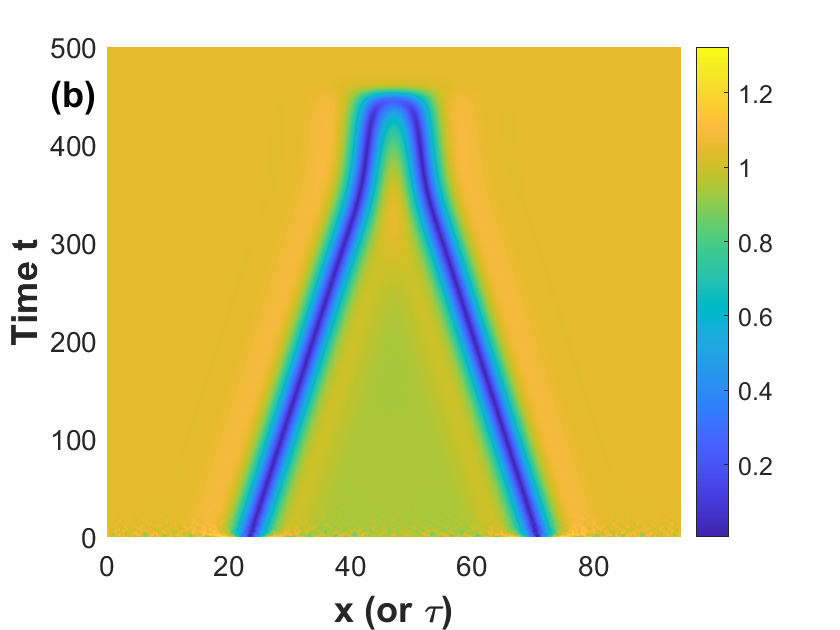}
	\caption{(a) Spatio-temporal evolution of the amplitude of the signal field $|A_1|$ from the numerical simulations of Eqs. (\ref{STDOPOequ}) for $k_0=k_1=1$, $E_{IN}=1.1$, $\theta_0=0$, $\theta_1=0$, $\beta_0=0.5$ and $\beta_1=1$ and with a signal seeding of $0.1$. (b) Same as (a) but for $k_1=0.1$.}
	\label{fig:DOPO_evolutions}
\end{figure}

It is relevant to mention that in broad/long cavities, it is possible to arrange random sequences of well separated and non interacting dark cavity soltions and create stationary spatial disorder. It is also possible to induce interactions of dark cavity solitons in 1D by slightly biasing one signal phase over the other via a small injection of the signal field in the cavity, known as seeding. In Fig. \ref{fig:DOPO_evolutions} we see the motion and the interaction of dark cavity solitons in the DOPO produced by small seeding of amplitude $0.1$ added to the signal equation. In the case of $k_0=k_1=1$ the two dark solitons move fast and lock to each other and create a new stable solitary structure with a central peak and two dark troughs, see Fig. \ref{fig:DOPO_evolutions}(a). In the case of $k_1=0.1$ the two dark solitons annihilate each other typical of dark cavity solitons of hyperbolic tangent profile, see Fig. \ref{fig:DOPO_evolutions}(b). As we have seen in Section \ref{sec:TDopo}, by decreasing $k_0$ with respect to $k_1$, i.e. the limit of a pump cavity of high finesse, temporal oscillations increase in amplitude. In the spatio-temporal DOPO it is exactly these spatial oscillations that allow for the locking of the cavity solitons as first demonstrated for DOPOs in 2D under the influence of local curvature \cite{Oppo99}.   

Although not presented here, dark cavity solitons in DOPOs can produce frequency combs very similar to those seen in Fig. \ref{fig:Spectra}(b) for grey cavity solitons of the LLE. A major advantage is that the DOPO frequency combs generated by cavity solitons comes automatically in couples, one at the frequency of the pump field $2\omega$ and one at the frequency of the signal field $\omega$. These are known as dual frequency combs and can present major advantages  in covering region of the spectrum that are difficult to reach (such as mid-IR) and in spectroscopy \cite{Coddington16}. Frequency combs in DOPOs have been recently reported via modulational instability instead of cavity solitons \cite{Mosca18}. Generalizations of the numerical simulations reported here are useful for the cases of non-degenerate OPOs \cite{Santagiustina02}, lasers with intracavity parameteric down conversion and cavity-enhanced second harmonic generation \cite{Miro16}.

\section{Conclusions}\label{sec:Conclusions}
The development and application of lasers and photonic devices has gone hand in hand with computer simulations of their structure, output and performance since the very first laser device in 1960. The reason is that the physical principles behind laser action and light-matter interaction are intrinsically nonlinear for the fundamental variables of the electric field, the material polarization field and the population inversion. Starting from the theoretical reproduction of the relaxation oscillations observed in the very first realization of a ruby laser by T. Maiman, computer simulations of photonic devices have spanned the last 65 years with incredible accuracy, predictive power and novel insights. Here we have reviewed only some of the most striking stationary, temporal and spatio-temporal dynamics of model equations that describe a variety of photonic devices via relatively simple numerical integration techniques.

Lasers can display not just relaxation oscillations but also sustained, quasi-periodic and chaotic oscillations. We have reviewed some of these features in lasers with modulated losses and in lasers with the injection of an external signal. We have also seen that the onset of these dynamical behaviours is related to nonlinear bifurcations of stationary and periodic states, the first and more important of all being the laser threshold corresponding to the emission of coherent, directional and monochromatic light. These kind of nonlinear features and complex temporal dynamics are not just constrained to laser systems but are also displayed by other nonlinear photonic devices such as saturable absorbers, Kerr cavities, optical parametric oscillators and second harmonic generators with applications to pulse generation, optical communications and neuromorphic networks. The numerical integration of all these model equations (apart from second harmonic cases) have been shown here to be relatively simple when using standard numerical techniques based on Runge-Kutta methods for ordinary differential equations.

We have then moved to spatio-temporal systems in photonics described by partial differential equations. The physical phenomena described here by the addition of the spatial derivatives are specifically diffraction and group velocity dispersion and come directly from Maxwell's equations for the electromagnetic field. New features appear such as solitons during propagation of light in Kerr media and cavity solitons for light interacting with a medium inside  an optical resonator. From their very first numerical prediction made at Strathclyde by Andrew Scroggie and Willie Firth for Kerr media in optical cavities in 1994, cavity solitons have grown enormously in popularity and application thanks also to their utilization in the generation of frequency combs that have revolutionised modern photonic research. Since 2010, the number of applications of photonic devices using cavity-soliton-based optical frequency combs has exploded: wave demultiplexing in optical communications, frequency standards, optical clocks, future GPS, astrocombs and quantum optic technologies. We expect these to continue to grow in the coming decades as technologies for integrated devices further develop. Cavity solitons are ideal for the processes of miniaturization and device integration.  

Finally we have seen the generalization of nonlinear oscillations and deterministic chaos to spatio-temporal configurations of photonic devices with stable and unstable Turing patterns, spatial disorder, spatio-temporal chaos and defect mediated turbulence in models of lasers and optical parametric oscillators. Photonic systems are not just relevant in science for their hundreds of engineering and technological applications but also for their accurate and unequalled ability to investigate fundamental physical phenomena. 

The aim of this review is to excite young scientists and new researchers to embrace computational physics applied to photonic devices by showing that the numerical methods are relatively simple and that the codes are easy to develop and even simpler to use on modern laptops. The history and success of lasers and other photonic devices are a testament to the development and application of powerful numerical techniques for small-scale computing that can provide insights into the behaviour of new lasers, new nonlinear and new quantum optical devices.  We are ready for further successes and societal benefits provided by computational photonics in the years to come. 

\newpage
\section{Acknowledgements and Declarations}
I would like first to thank my teachers and mentors Antonio Politi, Ray Kapral, Lorenzo Narducci, Tito Arecchi, and Willie Firth. Without their teachings and patience none of the research work presented here would have ever happened. During the years I have benefited from research collaborations and powerful insights from Andrew Scroggie, Graeme Harkness, Alison Yao, Gordon Robb, Thorsten Ackemann, Giampaolo D'Alessandro, Francesco Papoff, Gian-Luca Lippi, Andrew Kent, Damia Gomila, Graeme Campbell, Lewis Hill, Craig McIntyre, Stephane Coen, Miro Erkintalo, Stuart Murdoch, Julien Fatome, Alessia Pasquazi, Marco Peccianti, Luigi Lugiato, Giovanna Tissoni, Franco Prati, Massimo Brambilla, Pascal Del'Haye, Mansour Eslami, Jorge Tredicce, Massimo Giudici, Stephane Barland, Joceline Lega, Graham McDonald, Dmitry Skryabin, and many others. I would also like to thank the European Union for the financial support provided via several research networks (PIANOS, FunFACS, ColOpt, QStruct, QuantIM) and grant C299792458MS while also thanking European co-workers on these projects, too many to be named here. 

\newpage
\section{Appendix I}\label{sec:AppI}
In this Appendix we briefly review the standard numerical methods for the integration of nonlinear ordinary differential equations (ODEs). These are based on Runge-Kutta techniques \cite{NumRecipes}. ODEs are equations which involve the derivatives of a function (or functions) of a set of coupled variables arranged in an array $\vec{y}$.
A general form of ODEs is
\begin{equation}
\frac{d \vec{y}}{dt} = F \left( \vec{y}(t),\mu \right)
\label{ODEs}
\end{equation}
where $F$ is a nonlinear function of the variables, $\vec{y}(t)$ is a set of variables that are changing with $t$ the time, and $\mu$ is a set of parameters that are fixed to given numerical values. Note that $\vec{y}$ can also contain complex variables that can be split in their real and imaginary parts. It is possible to solve linear ODEs but in general exact solutions of nonlinear ODEs are not available. The solutions of nonlinear ODEs can change drastically upon changing the parameters $\mu$ and may even be effectively unpredictable, as in the case of deterministic chaos. This is why computational methods are necessary. 

We start with the easiest one. The Eurler's method. Given an initial condition of $\vec{y}(t)$ at time $t$, it is possible to approximate the solution at time $t+dt$ where $dt \ll 1$ by using
\begin{align}
\frac{d \vec{y}}{dt} &= F\left( \vec{y}(t),\mu \right) \approx \frac{\vec{y}(t+dt)-\vec{y}(t)}{dt} \nonumber \\
\vec{y}(t+dt) &= \vec{y}(t) + dt F\left( \vec{y}(t),\mu \right) + {\rm HOT}(dt^2) \label{Euler}
\end{align}
where ${\rm HOT}(dt^2)$ are Higher Order Terms in a power series starting with terms in $dt^2$. If $dt$ is small enough the ${\rm HOT}(dt^2)$ are negligible and the right hand side of Eq. (\ref{Euler}) can be numerically evaluated. 

As the Numerical Recipes book says: "There are several reasons that Euler's method is not recommended for practical use, among them, (i) the method is not very accurate when compared with other, fancier, methods run at the equivalent stepsize, and (ii) neither is very stable" \cite{NumRecipes}. We then proceed to the next order by using a Taylor expansion
\begin{align}
\vec{y}(t+dt) &= \vec{y}(t) + dt \frac{d \vec{y}}{dt} + \frac{dt^2}{2} \frac{d^2 \vec{y}}{dt^2} + {\rm HOT}(dt^3) = \nonumber \\
 &= \vec{y}(t) + dt \frac{d \vec{y}}{dt} + \frac{dt^2}{2} \frac{d F}{d \vec{y}} \frac{d \vec{y}}{dt}  + {\rm HOT}(dt^3)
\label{RK2a}
\end{align}
However, the term $d F/ d \vec{y}$ can be approximated via
\begin{equation}
\frac{d F}{d \vec{y}} \approx \frac{ F \left( \vec{y}(t) + \frac{dt}{2} F(\vec{y}(t)) \right) - F(\vec{y}(t))}{(dt/2) F(\vec{y}(t))}
\end{equation}
so that
\begin{equation}
F \left( \vec{y} \left(t+\frac{dt}{2} \right) \right) \approx F \left(\vec{y}(t) + \frac{dt}{2} \frac{d \vec{y}}{dt} \right) \approx F + \frac{dt}{2} \frac{d F}{d \vec{y}} F \, .
\end{equation}
Replacing the last term in Eq. (\ref{RK2a}) we get
\begin{equation}
\vec{y}(t+dt) = \vec{y}(t) + dt \, F \left( \vec{y}(t) + \frac{dt}{2} F(\vec{y}(t)) \right) + {\rm HOT}(dt^3) \, .
\label{RK2}
\end{equation}
This is the second order Runge-Kutta method. It shows that in order to increase accuracy, one needs to evaluate the derivatives of the vector $\vec{y}$, i.e. the function $F$, twice, first at time $t$ and then basically at time $t+dt/2$. We wrote the second order Runge-Kutta method explicitly because we are going to use it in Appendix II to implement solutions of PDEs via the split-step method.

In many applications of numerical solutions of ODEs one wants even higher accuracies. Suppose that $dt=10^{-3}$, a fourth order Runge-Kutta method can easily reach accuracies of around $10^{-12}$. Without entering into the details of the derivation, the formulas for the fourth order Runge-Kutta method are \cite{NumRecipes}
\begin{equation}
\vec{y}(t+dt) = \vec{y}(t) + dt \, \left( \frac{k_1}{6} + \frac{k_2}{3} + \frac{k_3}{3} + \frac{k_4}{6} \right) + {\rm HOT}(dt^5) 
\label{RK4}
\end{equation}
where
\begin{align}
k_1 &= F \left( \vec{y}(t),\mu \right) \;\;\;\;\;\;\;\;\;\;\;\;\;\;\;\;\;\;\;\;\;\;\;\;\;\;
k_2 = F \left( \vec{y}(t) + \frac{dt}{2} k_1 \, ,\mu \right) \nonumber \\
k_3 &= F \left( \vec{y}(t) + \frac{dt}{2} k_2 \, ,\mu \right) \;\;\;\;\;\;\;\;\;\;\;
k_4 = F \left( \vec{y}(t) + dt k_3 \, ,\mu \right) \, .
\end{align}
The fourth order Runge-Kutta method increases accuracy by evaluating the function $F$ four times. This is the method that we have used throughout Part A of this review paper and in the Python code below for the integration of the Lorenz-Laser  Eqs. (\ref{Lorenz}) to display deterministic chaos.

\noindent """ \\
Lorenz-Laser Equations
@author: gian-luca oppo  \\
""" \\
import numpy as np $\#$ For arrays \\
import matplotlib.pyplot as plt $\#$ For plotting \\
$\;$ \\
$\#$  Parameters for Lorenz equations \\
$\;$ \\
sigma = 10.0 $\#$ Parameter for dx/dt \\
rho = 28.0 $\#$ Parameter for dy/dt \\
$\#$ beta = 4.5 $\#$ Parameter for dz/dt (damped) \\
beta = 8/3 $\#$ Parameter for dz/dt (chaos) \\
t = 0 $\#$ Starting time \\
tf = 50 $\#$ Ending time \\
h = 0.01 $\#$ Step size for Runge-Kutta 4th order \\
$\;$  \\
$\#$ Derivative function for Runge-Kutta 4th order loop \\
def derivative(r,t): \\
\indent     x = r[0] \\
\indent     y = r[1] \\
\indent     z = r[2] \\
\indent     return np.array([sigma * (y - x), x * (rho - z) - y, (x * y) - (beta * z)]) \\
$\;$ \\
time = np.array([]) $\#$ Empty time array to fill for the x-axis \\
x = np.array([]) $\#$ Empty array for x values \\
y = np.array([]) $\#$ Empty array for y values \\
z = np.array([]) $\#$ Empty array for z values \\
r = np.array([-10.0, -10.0, 10.0]) $\#$ Initial conditions array \\
$\#$ r = np.array([10.001, 10.001, 10.001]) $\#$ Initial conditions array \\
$\;$ \\
while (t $<=$ tf ): \\
\indent         $\#$ Appending values to graph \\
\indent         time = np.append(time, t) \\
\indent         z = np.append(z, r[2]) \\
\indent         y = np.append(y, r[1]) \\
\indent         x = np.append(x, r[0]) \\
\indent         $\#$ Runge-Kutta 4th order method \\
\indent         k1 = h*derivative(r,t) \\
\indent         k2 = h*derivative(r+k1/2,t+h/2) \\
\indent         k3 = h*derivative(r+k2/2,t+h/2) \\
\indent         k4 = h*derivative(r+k3,t+h) \\
\indent         r += (k1+2*k2+2*k3+k4)/6 \\
\indent         $\#$ Updating time value with step size \\
\indent         t = t + h \\
$\;$  \\       
$\#$ Multiple graph plotting \\
fig, (ax1,ax2,ax3) = plt.subplots(1,3, figsize = (15, 5)) \\
ax1.plot(time, x) \\
ax1.set\_title($"$ X versus Time$"$ ) \\
ax2.plot(time, z) \\
ax2.set\_title($"$ Z versus Time$"$ ) \\
ax3.plot(x, z) \\
ax3.set\_title($"$ Z versus X$"$ ) \\
plt.show() \\

\newpage
\section{Appendix II}\label{sec:AppII}
In this Appendix we briefly review the standard numerical methods for the integration of nonlinear partial differential equations (PDEs). Our method of choice is the split-step method that is based on the evaluation of Fourier transforms and inverse Fourier transforms. This method applies successfully to PDEs where the spatial derivatives appear linearly. All the photonic devices described in this review article under the influence of diffraction or group velocity dispersion fall in this category. We consider PDEs of the general type
\begin{equation}
\partial_t \vec{u} = L(\vec{u}) + NL(\vec{u})
\label{PDE}
\end{equation}
where $\vec{u}$ is a set of variables that are changing with time $t$ and space $x$ and can contain complex variables that can be split in their real and imaginary parts, $L(\vec{u})$ is a linear operator that contains the linear terms of the equations and linear spatial derivatives of $\vec{u}$ while $NL(\vec{u})$ contains all the nonlinear and constant terms but no spatial derivative. In the case of the LLE (\ref{LLE}), for example, $L(E) = - (1+i\theta) E  + i  \partial^2_x E$, and $NL(E)= E_{IN} + i |E|^2 E$ for $a=\sigma=1$. A formal solution over the time $dt$ of Eq. (\ref{PDE}) is 
\begin{equation}
\vec{u}(t+dt) = \exp \left( (L+NL) dt \right) \vec{u}(t) \, .
\label{formal}
\end{equation}
If $Ldt$ and $NLdt$ were just numbers, say $Adt$ and $Bdt$ respectively, then we would have
\begin{equation}
\exp \left( (A+B) dt \right) \vec{u}(t) = \exp \left(Adt \right) \exp \left( Bdt \right) \vec{u}(t)
\end{equation}
because the commutator of two numbers $ [A,B]$ is always zero. But $Ldt$ and $NL dt$ are operators and they do not commute. We can then approximate the formal solution (\ref{formal}) with
\begin{equation}
\vec{u}(t+dt) = \exp \left( (L+NL) dt \right) \vec{u}(t) \approx \exp \left(L \, dt \right) \exp \left( NL \, dt \right) \vec{u}(t) \, .
\label{formal2}
\end{equation}
This means that we have assumed that $[L,NL]\vec{u}=0$ which introduces an error of order $dt$. The advantage is that we can now solve these equations in two parts, the linear part and then the nonlinear part. Hence the name of "split-step" for this method. The linear part for the example of the LLE is the equation
\begin{equation}
\partial_t E = - (1+i\theta) E  + i  \partial^2_x E
\end{equation}
that can be solved by taking the Fourier transform of both sides
\begin{equation}
\partial_t \tilde{E} = - (1+i\theta) \tilde{E}  - i  K^2 \tilde{E}
\end{equation}
where the tilde sign denotes Fourier transform of the given variable and $K$ is the spatial wavevector. Once this equation is solved over a time $dt$, one takes the inverse Fourier transform of the updated $E$ and solves the nonlinear equation
\begin{equation}
\partial_t E = E_{IN} + i |E|^2 E
\end{equation}
with, for example, a Runge-Kutta method of the kind we have seen in Appendix I. Since the splitting operator error is large, there is no need to use a high order Runge-Kutta method and a second order one is sufficient to maintain accuracy while performing rapidly on the spatial grid. The split-step method with Fourier transforms for the linear part and a Runge-Kutta method of the second order for the nonlinear part is the numerical technique that we have used throughout Part B of this review paper and in the Python code below for the integration of the LLE (\ref{LLE}) for $a=\sigma=1$ to produce a fabulous cavity soliton.

\noindent """ \\
Lugiato-Lefever Equation in 1D \\
@author: gian-luca oppo \\
""" \\
$\;$  \\ 
import numpy as np \\
from numpy.fft import fft, ifft \\
import matplotlib.pyplot as plt \\
$\;$  \\ 
$\#$ Size \\
N = 512 \\
$\#$ Parameters \\
Is = 1. \\
Delta = 3.6  $\#$ Detuning for one CS\\
$\#$ 0.0 (for Turing Pattern from zero),  \\
PP = np.sqrt(3) \\
P2 = PP * PP \\
$\#$ Coefficients \\
dt = 0.001 \\
kc = 1. \\
lambdac = 2 * np.pi / kc \\
domainwidth = 5 \\
L = domainwidth * lambdac \\
dx = L / N \\
dk = 2 * np.pi / L \\
xaxis = np.linspace(0, L, N) \\
Nout = 400  $\#$ Number of iterations before output \\
tend = 200  $\#$ Duration of the simulation or 200 \\
sz = xaxis.shape \\
U = np.sqrt(Is) * np.ones(N) + 0.02 * np.random.randn(*sz) \\
$\;$  \\
for jx in range(int(N/2-N/10), int(N/2+N/10)):  \\
\indent     U[jx] = 2.0 $\#$ Perturbation for CS. Remove for Turing patterns \\
Q = np.zeros(N, dtype=complex) \\
for j in range(N): \\
\indent    if j $<$ N / 2: \\
\indent       kx = (j) * dk \\
\indent    else: \\
\indent \indent        kx = (j - N) * dk \\
\indent    Q[j] = np.exp(dt * (-1. - 1j * Delta - 1j * (kx * kx))) \\
t = 0 \\
n = 0 \\
j = 1 \\
$\;$  \\ 
while t $<$ tend: \\
\indent    n += 1 \\
\indent    t += dt \\
\indent    $\#$ Fourier Transform \\
\indent    fu = fft(U) \\
\indent    fun = Q * fu \\
\indent    $\#$ Inverse Fourier Transform \\
\indent    U = ifft(fun) \\
\indent    $\#$ Second Order Runge-Kutta method
\indent    Uh = U \\
\indent    dUdt = PP + 1j * np.abs(Uh) ** 2 * Uh \\
\indent    Ut = Uh + dt * dUdt / 2. \\
\indent    dUdt = PP + 1j * np.abs(Ut) ** 2 * Ut \\
\indent    U = Uh + dt * dUdt \\
\indent    $\#$ Output \\ 
\indent    if n $\%$ Nout == 0 or n == 1: \\
\indent \indent        plt.clf() \\
\indent \indent        plt.plot(xaxis, np.abs(U) ** 2, $'$b$'$) \\
\indent \indent        plt.axis([0, L, np.min(np.abs(U) ** 2) - 0.2, np.max(np.abs(U) ** 2) + 0.2]) \\
\indent \indent        plt.xlabel($'$x$'$) \\
\indent \indent        plt.ylabel($'$ abs(U) ** 2 $'$) \\
\indent \indent        plt.pause(0.01) \\
$\;$  \\ 
input($"<$Hit Enter To Close Window$>"$) \\
np.save($'$finalU$'$, U) \\
$\;$  \\ 
print($'$Done.$'$) \\

\newpage

\section*{Statements and Declarations}

\noi {\it {\bf Funding.}} The author declares that no funds, grants, or other support were received during the preparation of this manuscript.\\

\noi {\it {\bf Competing interests.}} The author has no relevant financial or non-financial interests to disclose.\\

\noi {\it {\bf Author contributions.}}  G.-L.O.: conceptualization, data curation, investigation, writing—original draft, writing—review and editing. G.-L.O. gave final approval for publication and agreed to be held accountable for the work performed therein.\\

\noi {\it {\bf Data availability.}} Data underlying the results presented in this paper are not publicly available at this time but may be obtained from the author upon a reasonable request.\\

\noi {\it {\bf Declaration of AI use.}} No AI-assisted technology has ever been used. Ever.\\


\newpage
\begin{thebibliography}{1}

\bibitem{Maiman60}
T. H. Maiman, 
Stimulated Optical Radiation in Ruby,
Nature {\bf 187}, 493 (1960).

\bibitem{Statz60}
H. Statz and G. deMars,
Transients and Oscillation Pulses in Masers,
in {\it Quantum Electronics}, edited by C. H. Townes (Columbia University Press, New York, 1960), p. 650. 

\bibitem{Tang63}
C. L. Tang, H. Statz, and G. deMars,
Spectral Output and Spiking Behavior of Solid-State Lasers,
J. App. Phys. {\bf 34}, 2289 (1963).

\bibitem{Kikuchi59}
C. Kikuchi, J. Lambe, G. Makhov, and R. W. Terhune,
Ruby as a Maser Material,
J. Appl. Phys. {\bf 30}, 1061 (1959).

\bibitem{Chester58}
P. F. Chester, P. E. Wagner, and J. G. Castle Jr.,
Two-Level Solid-State Maser,
Phys. Rev. {\bf 110}, 281 (1958).

\bibitem{Collins60}
R. J. Collins, D. F. Nelson, A. L. Schawlow, W. Bond, C. G. B. Garrett, and W. Kaiser,
Coherence, Narrowing, Directionality, and Relaxation Oscillations in the Light Emission from Ruby, 
Phys. Rev. Letters {\bf 5}, 303 (1960).

\bibitem{Singer60}
See A. Yariv comment in the paper by J. R. Singer,
Maser Oscillator Line Shapes,
in {\it Quantum Electronics}, edited by C. H. Townes (Columbia University Press, New York, 1960), p. 528. 

\bibitem{Hellwarth61}
R. W. Hellwarth, in {\it Advances in Quantum Electronics}, edited by J. R. Singer (Columbia University Press, New York, 1961) p. 334; 
F. J. McClung and R. W. Hellwarth, 
Giant Optical Pulsations from Ruby,
J. Appl. Phys. {\bf 33}, 828 (1962).

\bibitem{Occhialini11}
See https://www.iop.org/about/awards/international-bilateral-awards/giuseppe-occhialini-medal-and-prize-recipients

\bibitem{LugiatoBook}
L. A. Lugiato, F. Prati and M. Brambilla,
{\it Nonlinear Optical Systems},
(Cambridge University Press, Cambridge, 2015).

\bibitem{NarducciAbrahamBook}
L. M. Narducci and N. B. Abraham,
{\it Laser Physics and Laser Instabilities}, 
(World Scientific, Singapore, 1988).

\bibitem{Oppo24}
G.-L. Oppo and W. J. Firth,
Theory and Application of Cavity Solitons in Photonic Devices,
Philosophical Transactions A {\bf 382}, 20230336 (2024).

\bibitem{Note1}
Note that all figures in this review have been obtained ex novo by running simple codes based on the model equations presented in the paper and on numerical methods explained in the Appendices I and II. 

\bibitem{StrogatzBook}
S. H. Strogatz,
{\it Nonlinear Dynamics and Chaos: with Applications to Physics, Biology, Chemistry, and Engineering}, Third Edition (CRC Press, Abingdon, 2024).

\bibitem{Maxwell_1868}
J. Clerk Maxwell, 
On Governors, 
Proceedings of the Royal Society of London, {\bf  16}, 270 (1868).

\bibitem{Turing52}
A. Turing,
The Chemical Basis of Morphogenesis, 
Phil. Trans. Royal Society of London B {\bf 237}, 37 (1952).

\bibitem{Swift77}
J. Swift and P. C. Hohenberg,
Hydrodynamic Fluctuations at the Convective Instability,
Phys. Rev. A {\bf 15}, 319 (1977).

\bibitem{MaxwellChaos}
J. Clerk Maxwell, 
Illustrations of the Dynamical Theory of Gases, 
Philosophical Magazine {\bf 19}, 19 (1860). See also 
B. R. Hunt and J. A. Yorke, 
Maxwell on Chaos, 
Nonlinear Science Today {\bf 3}, 2 (1993).

\bibitem{Poincare_1890}
H. Poincare, 
Sur le probleme des trois corps et les equations de la dynamique, 
Acta Mathematica {\bf 13}, 1 (1890).

\bibitem{Lorenz63}
E. Lorenz,
Deterministic Nonperiodic Flow,
J. Atmospheric Sciences {\bf 20}, 130 (1963).

\bibitem{Oppo84}
G.-L. Oppo and A. Politi,
Collision of Feigenbaum Cascades.
Phys. Rev. A {\bf 30}, 435 (1984).

\bibitem{Coullet87}
P. Coullet, C. Elphick, and D. Repaux,
Nature of Spatial Chaos,
Phys. Rev. Lett. {\bf 58}, 431 (1987). 

\bibitem{Haelterman92}
M. Haelterman, S. Trillo, and S. Wabnitz,
Low Dimensional Modulational Chaos in Diffractive Nonlinear Cavities,
Opt. Comm. {\bf 93}, 343 (1992).  

\bibitem{Gomila03}
D. Gomila and P. Colet,
Transition from Hexagons to Optical Turbulence,
Phys. Rev. A {\bf 68}, 011801(R)(2003).

\bibitem{Lugiato87}
L. A. Lugiato and R. Lefever,
Spatial Dissipative Structures in Passive Optical Systems,
Phys. Rev. Lett. {\bf 58}, 2209 (1987).

\bibitem{Geddes94}
J.B. Geddes, J. Lega, J.V. Moloney, R. A. Indik, E. M. Wright, and W. J. Firth,
Pattern Selection in Passive and Active Nonlinear Optical Systems,
Chaos, Solitons \& Fractals {\bf 4}, 1261 (1994).

\bibitem{Tlidi97}
M. Tlidi, P. Mandel and  M. Haelterman,
Spatiotemporal Patterns and Localized Structures in Nonlinear Optics,
Phys. Rev. E {\bf 56}, 6524 (1997).

\bibitem{DeKepper93}
J.-J. Perraud, A. De Wit, E. Dulos, P. De Kepper, G. Dewel, and P. Borckmans,
One-Dimensional "Spirals": Novel Asynchronous Chemical Wave Sources,
Phys. Rev. Lett. {\bf 71}, 1272 (1993).

\bibitem{Yang02}
L. Yang, M. Dolnik, A. M. Zhabotinsky, and I. R. Epstein,
Spatial Resonances and Superposition Patterns in a Reaction-Diffusion Model with Interacting Turing Modes,
Phys. Rev. Lett. {\bf 88}, 208303 (2002).

\bibitem{Kuptsov12}
P. V. Kuptsov, S. P. Kuznetsov, and A. Pikovsky,
Hyperbolic Chaos of Turing Patterns,
Phys. Rev. Lett. {\bf 108}, 194101 (2012).

\bibitem{Coullet89}
P. Coullet, L. Gil, and J. Lega,
Defect-mediated Turbulence,
Phys. Rev. Lett. {\bf 62}, 1619 (1989).

\bibitem{Chate94}
H. Chat\'e,
Spatiotemporal Intermittency Regimes of the One Dimensional Complex Ginzburg-Landau Equation,
Nonlinearity {\bf 7}, 185 (1994).

\bibitem{Lugiato23}
L. A. Lugiato and F. Prati, 
Fabry–Perot Cavities Made Easy,
Progress in Optics {\bf 68}, 329 (2023).

\bibitem{Maxwell_1865}
J. Clerk Maxwell,
A Dynamical Theory of the Electromagnetic Field, 
Phil. Transactions of the Royal Society of London {\bf 155}, 459 (1865).

\bibitem{Einstein17}
A. Einstein, 
The Quantum Theory of Radiation,
Physikalische Zeitschrift {\bf 18}, 121 (1917).

\bibitem{Haken75}
H. Haken,
Analogy Between Higher Instabilities in Fluids and Lasers, 
Phys. Lett. {\bf 53}A, 77 (1975).

\bibitem{Arecchi84}
F. T. Arecchi, G. L. Lippi, G. P. Puccioni, and J. R. Tredicce,
Deterministic Chaos in Lasers with Injected Signal,
Opt. Comm. {\bf 51}, 308 (1984).

\bibitem{Tredicce85_1}
J. R. Tredicce, F. T. Arecchi, G. L. Lippi, and G. P. Puccioni,
Instabilities in Lasers with an Injected Signal,
J. Opt. Soc. Am. B {\bf 2}, 173 (1985).

\bibitem{Shirley68}
For a comparison with the results of \cite{Statz60} we use Eq. (23) in J. H. Shirley, Dynamics of a Simple Maser Model,
Am. J. Phys. {\bf 36}, 949 (1968) where the time normalization coefficient is $\Gamma=2k/\gamma_\parallel$. 

\bibitem{OpPoliti85}
G.-L. Oppo and A. Politi,
Toda Potential in Laser Equations,
Z. Phys. B {\bf 59}, 111 (1985). See also "Spectral, spatial and temporal properties of lasers" 
by A. M. Ratner (Plenum Press, New York, 1972), where similar ideas have been discussed in a different framework.

\bibitem{Cialdi13}
S. Cialdi, F. Castelli, and F. Prati,
Lasers as Toda Oscillators: An Experimental Confirmation,
Opt. Comm. {\bf 287}, 176 (2013).

\bibitem{OpPoliti86}
G.-L. Oppo and A. Politi,
Improved Adiabatic Elimination in Laser Equations,
Eur. Phys. Lett. {\bf 1}, 549 (1986).

\bibitem{Henry82}
C. H. Henry, 
Theory of the Linewidth of Semiconductor Lasers, 
IEEE J. Quantum Electron. {\bf 18}, 259 (1982).

\bibitem{Oppo09}
G.-L. Oppo, A. M. Yao, F. Prati, and  G. J. de Valc\'arcel,
Long-term Spatiotemporal Dynamics of Solid-State Lasers and Vertical-Cavity Surface-Emitting Lasers,
Phys. Rev A {\bf 79}, 033824 (2009).

\bibitem{Tredicce85_2}
J. R. Tredicce, N. B. Abraham, G. P. Puccioni, and F. T. Arecchi,
On Chaos in Lasers with Modulated Parameters: a Comparative Analysis, 
Opt. Comm. {\bf 55}, 131 (1985).

\bibitem{Arecchi82}
F.T. Arecchi, R. Meucci, G. P. Puccioni and J. R. Tredicce, 
Experimental Evidence of Subharmonic Bifurcations, Multistability, and Turbulence in a Q-Switched Gas Laser
Phys. Rev. Lett. {\bf 49}, 1217 (1982). 

\bibitem{Lugiato83}
L.A. Lugiato, L.M. Narducci, D.K. Bandy, and C.A. Pennise,
Breathing, Spiking and Chaos in a Laser with Injected Signal,
Opt. Comm. {\bf 46}, 64 (1983).

\bibitem{Oppo86}
G.-L. Oppo, A. Politi, G. L. Lippi, and F. T. Arecchi,
Frequency Pushing in Lasers with Injected Signals,
Phys. Rev. A {\bf 34}, 4000 (1986).

\bibitem{Solari94}
H. Solari and G.-L. Oppo,
Laser with Injected Signal: Perturbation of an Invariant Circle,
Opt. Comm. {\bf 111}, 173 (1994).

\bibitem{Adler46}
R. Adler, 
A Study of Locking Phenomena in Oscillators,
Proceedings IRE {\bf 34}, 351 (1946). 

\bibitem{Zimmermann97}
M. G. Zimmermann, M. A. Natiello, and H. G. Solari,
Silnikov-saddle-node Interaction near a Codimension-2 Bifurcation: Laser with Injected Signal,
Physica D {\bf 109}, 293 (1997).

\bibitem{Zimmermann01}
M. G. Zimmermann, M. A. Natiello, and H. G. Solari,
Global Bifurcations in a Laser with Injected Signal: Beyond Adler's Approximation,
Chaos {\bf 11}, 500 (2001).

\bibitem{Wieczorek05}
S. Wieczorek, B. Krauskopf, T. B. Simpson, and D. Lenstra,
The Dynamical Complexity of Optically Injected Semiconductor Lasers,
Physics Reports {\bf 416}, 1 (2005).

\bibitem{Politi86}
A. Politi, G.-L. Oppo and R. Badii,
Coexistence of Conservative and Dissipative Behavior in Reversible Dynamical Systems,
Phys. Rev. A {\bf 33}, 4055 (1986).

\bibitem{Topaj02}
D. Topaj, and A. Pikovsky,
Reversibility vs. Synchronization in Oscillator Lattices,
Physica D {\bf 170}, 118 (2002).

\bibitem{Sprott14}
J. C. Sprott, W. G. Hoover, and C. Griswold Hoover,
Heat Conduction, and the Lack Thereof, in Time-reversible Dynamical Systems: Generalized Nose'-Hoover Oscillators with a Temperature Gradient,
Phys. Rev. E {\bf 89},042914 (2014).

\bibitem{Puccioni87}
G. P. Puccioni, M. V. Tratnik, J. E. Sipe, and G.-L. Oppo,
Low Instability Threshold in a Laser Operating in Both States of Polarization,
Opt. Lett. {\bf 12}, 242 (1987).

\bibitem{Oppo89}
G.-L. Oppo, J. R. Tredicce, and L. M. Narducci,
Dynamics of Vibro-rotational CO$_2$ Laser Transitions in a Two-dimensional Phase Space,
Opt. Comm. {\bf 69}, 393 (1989).

\bibitem{Ciofini93}
M. Ciofini, A. Politi, and R. Meucci
Effective Two-dimensional Model for CO$_2$ Lasers,
Phys. Rev. A {\bf 48}, 605 (1993).

\bibitem{Erzgraber08}
H. Erzgr\"aber, S. Wieczorek, and B. Krauskopf,
Dynamics of Two Laterally Coupled Semiconductor Lasers: Strong- and Weak-coupling Theory,
Phys. Rev. E {\bf 78}, 066201 (2008).

\bibitem{Pashotta09}
R. Paschotta,
Noise in Laser Technology Part 1: Intensity and Phase Noise,
Optik \& Photonik {\bf 2}, 48 (2009).

\bibitem{Kerr1875}
John Kerr, 
A New Relation between Electricity and Light: Dielectrified Media Birefringent, 
Philosophical Magazine and Journal of Science 4th Series, {\bf 50}, 337–348 (1875).

\bibitem{Coen16}
S. Coen and M. Erkintalo, 
Temporal Cavity Solitons in Kerr Media,
in {\it Nonlinear Optical Cavity Dynamics}, P. Grelu (ed.), (Wiley-VCH, Weinheim, 2016) p.~11.

\bibitem{Cole18}
D. C. Cole, A. Gatti, S. B. Papp, F. Prati, and L. Lugiato, 
Theory of Kerr Frequency Combs in Fabry-Perot Resonators,
Phys. Rev. A {\bf 98}, 013831 (2018).

\bibitem{Hill18}
M. T. M. Woodley, J. M. Silver, L. Hill, F. Copie, L. Del Bino, S. Zhang, G.-L. Oppo, and P. Del'Haye,
Universal Symmetry-breaking Dynamics for the Kerr Interaction of Counterpropagating Light in Dielectric Ring Resonators,
Phys. Rev. A {\bf 98}, 053863 (2018).

\bibitem{Zhu23}
G.-L. Zhu, C.-S. Hu, Y. Wu, and X.-Y. L\"u,
Cavity Optomechanical Chaos,
Fundamental Research {\bf 3}, 63 (2023). 

\bibitem{Biancalana11}
V. Grigoriev and F. Biancalana,
Resonant Self-pulsations in Coupled Nonlinear Microcavities,
Phys. Rev. A {\bf 83}, 043816 (2011).

\bibitem{Reynaud89}
S. Reynaud, C. Fabre, E. Giacobino, and A. Heidmann,
Photon Noise Reduction by Passive Optical Bistable Systems,
Phys. Rev. A {\bf 40}, 1440 (1989).

\bibitem{Franken61}
P. A. Franken, A. E. Hill, C. W. Peters, and G. Weinreich,
Generation of Optical Harmonics,
Phys. Rev. Lett. {\bf 7}, 118 (1961).

\bibitem{OPO_Fisher77}
R. Fischer and L. A. Kulevskii,
Optical Parametric Oscillators (Review),
Sov. J. Quantum Electron. {\bf 7}, 135 (1977).

\bibitem{OPO_Majid01}
M. Ebrahimzadeh and M. H. Dunn,
Optical Parametric Oscillators,
in {\it Handbook of Optics, Volume IV, Fiber Optics \& Nonlinear Optics}, M. Bass, J. M. Enoch, E. W. Van Stryland, and W. L. Wolfe editors, (McGraw-Hill, New York, 2021), Chapter 22. 

\bibitem{OPO_Melkonian21}
J.-M. Melkonian, J.-B. Dherbecourt, M. Raybaut and A. Godard,
Optical Parametric Oscillators, 
Photoniques {\bf 110}, 53 (2021).

\bibitem{Lugiato88}
L. A. Lugiato, C. Oldano, C. Fabre, E. Giacobino and R. J. Horowicz,
Bistability, Self-Pulsing and Chaos in Optical Parametric Oscillators,
Il Nuovo Cimento {\bf 10 D}, 959 (1988).

\bibitem{Santagiustina02}
M. Santagiustina, E. Hernandez-Garcia, M. San-Miguel, A. J. Scroggie, and G.-L. Oppo,
Polarization patterns and vectorial defects in type-II optical parametric oscillators,
Phys. Rev. E {\bf 65}, 036610 (2002).

\bibitem{Oppo01}
G.-L. Oppo, A. J. Scroggie, and W. J. Firth,
Characterization, Dynamics and Stabilization of Diffractive Domain Walls and Dark Ring Cavity Solitons in Parametric Oscillators,
Phys. Rev. E {\bf 63}, 066209 (2001).

\bibitem{Phua98}
P. B. Phua, K. S. Lai, R. F. Wu, and  T. C. Chong,
Coupled Tandem Optical Parametric Oscillator (OPO): an OPO within an OPO,
Opt. Lett. {\bf 23}, 1262 (1998).

\bibitem{Aceves97}
A. Aceves, D. D. Holm, G. Kovacic, I. Timofeyev,
Homoclinic Orbits and Chaos in a Second-harmonic Generating Optical Cavity, 
Physics Letters A {\bf 233}, 203 (1997).

\bibitem{Chaturvedi02}
S. Chaturvedi, K. Dechoum, and P. D. Drummond,
Limits to Squeezing in the Degenerate Optical Parametric Oscillator,
Phys. Rev. A {\bf 65}, 033805 (2002). 

\bibitem{AgrawalBook}
G. Agrawal, 
{\it Nonlinear Fiber Optics} (Academic Press, Amsterdam, 2006).

\bibitem{NLSE2005}
B. Malomed, 
{\it Nonlinear Schr\"odinger Equations}, in 
{\it Encyclopedia of Nonlinear Science} S. Alwyn (ed.), (Routledge, New York, 2005) pp. 639–643;
G. Fibich, {\it The Nonlinear Schr\"odinger Equation} (Springer, Heidelberg, 2015).

\bibitem{Schrod26}
E. Schr\"odinger,
An Undulatory Theory of the Mechanics of Atoms and Molecules, 
Phys. Rev. {\bf 28}, 1049 (1926).

\bibitem{Chiao64}
R. Y. Chiao, E. Garmire, and C. H. Townes,
Self-trapping of Optical Beams,
Phys. Rev. Lett. {\bf 13}, 479 (1964).

\bibitem{Gross61}
E. P. Gross,
Structure of a Quantized Vortex in Boson Systems,
Il Nuovo Cimento {\bf 20}, 454 (1961).

\bibitem{Lev61} 
L. P. Pitaevskii,
Vortex Lines in an Imperfect Bose Gas,
Sov. Phys. JETP {\bf 13}, 451 (1961).
 
\bibitem{Anderson95}
M. H. Anderson, J. R. Ensher, M. R. Matthews, C. E. Wieman, and E. A. Cornell,
Observation of Bose-Einstein Condensation in a Dilute Atomic Vapor,
Science {\bf 269}, 198 (1995).

\bibitem{Zakharov72}
V. E. Zakharov and A. B. Shabat,
Exact Theory of Two-dimensional Self-focusing and One-dimensional Self-modulation of Waves in Nonlinear Media, 
Soviet J. Exp. and Theo. Phys. {\bf 34} 62 (1972).

\bibitem{Mollenauer80}
L. F. Mollenauer, R. H. Stolen, and J. P. Gordon,
Experimental Observation of Picosecond Pulse Narrowing and Solitons in Optical Fibers,
Phys. Rev. Lett. {\bf 45}, 1095 (1980).

\bibitem{Barthelemy85}
A. Barthelemy, S. Maneuf, and C. Froehly,
Propagation Soliton et Auto-confinement de Faisceaux Laser par Non Linearit\'e Optique de Kerr (in French),
Opt. Comm. {\bf 55}, 201 (1985).

\bibitem{Aitchison90}
J. S. Aitchison, A. M. Weiner, Y. Silberberg, M. K. Oliver, J. L. Jackel, D. E. Leaird, E. M. Vogel, and P. W. E. Smith
Observation of Spatial Optical Solitons in a Nonlinear Glass Waveguide,
Opt. Lett. {\bf 15}, 471 (1990).

\bibitem{Assanto03}
G. Assanto and M. Peccianti,
Spatial Solitons in Nematic Liquid Crystals,
IEEE J. Q. Elect. {\bf 39}, 13 (2003).

\bibitem{Satsuma74}
J. Satsuma and N. Yajima,
Initial Value Problems of One-Dimensional Self-Modulation of Nonlinear Waves in Dispersive Media,
Supp. Prog. Theo. Phys. {\bf 55} 284 (1974).

\bibitem{Hasegawa73}
A. Hasegawa, and F. Tappert, 
Transmission of Stationary Nonlinear Optical Pulses in Dispersive Dielectric Fibers. II. Normal Dispersion,
Appl. Phys. Lett. {\bf 23}, 171 (1973.

\bibitem{Emplit87}
P. Emplit, J. P. Hamaide, F. Reynaud, C. Froehly, and A. Barthelemy, 
Picosecond Steps and Dark Pulses Through Nonlinear Single Mode Fibres,
Opt. Comm. {\bf 62}, 374 (1987).

\bibitem{Weiner88} 
A. M. Weiner, J. P. Heritage, R. J. Hawkins, R. N. Thurston, E. M. Krischner, D. E. Leaird, and W. J. Tom
linson, 
Experimental Observation of the Fundamental Dark Soliton in Optical Fibers,
Phys. Rev. Lett. {\bf 61}, 2445 (1988).

\bibitem{Kivshar98}
Y. S. Kivshar and B. Luther-Davies,
Dark Optical Solitons: Physics and Applications, 
Physics Reports {\bf 298}, 81 (1998).

\bibitem{Firth98}
W. J. Firth, and G. K. Harkness, 
Cavity Solitons,
Asian J. of Physics {\bf 7}, 665 (1998). 

\bibitem{Ackemann09}
T. Ackemann, W. J. Firth, and G.-L. Oppo,
Fundamentals and Applications of Spatial Dissipative Solitons in Photonic Devices,
Advances In Atomic, Molecular, and Optical Physics {\bf 57}, 323 (2009).

\bibitem{Leo10}
F. Leo, S. Coen, P. Kockaert, S.-P. Gorza, P. Emplit, and M. Haelterman, 
Temporal Cavity Solitons in One-dimensional Kerr Media as Bits in an All-optical Buffer,
Nature Photonics {\bf 4}, 471 (2010).

\bibitem{Firth10}
W. J. Firth, 
Temporal Cavity Solitons Buffering Optical Data, 
Nature Photonics {\bf 4}, 415 (2010).

\bibitem{Firth21}
W. J. Firth, J. B. Geddes, N. J. Karst, and G.-L. Oppo, 
Analytic Instability Thresholds in Folded Kerr Resonators of Arbitrary Finesse,
Phys. Rev. A {\bf 103}, 023510 (2021).

\bibitem{Haelterman92b}
M. Haelterman, S. Trillo, and S. Wabnitz, 
Dissipative Modulation Instability in a Nonlinear Dispersive Ring Cavity,
Opt. Comm. {\bf 91}, 401 (1992).

\bibitem{Castelli17}
F. Castelli, M. Brambilla, A. Gatti, F. Prati, and L. A. Lugiato, 
The LLE, Pattern Formation and a Novel Coherent Source,
Eur. Phys. J. D {\bf 71}, 84 (2017).

\bibitem{Lugiato18}
L. A. Lugiato, F. Prati, M. L. Gorodetsky, and T. J. Kippenberg, 
From the Lugiato–Lefever Equation to Microresonator-based Soliton Kerr Frequency Combs,
Phil. Trans. R. Soc. A {\bf 376}, 20180113 (2018).

\bibitem{Scroggie94}
A. J. Scroggie, W. J. Firth, G. S. McDonald, M. Tlidi, R. Lefever, and L. A. Lugiato, 
Pattern Formation in a Passive Kerr Cavity,
Chaos, Solitons and Fractals, {\bf 4}, 1323 (1994).

\bibitem{Firth96_PS}
W. J. Firth, A. Lord and A. J. Scroggie, 
Optical Bullet Holes,
Physica Scripta {\bf T67}, 12 (1996). 

\bibitem{Firth02}
W. J. Firth, G. K. Harkness, A. Lord, J. M. McSloy, D. Gomila and P. Colet, 
Dynamical properties of two-dimensional Kerr cavity solitons,
J. Opt. Soc. Am. B {\bf 19}, 747 (2002).

\bibitem{Pasquazi18}
A. Pasquazi, M. Peccianti, L. Razzari, D. J. Moss, S. Coen, M. Erkintalo, Y. K. Chembo, T. Hansson, S. Wabnitz, P. Del’Haye, X. Xue, A. M. Weiner, and R. Morandotti, 
Micro-combs: A Novel Generation of Optical Sources,
Phys. Rep. {\bf 729}, 1 (2018).

\bibitem{Harkness02}
G. K. Harkness, W. J. Firth, G.-L. Oppo, and J. M. McSloy, 
Computationally Determined Existence and Stability of Transverse Structures. I. Periodic Optical Patterns,
Phys. Rev. E {\bf 66}, 046605 (2002).

\bibitem{McSloy02}
J. M. McSloy, G. K. Harkness, G.-L. Oppo, and W. J. Firth, 
Computationally Determined Existence and Stability of Transverse Structures. II. Multipeaked Cavity Solitons,
Phys. Rev. E {\bf 66}, 046606 (2002).

\bibitem{Note2}
We note that in the CS literature dark CS are often grey CS without an exact zero of the local field intensity \cite{Oppo24,ParraRivas16}. In principle one should consider a broader class of dark CS containing both grey and black CS with black corresponding to the case of an exact zero of the local field intensity where real and imaginary parts are simultaneously equal to zero. This is however just a labelling issue. 

\bibitem{ParraRivas16}
P. Parra-Rivas, E. Knobloch, D. Gomila, and L. Gelens, 
Dark solitons in the Lugiato-Lefever Equation with Normal Dispersion,
Phys. Rev. A {\bf 93}, 063839 (2016).

\bibitem{HanschHall06}
T. W. H\"ansch, 
Nobel Lecture: Passion for Precision,
Rev. Mod. Phys. {\bf 78}, 1297 (2006); 
J. L. Hall, 
Nobel Lecture: Defining and Measuring Optical Frequencies,
Rev. Mod. Phys. {\bf 78}, 1279 (2006).

\bibitem{Pascal07}
P. Del’Haye, A. Schliesser, O. Arcizet, T. Wilken, R. Holzwarth, and T. J. Kippenberg, 
Optical Frequency Comb Generation from a Monolithic Microresonator,
Nature {\bf 450}, 1214 (2007).

\bibitem{Coen13}
S. Coen, H. G. Randle, T. Sylvestre, and M. Erkintalo, 
Modeling of Octave-spanning Kerr Frequency Combs Using a Generalized Mean-field Lugiato–Lefever Model,
Opt. Lett. {\bf 38}, 37 (2013).

\bibitem{Chembo13}
Y. K. Chembo and C. R. Menyuk, 
Spatiotemporal Lugiato-Lefever Formalism for Kerr-comb Generation in Whispering-gallery-mode Resonators,
Phys. Rev. A {\bf 87}, 053852 (2013).

\bibitem{Herr14}
T. Herr, V. Brasch, J. D. Jost, C. Y. Wang, N. M. Kondratiev, M. L. Gorodetsky and T. J. Kippenberg, 
Temporal Solitons in Optical Microresonators,
Nature Photonics {\bf 8}, 145 (2014).

\bibitem{Columbo21}
L. Columbo, M. Piccardo, F. Prati, L. A. Lugiato, M. Brambilla, A. Gatti, C. Silvestri, M. Gioannini, N. Opačak, B. Schwarz, and F. Capasso, 
Unifying Frequency Combs in Active and Passive Cavities: Temporal Solitons in Externally Driven Ring Lasers,
Phys. Rev. Lett. {\bf 126}, 173903 (2021).

\bibitem{Weiner17}
A. M. Weiner, 
Cavity Solitons Come of Age,
Nature Photonics {\bf 11}, 533 (2017).

\bibitem{Firth96PRL}
W. J. Firth and A. J. Scroggie, 
Optical Bullet Holes: Robust Controllable Localized States of a Nonlinear Cavity,
Phys. Rev. Lett. {\bf 76}, 1623 (1996).

\bibitem{Garbin17}
B. Garbin, Y. Wang, S. G. Murdoch, G.-L. Oppo, S. Coen, and M. Erkintalo, 
Experimental and Numerical Investigations of Switching Wave Dynamics in a Normally Dispersive Fibre Ring Resonator, Eur. Phys. J. D {\bf 71}, 240 (2017).

\bibitem{Garbin20}
B. Garbin, J. Fatome, G. L. Oppo, M. Erkintalo, G. S. Murdoch, and S. Coen, 
Asymmetric Balance in Symmetry Breaking, 
Phys. Rev. Res. {\bf 2}, 023244 (2020).

\bibitem{Garbin21}
B. Garbin, J. Fatome, G.-L. Oppo, M. Erkintalo, S. G. Murdoch, and S. Coen, 
Dissipative Polarization Domain Walls in a Passive Coherently Driven Kerr Resonator, 
Phys. Rev. Lett. {\bf 126}, 023904 (2021).

\bibitem{Xu21}
G. Xu, A. U. Nielsen, B. Garbin, L. Hill, G.-L. Oppo, J. Fatome, S, G. Murdoch, S. Coen, and M. Erkintalo, 
Spontaneous Symmetry Breaking of Dissipative Optical Solitons in a Two-component Kerr Resonator,
Nature Comm. {\bf 12}, 4023 (2021).

\bibitem{Lucas25}
E. Lucas, G. Xu, P. Wang, G.-L. Oppo, L. Hill, P. Del’Haye, B. Kibler, Y. Xu, S. G. Murdoch, M. Erkintalo, S. Coen, and J. Fatome,
Polarization Faticons: Chiral Localized Structures in Self-Defocusing Kerr Resonators,
Phys. Rev. Lett. {\bf 135}, 063803 (2025).

\bibitem{Fan20}
Z. Fan and D. V. Skryabin, 
Soliton Blockade in Bidirectional Microresonators, 
Opt. Lett. {\bf 45}, 6446 (2020).

\bibitem{Woodley21}
M. T. M. Woodley, L. Hill, L. Del Bino, G.-L. Oppo, and P. Del'Haye, 
Self-Switching Kerr Oscillations of Counterpropagating Light in Microresonators, 
Phys. Rev. Lett. 126, 043901 (2021).

\bibitem{Campbell22}
G. N. Campbell, S. Zhang, L. Del Bino, P. Del’Haye, and G.-L. Oppo,
Counterpropagating Light in Ring Resonators: Switching Fronts, Plateaus, and Oscillations,
Phys. Rev. A {\bf 106}, 043507 (2022).

\bibitem{Campbell23}
G. N. Campbell, L. Hill, P. Del’Haye, and G.-L. Oppo, 
Dark solitons in Fabry-Pérot resonators with Kerr media and normal dispersion,
Phys. Rev. A {\bf 108}, 033505 (2023).

\bibitem{Hill24}
L. Hill, E.-M. Hirmer, G. Campbell, T. Bi, P. Del’Haye, and G.-L. Oppo,
Symmetry Broken Vectorial Kerr Frequency Combs from Fabry-Pérot Resonators,
Comm. Phys. {\bf 7}, 82 (2024). 

\bibitem{Miller15}
S. A. Miller, Y. Okawachi, S. Ramelow, K. Luke, A. Dutt, A. Farsi, A. L. Gaeta, and M. Lipson,
Tunable Frequency Combs based on Dual Microring Resonators,
Optics Express {\bf 23}, 21527 (2015).

\bibitem{Komagata21}
K. Komagata, A. Tusnin, J. Riemensberger, M. Churaev, H. Guo,, A. Tikan, and T. J. Kippenberg,
Dissipative Kerr Cavity Solitons in a Photonic Dimer on Both Sides of Exceptional Point,
Commun. Phys. {\bf 4}, 159 (2021).

\bibitem{Pal24}
A. Pal, A. Ghosh, S. Zhang, L. Hill, H. Yan, H. Zhang, T. Bi, A. Alabbadi, and P. Del’Haye, 
Linear and Nonlinear Coupling of Light in Twin-resonators with Kerr Nonlinearity, 
Photon. Res. {\bf 12}, 2733 (2024).

\bibitem{Oppo22}
G.-L. Oppo, D. Grant, and M. Eslami, 
Temporal Cavity Solitons and Frequency Combs via Quantum Interference, 
Phys. Rev. A {\bf 105}, L011501 (2022).

\bibitem{Esalmi23}
M. Eslami, M. S. Hashemi, D. Grant, and G.-L. Oppo,
Control of Quantum Interference Frequency Combs: Multistable Temporal Cavity Solitons,
Phys. Rev. A {\bf 108}, 053511 (2023).

\bibitem{Jang16}
J. K. Jang, M. Erkintalo, K. Luo, G.-L. Oppo, S. Coen, and S. G. Murdoch, 
Controlled Merging and Annihilation of Localised Dissipative Structures in an AC-driven Damped Nonlinear Schr\"odinger System,
New J. Phys. {\bf 18}, 033034 (2016).

\bibitem{Seibold22}
K. Seibold, R. Rota, F. Minganti, and V. Savona
Quantum Dynamics of Kerr Cavity Solitons,
Phys. Rev. A {\bf 105}, 053530 (2022).

\bibitem{Bao19}
H. Bao, S. T. Chu, A. Cooper, M. Rowley, B. E. Little, G.-L. Oppo, R. Morandotti, D. J. Moss, B. Wetzel, M. Peccianti, and A. Pasquazi,
Laser Cavity-soliton Microcombs,
Nature Photon. {\bf 13}, 384 (2019).

\bibitem{Tanguy08}
Y. Tanguy, T. Ackemann, W. J. Firth, and R. J\"ager, 
Realization of a Semiconductor-based Cavity Soliton Laser, 
Phys. Rev. Lett. {\bf 100}, 013907 (2008).

\bibitem{Genevet08}
P. Genevet, S. Barland, M. Giudici, and J. R. Tredicce, 
Cavity Soliton Laser Based on Mutually Coupled Semiconductor Microresonators,
Phys. Rev. Lett. {\bf 101}, 123905 (2008).

\bibitem{Scroggie09}
A. J. Scroggie, W. J. Firth, and G.-L. Oppo, 
Cavity-soliton Laser with Frequency Selective Feedback, 
Phys. Rev. A {\bf 80}, 013829 (2009).

\bibitem{Marconi14}
M. Marconi, J. Javaloyes, S. Balle, and M. Giudici, 
How Lasing Localized Structures Evolve Out of Passive Mode Locking, 
Phys. Rev. Lett. {\bf 112}, 223901 (2014).

\bibitem{Gustave17}
F. Gustave, N. Radwell, C. McIntyre, J. P. Toomey, D. M. Kane, S. Barland, W. J. FIrth, G.-L. Oppo, and T. Ackemann,
Observation of Mode-locked Spatial Laser Solitons, 
Phys. Rev. Lett. {\bf 118}, 044102 (2017).

\bibitem{Rowley22}
M. Rowley, P.-H. Hanzard, A. Cutrona, H. Bao, S. T. Chu, B. E. Little, R. Morandotti, D. J. Moss, G.-L. Oppo, J. S. Totero-Gongora, M. Peccianti, and A. Pasquazi,
Self-emergence of Robust Solitons in a Microcavity,
Nature {\bf 608}, 303 (2022).

\bibitem{Cutrona21}
A. Cutrona, P.-H. Hanzard, M. Rowley, J. S. Totero-Gongora, M. Peccianti, B. A. Malomed, G.-L. Oppo, and A. Pasquazi,
Temporal Cavity Solitons in a Laser-based Microcomb: a Path to a Self-starting Pulsed Laser Without Saturable Absorption,
Optics Express {\bf 29}, 6629 (2021).

\bibitem{Bao20}
H. Bao, L. Olivieri, M. Rowley, S. T. Chu, B. E. Little, R. Morandotti, D. J. Moss, J. S. Totero-Gongora, M. Peccianti, and A. Pasquazi,
Turing Patterns in a Fiber Laser with a Nested Microresonator: Robust and Controllable Microcomb Generation,
Phys. Rev. Research {\bf 2}, 023395 (2020).

\bibitem{Cutrona23}
A. Cutrona, M. Rowley, A. Bendahmane, V. Cecconi, L. Peters, L. Olivieri, B. E. Little, S. T. Chu, S. Stivala, R. Morandotti, D. J. Moss, J. S. Totero-Gongora,M. Peccianti, and A. Pasquazi, 
Stability of laser cavity-solitons for metrological applications, 
Applied Physics Letters {\bf 122}, 121104 (2023).

\bibitem{Pfeifle15}
J. Pfeifle, A. Coillet, R. Henriet, K. Saleh, P. Schindler, C. Weimann, W. Freude, I. V. Balakireva, L. Larger, C. Koos, and Y. K. Chembo, 
Optimally Coherent Kerr Combs Generated with Crystalline Whispering Gallery Mode Resonators for Ultrahigh Capacity Fiber Communications,
Phys. Rev. Lett. {\bf 114}, 093902 (2015).

\bibitem{Chate99}
H. Chate', A. Pikovsky, and O. Rudzick,
Forcing Oscillatory Media: Phase Kinks vs. Synchronization,
Physica D {\bf 131}, 17 (1999).

\bibitem{Gibson16}
C. J. Gibson, A. M. Yao, and G.-L. Oppo,
Optical Rogue Waves in Vortex Turbulence,
Phys. Rev. Lett. {\bf 116}, 043903 (2016).

\bibitem{Oppo22csf}
G.-L. Oppo,
Defect-mediated Turbulence in Lasers with Injected Signals,
Chaos, Solitons and Fractals {\bf 155}, 111750 (2022).

\bibitem{Oppo94}
G.-L. Oppo, M. Brambilla, and L. A. Lugiato, 
Formation and Evolution of Roll Patterns in Optical Parametric Oscillators, 
Phys. Rev. A {\bf 49}, 2028 (1994).

\bibitem{Oppo99}
G.-L. Oppo, A. J. Scroggie, and W. J. Firth,
From Domain Walls to Localized Structures in Degenerate Optical Parametric Oscillators,
J. Opt. B: Quantum Semiclass. Opt. {\bf 1}, 133 (1999).

\bibitem{Parra19}
P. Parra-Rivas, L. Gelens, T. Hansson, S. Wabnitz, and F. Leo, 
Frequency Comb Generation Through the Locking of Domain Walls in Doubly Resonant Dispersive Optical Parametric Oscillators, 
Opt. Lett. {\bf 44}, 2004 (2019).

\bibitem{Trillo97}
S. Trillo, M. Haelterman, and A. Sheppard,
Stable Topological Spatial Solitons in Optical Parametric Oscillators,
Optics Lett. {\bf 22}, 970 (1997).

\bibitem{Coullet90}
P. Coullet, J. Lega, B. Houchmandzadeh, and J. Lajzerowicz
Breaking Chirality in Nonequilibrium Systems,
Phys. Rev. Lett. {\bf 65}, 1352 (1990).

\bibitem{Coddington16}
I. Coddington, N. Newbury, and W. Swann, 
Dual-comb Spectroscopy,
Optica {\bf 3}, 414 (2016).

\bibitem{Mosca18}
S. Mosca, M. Parisi, I. Ricciardi, F. Leo, T. Hansson, M. Erkintalo, P. Maddaloni, P. De Natale, S. Wabnitz, and M. De Rosa, 
Modulation Instability Induced Frequency Comb Generation in a Continuously Pumped Optical Parametric Oscillator, 
Phys. Rev. Lett. {\bf 121}, 093903 (2018).

\bibitem{Miro16}
F. Leo, T. Hansson, I. Ricciardi, M. De Rosa, S. Coen, S. Wabnitz, and M. Erkintalo,
Walk-Off-Induced Modulation Instability, Temporal Pattern Formation, and Frequency Comb Generation in Cavity-Enhanced Second-Harmonic Generation,
Phys. Rev. Lett. {\bf 116}, 033901 (2016).

\bibitem{NumRecipes}
W. H. Press, S. A. Teukolsky, W. T. Vetterling, and B. P. Flannery. 
{\it Numerical Recipes: the Art of Scientific Computing}, 
Third Edition (Cambridge University Press, Cambridge, 2007).

\end{thebibliography}
\end{document}